\documentclass[11pt]{article}
\usepackage{natbib}
\usepackage{amsmath,amstext,amssymb,amsfonts,amscd,color}
\usepackage{graphicx}
\usepackage{enumitem,epsfig}
\usepackage[hmargin={1.0in, 1.0in}, vmargin={1.2in, 1.2in}]{geometry}
\usepackage{threeparttable}
\usepackage{cases,color}
\usepackage[table,dvipsnames]{xcolor}
\usepackage{verbatim}
\usepackage{booktabs}
\usepackage{multirow}
\usepackage{multicol}
\usepackage{diagbox}
\usepackage{bbm}
\usepackage{amsthm}
\usepackage{bm}
\usepackage{authblk}
\usepackage{subcaption}
\usepackage{algorithm}
\usepackage{algpseudocode}

\usepackage[colorlinks,citecolor=blue,urlcolor=blue,linkcolor=blue]{hyperref}
\allowdisplaybreaks[4]

\newtheorem{theorem}{Theorem}[section]
\newtheorem{cor}{Corollary}[section]
\newtheorem{example}{Example}[section]
\newtheorem{rmk}{Remark}[section]
\newtheorem{lemma}{Lemma}[section]
\newtheorem{defi}{Definition}[section]

\newtheorem{prop}{Proposition}[section]
\newtheorem{assumpt}{Assumption}[section]

\newenvironment{Proof}{{\noindent \it Proof:\quad}}{\hfill $\fbox{}$ \vspace*{5mm}}

\numberwithin{equation}{section}
\numberwithin{figure}{section}
\numberwithin{table}{section}

\newcommand{\ca}{\mathcal{A}}
\newcommand{\cb}{\mathcal{B}}

\newcommand{\cd}{\mathcal{D}}

\newcommand{\cf}{\mathcal{F}}
\newcommand{\cg}{\mathcal{G}}
\newcommand{\ch}{\mathcal{H}}
\newcommand{\ci}{\mathcal{I}}

\newcommand{\cl}{\mathcal{L}}

\newcommand{\cn}{\mathcal{N}}
\newcommand{\cp}{\mathcal{P}}

\newcommand{\cs}{\mathcal{S}}
\newcommand{\ct}{\mathcal{T}}

\newcommand{\cv}{\mathcal{V}}

\newcommand{\cx}{\mathcal{X}}
\newcommand{\cy}{\mathcal{Y}}
\newcommand{\cz}{\mathcal{Z}}

\newcommand{\bd}{\mathbb{D}}
\newcommand{\be}{\mathbb{E}}
\newcommand{\bg}{\mathbb{G}}
\newcommand{\bh}{\mathbb{H}}
\newcommand{\bn}{\mathbb{N}}
\newcommand{\bp}{\mathbb{P}}

\newcommand{\br}{\mathbb{R}}

\newcommand{\bone}{\mathbb{I}}

\newcommand{\Var}{\mbox{Var}}

\newcommand{\Cov}{\mbox{Cov}}

\newcommand{\pell}{(\ell)}
\newcommand{\pmell}{(-\ell)}

\newcommand{\indept}{\perp\!\!\!\perp}
\newcommand{\stra}[1]{\stackrel{#1}{\longrightarrow}}

\newcommand{\stsim}[1]{\stackrel{#1}{\sim}}

\newcommand{\bbe}[1]{\mathbb{E}\big[{#1}\big]}

\newcommand{\blre}[1]{\mathbb{E}\left[{#1}\right]}
\newcommand{\blreft}[2]{\mathbb{E}_{#1}\left[{#2}\right]}

\newcommand{\bbp}[1]{\mathbb{P}\big({#1}\big)}

\newcommand{\blrp}[1]{\mathbb{P}\left({#1}\right)}

\newcommand{\bigp}[1]{\big({#1}\big)}
\newcommand{\lrp}[1]{\left({#1}\right)}
\newcommand{\bigcp}[1]{\big\{{#1}\big\}}
\newcommand{\lrcp}[1]{\left\{{#1}\right\}}

\newcommand{\lrbk}[1]{\left[{#1}\right]}
\newcommand{\lrag}[1]{\left\langle{#1}\right\rangle}
\newcommand{\lrabs}[1]{\left|{#1}\right|}
\newcommand{\lrnorm}[1]{\left\|{#1}\right\|}
\newcommand{\lrfl}[1]{\lfloor{#1}\rfloor}

\newcommand{\BE}{\begin{equation}}
\newcommand{\EE}{\end{equation}}
\newcommand{\BEqn}{\begin{eqnarray*}}
\newcommand{\EEqn}{\end{eqnarray*}}
\newcommand{\benu}{\begin{enumerate}[label=(\roman*)]}
\newcommand{\eenu}{\end{enumerate}}
\DeclareMathOperator*{\argmax}{argmax}

\def\spacingset#1{\renewcommand{\baselinestretch}%
{#1}\small\normalsize}
\spacingset{1.25}

\begin{document}
\title{Testing Equality of Conditional Distributions via Generative Models}

\author[1]{Hanjia Gao}
\author[2]{Linjun Huang}
\author[3]{Yun Yang}
\author[4]{Xiaofeng Shao}
\affil[1]{Department of Statistics, University of California, Irvine}
\affil[2]{Department of Statistics, University of Illinois at Urbana-Champaign} 
\affil[3]{Department of Mathematics, University of Maryland, College Park}
\affil[4]{Department of Statistics and Data Science, and Department of Economics, Washington University in St Louis}
\date{}	
\maketitle

\begin{abstract}
We study the problem of testing whether two conditional distributions are equal using generative models. The proposed method learns a conditional generator from each sample and uses it to create responses at covariate values observed in the other sample, allowing generated and observed responses to be compared directly. By aligning covariates through cross-generation, the approach avoids conditional density-ratio estimation and local smoothing over high-dimensional covariates. The population version of this construction yields a conditional discrepancy that characterizes equality of the two conditional distributions under suitable overlap conditions, while the sample version leads to a test statistic defined as the supremum of an RKHS-indexed empirical process with multiplier bootstrap calibration. A computationally efficient algorithm for evaluating the statistic and its bootstrap analogue is developed based on alternating maximization and the kernel trick. Theoretically, we derive the limiting distribution of the test statistic under both the null and alternative hypotheses, prove bootstrap validity and consistency of the resulting test, and show that the proposed procedure attains a double-robustness property with respect to conditional generator estimation errors. Simulations and real data applications suggest that the proposed method performs well for multivariate responses and high-dimensional covariates.
\end{abstract}

\section{Introduction}

In this article, we study the problem of testing the equality of two conditional distributions. Specifically, let $(X_1,Y_1) \sim P_{X_1,Y_1}$ and $(X_2,Y_2) \sim P_{X_2,Y_2}$ be two joint distributions with $X_1,X_2\in\cx\subseteq\br^{p}$ and $Y_1,Y_2\in\cy\subseteq\br^{q}$. Let $P_{Y_1|X_1}(\cdot|x)$ and $P_{Y_2|X_2}(\cdot|x)$ denote the conditional distributions of $Y_1$ given $X_1=x$ and $Y_2$ given $X_2=x$, respectively. Our goal is to test whether these two conditional distributions coincide, i.e.
\begin{equation*}
    H_0: P_{Y_1|X_1}(\cdot|x) = P_{Y_2|X_2}(\cdot|x)
    \mbox{ for almost every } x\in\cx
    \quad\mbox{versus}\quad
    H_1: \mbox{otherwise}.
\end{equation*}
Testing the equality of two conditional distributions is a fundamental problem that arises in many areas of modern statistics, econometrics, and machine learning. In causal inference, invariant causal prediction \cite{PetersBuhlmannMeinshausen2016ICP} exploits the principle that the conditional distribution of an outcome given its causal parents remains invariant across different environments. In machine learning, closely related questions appear in the study of distributional fairness, where one seeks to assess whether predictive distributions differ across protected groups after conditioning on relevant features \cite{cherian2024statistical}. In transfer learning and domain adaptation, comparing conditional distributions across environments provides a principled way to detect distribution shift and evaluate whether models trained in one domain remain valid in another \cite{Kouw2018}. Similar questions also arise in the validation of scientific simulators, where one aims to determine whether simulated data reproduce the conditional behavior of real observations, as well as in model diagnostics for regression and generative models \cite{chatterjee2024}.

In the early literature, one common strategy is to pursue a weaker hypothesis by testing the equality of conditional moments; see \cite{hall1990bootstrap, kulasekera1995comparison, kulasekera1997smoothing, fan1998test, neumeyer2003nonparametric}. While such methods are computationally convenient, they may fail to capture discrepancies in higher-order moments or other aspects of the conditional distributions, such as tail behavior, multimodality, dependence structure, or support. Another related strategy is to generalize unconditional two-sample distribution tests to the conditional setting by formulating the problem as a goodness-of-fit test for conditional distributions; see \cite{andrews1997conditional, zheng2000consistent, fan2006nonparametric, delgado2008distribution}. These methods often require smoothness assumptions on the conditional density or regression functions and involve nonparametric density estimation as an intermediate step, which can suffer from the curse of dimensionality when the dimension of the conditioning variable is moderately high.

In many contemporary pipelines, data are generated or evaluated conditional on high-dimensional contextual variables or learned representations \cite{zhao2021diagnostics,dey2022calibrated}. This makes it essential to assess whether conditional laws coincide, both for validating distributional alignment and for evaluating model robustness. Multivariate responses are also common in machine learning and econometrics applications. Despite its broad relevance, testing the equality of conditional distributions remains challenging in practice, especially when the responses are multivariate and the conditioning variables are high-dimensional. These challenges call for new testing procedures that are statistically principled, computationally scalable, and effective in modern high-dimensional settings.

Recently, \cite{hu2024two} proposed a nonparametric testing procedure under the conformal prediction framework, based on a weighted rank-sum statistic. The key idea is to construct a conformity score that measures how likely a sample from one joint distribution is to have been generated from the other. This construction relies on estimating both the marginal density ratio between $X_1$ and $X_2$ and the conditional density ratio between $Y_1 \mid X_1$ and $Y_2 \mid X_2$ using classification-based approaches. However, density-ratio estimation, especially in the conditional setting, can be challenging when the conditioning variable is high-dimensional, and the performance of the test may deteriorate when these quantities are poorly estimated. This issue is especially relevant under covariate shift in the marginal distributions, where some regions of the covariate space $\mathcal X$ may have substantial probability mass under one distribution but only sparse observations under the other, or when the data distribution has a singular structure so that the density ratio may not be well defined; see Section~\ref{Sec:RealData} for numerical evidence. In addition, the rank-based test proposed by \cite{hu2024two} is tailored to univariate responses with independent conditioning variables $X_1$ and $X_2$, and its extension to multivariate responses appears nontrivial.

Along a different line, kernel- and distance-based conditional energy-distance tests have been developed by \cite{chatterjee2024,yan2025distance}. These methods characterize equality of conditional distributions through population discrepancies such as integrated conditional maximum mean discrepancy (MMD) or energy distance. While they avoid explicit conditional density estimation, they typically rely on smoothing or local-neighborhood construction in the space of the conditioning variable. As the dimension of the conditioning variable increases, for example when $X$ consists of moderate- or high-dimensional learned features from modern representation models, such procedures may suffer from unstable calibration and loss of power. This limitation motivates us to seek a formulation in which the comparison is carried out in the response space, without local smoothing over the covariate space.

In this work, we adopt such a perspective by combining kernel- and distance-based testing with samples from learned conditional generators. The main idea is to use conditional generative models to approximate the laws of $Y_1 \mid X_1$ and $Y_2 \mid X_2$, and then compare the resulting generated responses through kernel or distance discrepancies. Concretely, we use the fitted conditional generator for $Y_2 \mid X_2$ to generate new responses $Y_2^{\ast}$ at the observed covariate values from $X_1$. In this way, the generated responses follow the estimated conditional law of $Y_2 \mid X_2$ but are evaluated at the same covariates as $Y_1 \mid X_1$, so the two response distributions can be compared directly in the response space without smoothing over $\mathcal X$ to match the covariates. We apply the same construction in the reverse direction, generating $Y_1^{\ast}$ at the observed covariates from $X_2$ using the fitted generator for $Y_1 \mid X_1$. Finally, we combine these two comparisons to obtain a ``double-robustness'' property under the null: at the $\sqrt{n}$-scale, the approximation error of each estimated generator only needs to decay faster than $n^{-1/4}$ to eliminate the impact of slow nonparametric convergence of generator approximation errors on the testing size; see Remark~\ref{rmk:double_ro}.

This construction avoids explicit smoothing or local-neighborhood construction over the conditioning variable. The covariates are used only as conditioning inputs to the fitted generators, while the final discrepancy is computed from Monte Carlo samples in the response space. This is particularly useful when $X$ is high-dimensional but the conditional distribution of the response can still be learned effectively from data. To this end, we use conditional generative models, including mixture density networks (MDNs) \cite{bishop1994mixture,papamakarios2016fast} and conditional diffusion models (CDMs) \cite{zhang2023adding}, to learn conditional generators for $P_{Y_1\mid X_1}$ and $P_{Y_2\mid X_2}$. Both models use neural network architectures to capture complex and potentially multimodal conditional distributions. Specifically, MDNs parameterize the conditional density through a mixture model, while CDMs generate samples through a denoising process guided by the conditioning variables. Although we focus on MDNs and CDMs for concreteness, our framework is model-agnostic and can accommodate other conditional generative models, such as generative adversarial networks (GANs) \cite{goodfellow2014generative}, generative moment matching networks (GMMNs) \cite{dziugaite2015training,li2015generative}, Engression \cite{shen2025engression}, and Wasserstein generative regression \cite{song2025wasserstein}.

Generative modeling techniques have recently been used to facilitate statistical inference involving conditional distributions in a variety of settings, including conditional independence testing \cite{zhang2024doubly,shi2021double,he2025conditional}, variable selection \cite{romano2020deep}, feature-importance evaluation \cite{blesch2025conditional}, and causal inference \cite{chen2025enhancing}. For example, \cite{zhang2024doubly} proposed a doubly robust conditional independence testing procedure based on learned conditional distributions, extending the earlier work of \cite{shi2021double}. While our work also combines generative models with kernel-based statistics and bootstrap calibration, it differs from these works in several important ways. 
First, conditional independence testing is a special case of conditional distribution testing with a shared conditioning variable, since $X \indept Y \mid Z$ if and only if $P_{(X,Y)\mid Z} = P_{X \mid Z} \otimes P_{Y \mid Z}$. By contrast, our formulation allows distinct conditioning variables $X_1$ and $X_2$, which leads to additional identifiability and theoretical challenges.
Second, our test statistic can be viewed as the supremum of an empirical process indexed by an infinite-dimensional RKHS class. In contrast, the test statistic in \cite{shi2021double} is constructed as the maximum of the generalized covariance measures associated with a finite collection of transformation functions, and the test statistic in \cite{zhang2024doubly} extends the classical MMD construction and can be equivalently expressed as a $U$-statistic associated with pre-specified reproducing kernels.
Third, these differences in the statistical construction lead to substantially different theoretical analyses. Our analysis relies on weak convergence and multiplier bootstrap theory for RKHS-indexed empirical processes, while \cite{shi2021double} mainly relies on the argument for Gaussian approximation \cite{chernozhukov2012gaussian} and \cite{zhang2024doubly} primarily relies on $U$-statistic theory.

Our main methodological and theoretical contributions are as follows:
\begin{enumerate}[label=(\roman*)]
    \item First, we introduce a conditional analogue of the MMD in \cite{gretton2012kernel} for testing the equality of two conditional distributions by using conditional generators to create cross-generated responses at the observed covariate values from the opposite sample. Unlike the classical MMD, which directly compares two unconditional distributions, the proposed population discrepancy first aligns the covariates through the cross-generation step and then compares the resulting response distributions. We establish that, under suitable overlap conditions on the covariate distributions, this discrepancy is zero if and only if the two conditional distributions are equal.
    
    \item Second, we propose a feasible test statistic and its multiplier bootstrap counterpart based on the conditional discrepancy metric. Both statistics are defined as suprema of empirical processes indexed by infinite-dimensional RKHS classes. The computation is further complicated by the simultaneous maximization over multiple RKHS kernel blocks, leading to no closed-form solutions. To address this challenge, we develop an alternating maximization algorithm based on the reproducing property of RKHS kernels and the kernel trick, which enables efficient computation of both the feasible statistic and its bootstrap analogue.
    
    \item Third, we quantify the plug-in error induced by estimating the conditional generators for both the feasible statistic and its bootstrap counterpart. Under the null hypothesis, the leading first-order perturbations from the two estimated generators cancel, so that the remaining approximation error only appears through a higher-order interaction term. Consequently, we establish a double-robustness phenomenon that the approximation error of each generator only needs to decay faster than $n^{-1/4}$ on average for the feasible statistic to be asymptotically equivalent to its oracle counterpart at $\sqrt{n}$-scale. In contrast, the bootstrap statistic enjoys asymptotically negligible plug-in error under both the null and alternative hypotheses due to the centering effect of the multiplier bootstrap. These results justify the asymptotic validity of the bootstrap calibration while preserving power against local alternatives.
    
    \item Fourth, we study the asymptotic properties of the proposed supremum statistic and its multiplier bootstrap analogue. The statistic is the supremum of an empirical process indexed by three infinite-dimensional RKHS function classes, so its theoretical analysis requires uniform control of a rich compositional function class rather than a finite-dimensional or closed-form statistic. We derive Rademacher complexity bounds under suitable entropy growth conditions \cite{talagrand1994sharper,talagrand1996new,gine2001consistency}, and use them to establish the weak convergence of the statistic while keeping track of the perturbation error due to conditional generator estimation. Combining these ingredients with classical multiplier bootstrap theory \cite{van1996weak,kosorok2008introduction}, we prove the asymptotic validity of the bootstrap procedure.
    
\end{enumerate}

The article is organized as follows. In Section~\ref{Sec:Method}, we introduce a population-level discrepancy measure for conditional distribution testing and develop the proposed test statistic. Section~\ref{Sec:Method_Computation} provides the bootstrap calibration and the full testing procedure, along with the computation algorithm. Section~\ref{Sec:Theory} establishes the theoretical properties, including the approximation analysis of the test statistic with its oracle counterpart and the asymptotic behavior of the proposed test. In Section~\ref{Sec:Simulation}, we evaluate finite-sample performance through simulation studies, and in Section~\ref{Sec:RealData}, we present real data applications. Section~\ref{Sec:Discussion} concludes with a discussion of limitations and future directions.

\smallskip
\noindent {\bf Notations.} Throughout the article, we use $\stra{d}$ and $\stra{p}$ to denote convergence in distribution and convergence in probability, respectively. Let $\{X_n\}_{n=1}^{\infty}$ be a sequence of random variables. We write $X_n = O_p(r_n)$ if, for any $\varepsilon>0$, there exist constants $M>0$ and $N\geq 1$ such that $\blrp{|X_n/r_n| \geq M} \le \varepsilon$ for all $n\geq N$, and write $X_n = o_p(r_n)$ if $|X_n|/r_n \stra{p} 0$. We write $X_n = \omega(r_n)$ if, for any $M\geq 0$, there exists $N\geq 1$ such that $X_n > M r_n$ for all $n\geq N$. We use $\odot$ to denote the Hadamard product.

\section{Methodology}\label{Sec:Method}

In Section~\ref{Sec:Method_MMD}, we introduce a population-level discrepancy metric that fully characterizes the equality of two conditional distributions. Section~\ref{Sec:Method_TestStat} then develops the corresponding feasible test statistic at the sample level.

\subsection{A Cross-generated RKHS Discrepancy}\label{Sec:Method_MMD}

Throughout this article, we allow $X_1$ and $X_2$ to be dependent, while assuming that $Y_1\indept Y_2 \mid (X_1,X_2)$. By the noise-outsourcing lemma in probability theory (see, e.g., Theorem 5.1 of \cite{kallenberg2002foundations}, Lemma 3.1 of \cite{austin2015exchangeable}, and Lemma 2.1 of \cite{zhou2022deep}), there exist measurable functions $G_j:\br^{p}\times\br^m\to\br^q$ and random vectors $Z_j^{\ast}\in\br^m$ independent of $X_j$, for $j=1,2$, such that 
\begin{equation*}
    (X_1,Y_1) =^d (X_1, G_1^{\ast}(X_1,Z_1^{\ast})), \quad \mbox{and}\quad 
    (X_2,Y_2) =^d (X_2, G_2^{\ast}(X_2,Z_2^{\ast})).
\end{equation*}
Here, $Z_j^{\ast}$ denotes an auxiliary random vector independent of $X_j$, drawn from a specified reference distribution such as the standard multivariate normal distribution.
Thus, $G_1^{\ast}$ and $G_2^{\ast}$ can be interpreted as oracle generators for the conditional distributions $P_{Y_1|X_1}(\cdot|x)$ and $P_{Y_2|X_2}(\cdot|x)$, respectively. In particular, for $P_{X_1}$-almost every $x$ and $P_{X_2}$-almost every $x$, respectively,
\begin{equation*}
    G_1^{\ast}(x,Z_1^{\ast})\sim P_{Y_1|X_1}(\cdot|x),  \quad \mbox{and}\quad 
    G_2^{\ast}(x,Z_2^{\ast})\sim P_{Y_2|X_2}(\cdot|x).
\end{equation*}

To motivate the proposed discrepancy, we first consider the simpler setting in which the two samples share the same conditioning variable, namely $X_1=X_2=:X$. Let $\cf$ denote a sufficiently rich class of measurable functions. In this case, testing the equality of $P_{Y_1|X}$ and $P_{Y_2|X}$ is equivalent to testing the equality of the joint distributions $P_{X,Y_1}$ and $P_{X,Y_2}$, because the marginal distribution of $X$ is common to both samples. Therefore, the conditional distribution testing problem reduces to an ordinary two-sample testing problem for the joint distributions.
When $\cf$ is chosen as the unit ball of an RKHS, the discrepancy between $P_{X,Y_1}$ and $P_{X,Y_2}$ can be measured by the maximum mean discrepancy (MMD), $\sup_{f\in\cf}\lrabs{\be[f(X,Y_1)]-\be[f(X,Y_2)]}$. Under standard conditions on the RKHS kernel, this quantity is zero if and only if the two joint distributions are identical. Since the two joint distributions share the same marginal distribution of $X$, this is further equivalent to the equality of the two conditional distributions. That is,
\begin{equation*}
    \sup_{f\in\cf} \lrabs{\be[f(X,Y_1)]-\be[f(X,Y_2)]} = 0
\end{equation*}
if and only if $P_{Y_1|X}(\cdot|x)=P_{Y_2|X}(\cdot|x)$ for $P_X$-almost every $x$; see \cite{gretton2012kernel} for details.

However, this reduction no longer applies when $X_1$ and $X_2$ are not identically distributed. Even under the null hypothesis $P_{Y_1|X_1}=P_{Y_2|X_2}$ and on the common covariate support, the joint distributions $P_{X_1,Y_1}$ and $P_{X_2,Y_2}$ may still differ because the marginal distributions of $X_1$ and $X_2$ may differ. Consequently, directly applying the classical MMD to the joint samples $(X_1,Y_1)$ and $(X_2,Y_2)$ would confound two sources of discrepancy: differences in the marginal covariate distributions and differences in the conditional response distributions. 

To isolate discrepancies in the conditional distributions, we use the oracle generators to construct cross-generated samples. The basic idea is that, under the null hypothesis, the generator learned from one population should also generate valid responses when evaluated at covariates from the other population. Specifically, let $Z_1^{c \ast}$ and $Z_2^{c \ast}$ be independent copies of the generator noises, independent of $(X_1,Y_1,X_2,Y_2)$, and define
\begin{equation*}
    Y_1^{c\ast} = G_2^{\ast}(X_1,Z_1^{c \ast}), \quad \mbox{and}\quad 
    Y_2^{c\ast} = G_1^{\ast}(X_2,Z_2^{c \ast}).
\end{equation*}
Thus, $Y_1^\ast$ is generated from the second conditional law but evaluated at the covariate $X_1$, while $Y_2^\ast$ is generated from the first conditional law but evaluated at the covariate $X_2$. Under the null hypothesis, these cross-generated responses have the same conditional distributions as the observed responses at the corresponding covariates, and hence
\begin{equation*}
    (X_1, Y_1) =^d (X_1, Y_1^{c\ast}),
    \quad \mbox{and}\quad 
    (X_2, Y_2) =^d (X_2, Y_2^{c\ast}).
\end{equation*}
Let $\ch$ be a sufficiently rich class of measurable functions. Under the null hypothesis, for every $f,g\in\ch$, the centered discrepancies $f(X_1,Y_1)-f(X_1,Y_1^{c\ast})$ and $g(X_2,Y_2)-g(X_2,Y_2^{c\ast})$ both have conditional mean zero given $(X_1,X_2)$. This motivates the covariance-type product 
\begin{equation*}
    \lrp{f(X_1,Y_1)-f(X_1,Y_1^{c\ast})}\cdot\lrp{g(X_2,Y_2)-g(X_2,Y_2^{c\ast})},
\end{equation*}
which has vanishing expectation under the null, while it carries a signal when the two conditional distributions differ.

When $X_1$ and $X_2$ are dependent, however, the unweighted product above may not fully capture all conditional distribution discrepancies. To account for the joint dependence structure of the conditioning variables, we further introduce an interaction function $h(X_1,X_2)$, where $h$ is chosen from another rich function class $\ch'$. This leads to the population-level discrepancy
\begin{equation}\label{Equ:Expectation_Uast}
    \sup\limits_{f,g\in\ch,h\in\ch'} 
    \big|\,\bbe{\bigp{f(X_1,Y_1) - f(X_1,Y_1^{c\ast})} \cdot 
                \bigp{g(X_2,Y_2) - g(X_2,Y_2^{c\ast})} \cdot 
                h(X_1,X_2)}\,\big|.
\end{equation}
The discrepancy in \eqref{Equ:Expectation_Uast} extends the classical MMD idea to the conditional setting through the cross-generated construction. The centered terms compare observed and cross-generated responses at the same covariate values, thereby removing the direct effect of differences between the marginal distributions of $X_1$ and $X_2$. The interaction function $h(X_1,X_2)$ then weights these comparisons according to the joint structure of the conditioning variables, which is essential when $X_1$ and $X_2$ are dependent.

To obtain a discrepancy measure that is rich enough to characterize equality of conditional distributions and, at the same time, computationally tractable, we take $\ch$ and $\ch'$ to be unit balls of reproducing kernel Hilbert spaces (RKHSs). Specifically, let $K:\br^{p+q}\times\br^{p+q}\mapsto\br$ and $K':\br^{2p}\times\br^{2p}\mapsto\br$ be two positive definite kernels. Let $\bh$ denote the RKHS associated with $K$, equipped with inner product $\lrag{\cdot,\cdot}_{\bh}$, and let $\bh'$ denote the RKHS associated with $K'$, equipped with inner product $\lrag{\cdot,\cdot}_{\bh'}$. By the reproducing property, for any $f\in\bh$ and $h\in\bh'$,
\begin{equation}\label{Equ:kernel_trick}
    f(u) = \lrag{f, K(u,\cdot)}_{\bh}, \quad \mbox{and}\quad 
    h(v) = \lrag{h, K'(v,\cdot)}_{\bh'},
    \qquad \forall~ u\in\br^{p+q}, v\in\br^{2p}.
\end{equation}
We then define $\ch$ and $\ch'$ as the corresponding unit balls,
\begin{equation}\label{Equ:UnitBall}
    \ch = \{f\in\bh: \|f\|_{\bh} \le 1\},
    \quad \mbox{and}\quad 
    \ch' = \{h\in\bh': \|h\|_{\bh'}\le 1\}.
\end{equation}
For any $f,g\in\ch$ and $h\in\ch'$, define
{\small\begin{equation}\label{Equ:phi}
    \phi_{f,g,h}(x_1, y_1, y_1', x_2, y_2, y_2') 
  = \bigp{f(x_1,y_1) - f(x_1,y_1')} \cdot \bigp{g(x_2,y_2) - g(x_2,y_2')} \cdot h(x_1, x_2).
\end{equation}}
Let $\Phi=\{\phi_{f,g,h}: f,g\in\ch, h\in\ch'\}$ denote the resulting class of functions, then the population discrepancy in \eqref{Equ:Expectation_Uast} can be written as $\sup_{\phi_{f,g,h}\in\Phi} \big|\be\big[\phi_{f,g,h}(X_1,Y_1,Y_1^\ast,X_2,Y_2,Y_2^\ast)\big]\big|$. This RKHS formulation is useful for two reasons. First, with suitable kernels, the function classes are sufficiently rich to distinguish different conditional distributions. Second, the reproducing property leads to a kernel representation of the empirical statistic, which makes the resulting test computationally feasible. Under some mild conditions, this discrepancy is zero if and only if the null hypothesis holds (c.f.~Theorem~\ref{Thm:Validity}).

\begin{assumpt}\label{Assumpt:kernel}
Let $K: \br^{p+q}\times\br^{p+q}\rightarrow\br$ and $K': \br^{2p}\times\br^{2p}\rightarrow\br$ be continuous kernels. We assume that
\begin{enumerate}[label=(\roman*)]    
    \item \label{Assumpt:kernel_bound}
    The kernels $K$ and $K'$ are uniformly bounded, i.e., 
    \begin{equation*}
        u_{K,K'}
    :\,= \max\{\sup_{z\in\br^{p+q}} K(z,z),
               \sup_{z'\in\br^{2p}} K'(z',z')\} 
    < \infty.
    \end{equation*}
    
    \item \label{Assumpt:kernel_characteristic}
    For every $x\in\cx$, the kernel $K_x(y,y') = K((x,y),(x,y'))$ is characteristic on $\cy$, that is, the kernel mean embedding $\mu_P(\cdot) = \int_{\cy} K_x(y,\cdot) dP(y)$ is injective over all probability measures $P$ on $\cy$.
    
    \item \label{Assumpt:kernel_L2_dense}
    The RKHS $\bh'$ is dense in $L_2(P_{X_1,X_2})$, i.e., for any $\ell \in L_2(P_{X_1,X_2})$ and any $\varepsilon > 0$, there exists $h \in \bh'$ such that $\|\ell-h\|_{L_2(P_{X_1,X_2})} < \varepsilon$.
\end{enumerate}
\end{assumpt}

\begin{rmk}
Assumption~\ref{Assumpt:kernel}\ref{Assumpt:kernel_bound} requires the kernels $K$ and $K'$ to be uniformly bounded on the diagonal, which is standard in RKHS-based empirical process analysis. By the reproducing property, for any $f\in\ch$ and $z\in\br^{p+q}$,
\begin{equation*}
    |f(z)|
\le \|f\|_{\bh} \|K(z,\cdot)\|_{\bh}
  = \|f\|_{\bh} K^{1/2}(z,z).
\end{equation*}
Since $\ch$ is the unit ball of $\bh$, we have $\sup_{f\in\ch} \|f\|_{\infty} \le u_{K,K'}^{1/2}$. Similarly, $\sup_{h\in\ch'}\|h\|_\infty \le u_{K,K'}^{1/2}$. Therefore, Assumption~\ref{Assumpt:kernel}\ref{Assumpt:kernel_bound} implies uniform boundedness of the RKHS classes $\ch$ and $\ch'$.
Assumption~\ref{Assumpt:kernel}\ref{Assumpt:kernel_characteristic}
and Assumption~\ref{Assumpt:kernel}\ref{Assumpt:kernel_L2_dense}
impose two different richness conditions on the RKHS classes
$\ch$ and $\bh'$, respectively. Both conditions are satisfied by many commonly used kernels, including Gaussian and Laplacian kernels; see \cite{fukumizu2007kernel,sriperumbudur2010hilbert,sriperumbudur2011universality} for related discussions.
Specifically, Assumption~\ref{Assumpt:kernel}\ref{Assumpt:kernel_characteristic} requires the conditional kernel $K_x$ to be characteristic on $\cy$ for every fixed $x\in\cx$. This guarantees that the corresponding RKHS embedding uniquely determines the conditional distribution $P_{Y|X=x}$, which is essential for identifying the conditional distribution equality through the discrepancy measure. 
In contrast, Assumption~\ref{Assumpt:kernel}\ref{Assumpt:kernel_L2_dense} is an approximation condition on the interaction RKHS $\bh'$. It requires $\bh'$ to be dense in $L_2(P_{X_1,X_2})$ so that the interaction function $h(X_1,X_2)$ is sufficiently rich to capture the dependence structure between $X_1$ and $X_2$. Unlike the characteristic property, which concerns the injectivity of kernel mean embeddings, the $L_2$-density condition concerns the approximation richness of the RKHS.
\end{rmk}

Recall that $X_1, X_2 \in \cx$ are allowed to be dependent throughout the article. We impose the following overlap conditions on their marginal supports and joint dependence structure.

\begin{assumpt}\label{Assumpt:X1X2}
\begin{enumerate}[label=(\roman*)]
    \item \label{Assumpt:mutual_abs_continuity} The marginal distributions $P_{X_1}$ and $P_{X_2}$ are mutually absolutely continuous, i.e., $P_{X_1}\ll P_{X_2}$ and $P_{X_2}\ll P_{X_1}$.
    \item \label{Assumpt:joint_overlap} For any measurable set $\ca\subseteq\cx$, if $\bp(X_1\in\ca)>0$ and $\bp(X_2\in\ca)>0$, then $\bp(X_1\in\ca, X_2\in\ca)>0$.
\end{enumerate}
\end{assumpt}

\begin{rmk}
Assumption~\ref{Assumpt:X1X2}\ref{Assumpt:mutual_abs_continuity} ensures that $X_1$ and $X_2$ have the same support, up to null sets; see similar setting in \cite{hu2024two,yan2025distance}. Assumption~\ref{Assumpt:X1X2}\ref{Assumpt:joint_overlap} further requires that any region of the covariate space that is marginally possible for both $X_1$ and $X_2$ can also occur jointly with positive probability. This condition is satisfied in standard settings, for example, when $X_1=X_2$, when $X_1$ and $X_2$ are independent, or when their joint distribution admits a density that is strictly positive on the common support.
\end{rmk}

Under Assumptions~\ref{Assumpt:kernel}--\ref{Assumpt:X1X2}, the population discrepancy in \eqref{Equ:Expectation_Uast} fully characterizes equality of two conditional distributions.

\begin{theorem}\label{Thm:Validity}
Under Assumption \ref{Assumpt:kernel}--\ref{Assumpt:X1X2}, 
\begin{equation*}
    \sup\limits_{\phi_{f,g,h}\in\Phi} 
    \big|\,\bbe{\phi_{f,g,h}(X_1,Y_1,Y_1^{c\ast},X_2,Y_2,Y_2^{c\ast})}\,\big| 
  = 0
\end{equation*}
if and only if $P_{Y_1|X_1}(\cdot|x) = P_{Y_2|X_2}(\cdot|x)$ for both $P_{X_1}$- and $P_{X_2}$-almost every $x\in\cx$.
\end{theorem}

\begin{rmk}\label{Rmk:Counterexample}
When $X_1$ and $X_2$ are dependent and have different distributions, both the interaction function $h(X_1,X_2)$ and the overlap condition in Assumption~\ref{Assumpt:X1X2} are essential for the characterization property. In the special cases where $X_1$ and $X_2$ are independent or identical, one may simply choose $h(x_1,x_2)\equiv 1$. However, such a simplified discrepancy generally fails to characterize the equality of conditional distributions when the conditioning variables are dependent; see the supplement for two counterexamples that demonstrate the necessity of including $h(X_1,X_2)$ and the overlap condition.
\end{rmk}

\subsection{Proposed Test Statistic}\label{Sec:Method_TestStat}

Given two random samples $\cd_1 := \{(X_{1i},Y_{1i})\}_{i=1}^{n} \stsim{iid} P_{X_1,Y_1}$ and $\cd_2 := \{(X_{2i},Y_{2i})\}_{i=1}^{n} \stsim{iid} P_{X_2,Y_2}$ in $\br^{p+q}$, we define 
\begin{equation}\label{Equ:Y_ast}
    Y_{1i}^{c\ast} = G_2^{\ast}(X_{1i}, Z_{1i}^{c \ast}), \quad \mbox{and}\quad
    Y_{2i}^{c\ast} = G_1^{\ast}(X_{2i}, Z_{2i}^{c \ast}), \quad i=1,\cdots,n,
\end{equation}
where $\{(Z_{1i}^{c \ast},Z_{2i}^{c \ast})\}_{i=1}^{n}$ are iid auxiliary random vectors used by the oracle conditional generators $G_1^{\ast},G_2^{\ast}$, independent of the given random samples $\cd_1,\cd_2$.

Our proposed framework accommodates both paired and unpaired samples. When $(X_1,Y_1)$ and $(X_2,Y_2)$ are independent, the pairing structure is irrelevant, and the two samples may be viewed as unpaired. When $(X_1,Y_1)$ and $(X_2,Y_2)$ are dependent, we naturally regard the observations as paired, in the sense that $\{(X_{1i},Y_{1i},X_{2i},Y_{2i})\}_{i=1}^{n}$ are independently sampled from a common joint distribution and the dependence between $X_1$ and $X_2$ is observed through the pairing.

Motivated by the population-level cross-generated RKHS discrepancy in \eqref{Equ:Expectation_Uast}, we first define an oracle test statistic as its sample analogue constructed using $G_1^{\ast}$ and $G_2^{\ast}$:
\begin{equation}\label{Equ:Uast}
    U^{\ast}
  = \sup\limits_{\phi_{f,g,h} \in \Phi}
    \lrabs{\,\frac{1}{n} \sum\limits_{i=1}^{n} \phi_{f,g,h}(X_{1i},Y_{1i},Y_{1i}^{c\ast},X_{2i},Y_{2i},Y_{2i}^{c\ast})\,}.
\end{equation}

\begin{rmk}\label{Rmk:oracle_U_limit}
Note that $U^{\ast}$ is the supremum of an empirical process indexed by the function class $\Phi$. Under the null hypothesis, the empirical process is centered, and we will later show that $\sqrt{n} U^{\ast}$ converges weakly to the supremum of a tight centered Gaussian process; see Theorem~\ref{Thm:Uast_null}. However, the resulting limiting distribution depends on the unknown joint distribution of $(X_1, Y_1, Y_1^{c\ast}, X_2, Y_2, Y_2^{c\ast})$ and is therefore non-pivotal. Under the alternative hypothesis, the population-level discrepancy no longer vanishes, so the empirical process is no longer centered. Consequently, the limiting behavior of $U^{\ast}$ becomes substantially more intricate due to the presence of a nonzero mean structure.
\end{rmk}

\begin{rmk}
For the standard MMD, the reproducing property allows the population discrepancy of two distributions to be expressed as the RKHS distance between two kernel mean embeddings. Consequently, the sample counterpart admits a closed-form representation as a quadratic form and can be efficiently computed through a $U$-statistic; see, for example, Section 2.3 of \cite{zhang2024doubly}. In contrast, the discrepancy considered here is defined as the supremum over the compositional function class $\Phi$, which involves multiple RKHS components and an interaction function. The resulting supremum does not admit a closed-form kernel mean embedding representation, and therefore cannot be reduced to a standard $U$-statistic. As a consequence, both the computation and asymptotic analysis differ substantially from those of the classical MMD.
\end{rmk}

The oracle test statistic~\eqref{Equ:Uast} involves oracle conditional generators, which are infeasible in practice. We therefore approximate them using conditional generative models, such as mixture density networks (MDNs) \cite{bishop1994mixture,zhou2022deep} and conditional diffusion models (CDMs) \cite{zhang2023adding}, to learn the conditional distributions. A brief review of these methods and their implementation details is provided in the supplementary materials. 

To avoid overfitting and to maintain independence between the training samples used to learn the conditional generators and the evaluation samples used to construct the test statistic, we use sample-splitting and cross-fitting techniques. Let $L\geq 2$ be a fixed number of splits. We randomly partition the index set $\{1,2,\cdots,n\}$ into $L$ disjoint groups $\ci_1,\cdots,\ci_L$. For simplicity, we assume that all splits have equal size $n_0=n/L$.
Let $\cd_1^{\pell}$ and $\cd_1^{\pmell}$ denote the $\ell$-th split of $\cd_1$ and its complement, respectively. Analogously, we define $\cd_2^{\pell}$ and $\cd_2^{\pmell}$ for $\cd_2$. Moreover, define
\begin{equation*}
    \cd^{(\ell)} = \{(X_{1i},Y_{1i},X_{2i},Y_{2i})\}_{i\in\ci_\ell}, \quad
    \cd^{(-\ell)} = \{(X_{1i},Y_{1i},X_{2i},Y_{2i})\}_{i\notin\ci_\ell},
\end{equation*}
and $\cd := \{(X_{1i},Y_{1i},X_{2i},Y_{2i})\}_{i=1}^{n}$.
For each split $\ell$, we train the empirical conditional generators $\widehat{G}_1^{\pell}$ and $\widehat{G}_2^{\pell}$ using the training samples $\cd_1^{\pmell}$ and $\cd_2^{\pmell}$, respectively. We then apply the cross-generation procedure to construct synthetic responses for the held-out split. Specifically, for each $\ell=1,\cdots,L$ and $i\in\ci_\ell$, we define
\begin{equation}\label{Equ:Y_hat}
    \widehat{Y}_{1i} = \widehat{G}_2^{\pell}(X_{1i},Z_{1i}),
    \qquad
    \widehat{Y}_{2i} = \widehat{G}_1^{\pell}(X_{2i},Z_{2i}),
\end{equation}
where $\cz=\{(Z_{1i},Z_{2i})\}_{i=1}^{n}$ are iid random vectors serving as input noise to the generators and are independent of $\cd$. The distribution of $\cz$ is specified in the supplementary materials.

We use $\{(\widehat{Y}_{1i},\widehat{Y}_{2i})\}_{i=1}^{n}$ as feasible approximations to the oracle cross-generated samples $\{(Y_{1i}^{c\ast},Y_{2i}^{c\ast})\}_{i=1}^{n}$, and define the proposed feasible test statistic as
\begin{equation}\label{Equ:U}
    \widehat{U} 
  = \sup\limits_{\phi_{f,g,h} \in \Phi} 
    \lrabs{\frac{1}{n}\sum_{i=1}^{n}
    \phi_{f,g,h}(X_{1i},Y_{1i},\widehat{Y}_{1i},X_{2i},Y_{2i},\widehat{Y}_{2i})}.
\end{equation}


\begin{rmk}\label{Rmk:feasible_U_limit}
By construction, $\widehat{U}$ is a feasible approximation of $U^{\ast}$, and its approximation accuracy depends on the estimation quality of the conditional generators $\widehat{G}_1^{\pell}$ and $\widehat{G}_2^{\pell}$ for each $\ell=1,\cdots,L$, which is summarized in Assumption~\ref{Assumpt:double_robustness}. To quantify the plug-in error of $\widehat{U}$, we need to control the discrepancy between two empirical-process suprema indexed by $\Phi$. This requires entropy control of $\Phi$ together with suitable empirical process complexity bounds; see Theorem~\ref{Thm:Double-robustness} for explicit approximation rates between $\widehat{U}$ and $U^{\ast}$.
\end{rmk}

Larger values of $\widehat{U}$ correspond to stronger evidence against the null hypothesis whenever the plug-in error remains sufficiently small. As discussed in Remark~\ref{Rmk:oracle_U_limit}, the limiting distribution of $U^{\ast}$ is non-pivotal, which motivates us to approximate the critical value of the test by the multiplier bootstrap procedure.

\section{Bootstrap Calibration and Computation}\label{Sec:Method_Computation}
In Section~\ref{Sec:Method_Bootstrap}, we first describe the multiplier bootstrap procedure used to calibrate the rejection threshold. Next, in Section~\ref{Sec:Method_Algorithm}, we provide an iterative algorithm for computing the proposed test statistic $\widehat{U}$ and an end-to-end algorithm summarizing the full testing procedure.

\subsection{Gaussian Multiplier Bootstrap Procedures}\label{Sec:Method_Bootstrap}

As discussed in Remark~\ref{Rmk:oracle_U_limit}, under the null hypothesis, $\sqrt{n}U^{\ast}$ converges weakly to the supremum of a tight centered Gaussian process whose limiting distribution is non-pivotal. Moreover, Theorem~\ref{Thm:Double-robustness} later shows that the feasible statistic $\widehat{U}$ is asymptotically equivalent to $U^{\ast}$ under suitable generator approximation conditions. Consequently, $\widehat{U}$ inherits the same non-pivotal limiting behavior under the null hypothesis, which motivates the use of bootstrap calibration for approximating the critical value.
The bootstrap strategies have also been considered in \cite{shi2021double} for maximum-type statistics over finitely many transformations. In contrast, our test statistic is defined as the supremum over an infinite-dimensional function class $\Phi$, and its theoretical justification relies on the multiplier central limit theorem for empirical processes; see \cite{van1996weak,kosorok2008introduction,chernozhukov2012gaussian}.


Let $\{(X_{1i}, Y_{1i}, \widehat{Y}_{1i}, X_{2i}, Y_{2i}, \widehat{Y}_{2i})\}_{i=1}^{n}$ denote the data used in \eqref{Equ:U} to compute $\widehat{U}$. We generate an independent random sample $\{\varepsilon_i\}_{i=1}^{n}$ of standard normal random variables, and the oracle bootstrap test statistic is given by
\begin{equation*}
    U^{b\ast}
  = \sup\limits_{\phi_{f,g,h}\in\Phi}
    \lrabs{\frac{1}{n} \sum\limits_{i=1}^{n} 
    \phi_{f,g,h}(X_{1i}, Y_{1i}, Y_{1i}^{c\ast}, X_{2i}, Y_{2i}, Y_{2i}^{c\ast})
    (\varepsilon_i - \bar\varepsilon)},
\end{equation*}
where $\bar\varepsilon = \frac{1}{n}\sum_{i=1}^{n}\varepsilon_i$. Similarly, by replacing $(Y_{1i}^{c \ast}, Y_{2i}^{c \ast})$ with $(\widehat{Y}_{1i},\widehat{Y}_{2i})$, we define the feasible bootstrap test statistic
\begin{equation}\label{Equ:U_boot}
    \widehat{U}^{b}
 =  \sup\limits_{\phi_{f,g,h}\in\Phi}
    \lrabs{\frac{1}{n} \sum\limits_{i=1}^{n}
    \phi_{f,g,h}(X_{1i},Y_{1i},\widehat{Y}_{1i},X_{2i},Y_{2i},\widehat{Y}_{2i}) 
    (\varepsilon_i - \bar{\varepsilon})}.
\end{equation}

\begin{rmk}\label{Rmk:U_boot}
The bootstrap statistics $U^{b\ast}$ and $\widehat{U}^b$ are constructed as the suprema of centered multiplier processes. The centering step is important because $\phi_{f,g,h}$ is not mean-zero in general. After centering, the bootstrap process captures the stochastic fluctuation around the empirical mean rather than the deterministic discrepancy component. Provided that the conditional generators satisfy suitable approximation rates, the error between $U^{b\ast}$ and $\widehat{U}^{b}$ is asymptotically negligible under both the null and the alternative (cf. Proposition~\ref{Prop:Double-robustness-boot}). 
\end{rmk}

\begin{rmk}
The multipliers $\{\varepsilon_i\}_{i=1}^n$ are generated independently of the observed data used to compute $\widehat{U}$. The standard normal distribution is used throughout the paper, but it can be replaced by other mean-zero, unit-variance multiplier distributions satisfying suitable tail conditions, such as the Rademacher distribution.
\end{rmk}

Multiplier bootstrap procedures are standard for approximating the distributions of suprema of empirical processes. When the function class is Donsker, the unconditional and conditional weak convergence of the multiplier process follows from classical empirical process theory; see Section~2.9 of \cite{van1996weak} and Section~10.1 of \cite{kosorok2008introduction}. Related Gaussian and multiplier approximation results for suprema of empirical processes have also been developed in \cite{chernozhukov2012gaussian}.

In our setting, $\sqrt{n} U^{b\ast}$ converges conditionally to the same null limiting distribution as $\sqrt{n} U^{\ast}$; see Theorem~\ref{Thm:Uast_null}. When the approximation errors between $\widehat{U}$ and $U^{\ast}$, and between $\widehat{U}^{b}$ and $U^{b\ast}$, are both negligible, the conditional distribution of $\sqrt{n}\widehat{U}^{b}$ given $(\cd,\cz)$ approximates the null distribution of $\sqrt{n}\widehat{U}$, yielding the bootstrap consistency established in Theorem~\ref{Thm:Bootstrap_null}.

Let $\gamma_{1-\alpha}$ denote the conditional $(1-\alpha)$ quantile of $\sqrt{n}\widehat{U}^{b}$ given $(\cd,\cz)$, which serves as the ideal bootstrap critical value. However, the conditional distribution of $\sqrt{n}\widehat{U}^{b}$ is infeasible in practice. To approximate $\gamma_{1-\alpha}$, we generate $B$ independent multiplier samples $({\varepsilon^{(\beta)}})_{\beta=1}^{B}$ and compute the corresponding bootstrap statistics $({\widehat{U}^{b(\beta)}})_{\beta=1}^{B}$. We then estimate the critical value by the empirical $(1-\alpha)$ quantile of $({\sqrt{n}\widehat{U}^{b(\beta)}})_{\beta=1}^{B}$, denoted by $\widehat{\gamma}_{1-\alpha}$. Finally, we reject the null hypothesis whenever $\sqrt{n}\widehat{U} > \widehat{\gamma}_{1-\alpha}$. The asymptotic results in Section~\ref{Sec:Theory} are stated for the ideal critical value $\gamma_{1-\alpha}$ and do not account for the additional Monte Carlo error induced by finite $B$, which is quite common in the bootstrap literature.


\subsection{Computation of the Test Statistic}\label{Sec:Method_Algorithm}
The computation of $\widehat{U}$ and $\widehat{U}^{b}$ involves finding the suprema over the function class $\Phi$, which does not admit a closed-form expression due to the simultaneous supremum over $f,g\in\ch$ and $h\in\ch'$. Fortunately, by leveraging the reproducing property of RKHS in \eqref{Equ:kernel_trick} and the kernel trick, the maximization over each of $f$, $g$, and $h$ admits a closed-form solution when the other two functions are fixed. This motivates an alternating maximization algorithm to approximate $\widehat{U}$ and $\widehat{U}^{b}$.

Specifically, let $\alpha_f = (\alpha_{f,1},\cdots,\alpha_{f,n})^{\top}$, $\beta_g = (\beta_{g,1},\cdots,\beta_{g,n})^{\top}$ and $\gamma_h = (\gamma_{h,1},\cdots,\gamma_{h,n})^{\top}$, where for each $i=1,\cdots,n$, 
\begin{equation*}
    \alpha_{f,i} = f(X_{1i},Y_{1i}) - f(X_{1i},\widehat{Y}_{1i}), \quad
    \beta_{g,i} = g(X_{2i},Y_{2i}) - g(X_{2i},\widehat{Y}_{2i}), \quad
    \gamma_{h,i} = h(X_{1i},X_{2i}).
\end{equation*}
In light of \eqref{Equ:phi} and \eqref{Equ:U}, we have that
\begin{equation*}
    \widehat{U}
  = \sup\limits_{f,g\in\ch,~h\in\ch'} 
    \lrabs{\frac{1}{n} 
           \sum\limits_{i=1}^{n} \alpha_{f,i} \beta_{g,i} \gamma_{h,i}}
  = \lrabs{\frac{1}{n} 
           \sum\limits_{i=1}^{n}
           \alpha_{f_0,i} \beta_{g_0,i} \gamma_{h_0,i}},
\end{equation*}
where $(f_0,g_0,h_0)$ denotes a maximizer whenever the supremum is attained. To evaluate $\widehat{U}$, we approximate the corresponding maximizing vectors $\alpha_{f_0},\beta_{g_0}$ and $\gamma_{h_0}$ iteratively. At each iteration, we maximize over one RKHS component while fixing the remaining two components, which induces updates of the vectors $\alpha_f, \beta_g, \gamma_h$ in alternating orders.

\begin{prop}\label{Prop:Update_U}
Let $\alpha_f^{(r-1)}, \beta_g^{(r-1)}, \gamma_h^{(r-1)}$ be the updates of the $(r-1)$-th iteration to approximate $\widehat{U}$. We have that
\BEqn
    \alpha_f^{(r)}
&=& \argmax\limits_{\alpha_f:~f\in\ch} 
    \lrabs{\frac{1}{n}
           \sum\limits_{i=1}^{n} \alpha_{f,i} \beta_{g,i}^{(r-1)} \gamma_{h,i}^{(r-1)}}
 =  \frac{\Omega_1 p_f^{(r)}}{\sqrt{p_f^{(r)\top} \Omega_1 p_f^{(r)}}}, \qquad
    p_f^{(r)} = \beta_g^{(r-1)} \odot \gamma_h^{(r-1)}, \\
    \beta_g^{(r)}
&=& \argmax\limits_{\beta_g:~g\in\ch} 
    \lrabs{\frac{1}{n}
           \sum\limits_{i=1}^{n} \alpha_{f,i}^{(r)} \beta_{g,i} \gamma_{h,i}^{(r-1)}}
 =  \frac{\Omega_2 p_g^{(r)}}{\sqrt{p_g^{(r)\top} \Omega_2 p_g^{(r)}}}, \qquad
    p_g^{(r)} = \alpha_f^{(r)} \odot \gamma_h^{(r-1)}, \\
    \gamma_h^{(r)}
&=& \argmax\limits_{\gamma_h:~h\in\ch'} 
    \lrabs{\frac{1}{n}
           \sum\limits_{i=1}^{n} \alpha_{f,i}^{(r)} \beta_{g,i}^{(r)} \gamma_{h,i}}
 =  \frac{\Omega_3 p_h^{(r)}}{\sqrt{p_h^{(r)\top} \Omega_3 p_h^{(r)}}}, \qquad
    p_h^{(r)} = \alpha_f^{(r)} \odot \beta_g^{(r)},
\EEqn
where $\Omega_1,\Omega_2,\Omega_3$ are $n\times n$ matrices with the $(i_1,i_2)$-th entry given by
\BEqn
    \Omega_{1,i_1i_2}
&=& K((X_{1i_1},Y_{1i_1}), (X_{1i_2},Y_{1i_2})) 
  - K((X_{1i_1},Y_{1i_1}), (X_{1i_2},\widehat{Y}_{1i_2})) \\
& & \hspace{2em}
  - K((X_{1i_1},\widehat{Y}_{1i_1}), (X_{1i_2},Y_{1i_2})) 
  + K((X_{1i_1},\widehat{Y}_{1i_1}), (X_{1i_2},\widehat{Y}_{1i_2})), \\
    \Omega_{2,i_1i_2}
&=& K((X_{2i_1},Y_{2i_1}), (X_{2i_2},Y_{2i_2})) 
  - K((X_{2i_1},Y_{2i_1}), (X_{2i_2},\widehat{Y}_{2i_2})) \\
& & \hspace{2em}
  - K((X_{2i_1},\widehat{Y}_{2i_1}), (X_{2i_2},Y_{2i_2})) 
  + K((X_{2i_1},\widehat{Y}_{2i_1}), (X_{2i_2},\widehat{Y}_{2i_2})), \\
    \Omega_{3,i_1i_2}
&=& K'((X_{1i_1},X_{2i_1}), (X_{1i_2},X_{2i_2})).
\EEqn
If the denominator in any of the above updates is zero, then the corresponding blockwise maximizer is non-unique, and we set the updated vector equal to the previous iterate by convention.
\end{prop}

\begin{rmk}
Although the optimization is carried out over infinite-dimensional RKHSs, Proposition~\ref{Prop:Update_U} shows that the updating formulas reduce to finite-dimensional operations involving only kernel matrices and vector products, and each blockwise update admits a closed-form solution by the reproducing property of RKHS.
\end{rmk}

Based on the updating formulas stated in Proposition~\ref{Prop:Update_U}, we follow the alternating maximization Algorithm \ref{Alg:stat} to approximate $\widehat{U}$.

\begin{algorithm}[h!]
\caption{Alternating maximization for approximating $\widehat{U}$}\label{Alg:stat}

\begin{algorithmic}[1]
\Require $\{(X_{1i},Y_{1i},\widehat{Y}_{1i},X_{2i},Y_{2i},\widehat{Y}_{2i})\}_{i=1}^{n} \in \br^{2p+4q}$, tolerance level $\delta$, maximum iteration number $R$
\Ensure the feasible test statistic $\widehat{U}$

\vspace{3mm}
\Procedure{Test Statistic $\widehat{U}$}{}
\State Initialize $r=0$
\State Initialize $\alpha_f^{(0)},\beta_g^{(0)},\gamma_h^{(0)} \in \br^{n}$
\State Initialize $\widehat{U}^{(-1)} = -\infty$ and $\widehat{U}^{(0)} = \lrabs{ \frac{1}{n} \sum\limits_{i=1}^{n} \alpha_{f,i}^{(0)} \beta_{g,i}^{(0)} \gamma_{h,i}^{(0)} }$.
\State Compute $\Omega_1, \Omega_2, \Omega_3 \in \br^{n\times n}$ given by Proposition~\ref{Prop:Update_U}
\While{$|\widehat{U}^{(r)} - \widehat{U}^{(r-1)}| > \delta$ and $r\le R$}
    \State $r \leftarrow r+1$
    \State Compute $\alpha_f^{(r)}, \beta_g^{(r)}, \gamma_h^{(r)}$ by the updating formulas in Proposition~\ref{Prop:Update_U}.
    \Comment{Blockwise update}
    \State Evaluate $\widehat{U}^{(r)} = \lrabs{ \frac{1}{n} \sum\limits_{i=1}^{n} \alpha_{f,i}^{(r)} \beta_{g,i}^{(r)} \gamma_{h,i}^{(r)} }$.
    \Comment{Check for early stopping}
    
    
\EndWhile

\State Approximate $\widehat{U} = \lrabs{ \frac{1}{n} \sum\limits_{i=1}^{n} \alpha_{f,i}^{(r)} \beta_{g,i}^{(r)} \gamma_{h,i}^{(r)} }$
\EndProcedure
\end{algorithmic}
\end{algorithm}

\begin{rmk}
Since the objective is jointly nonconvex in $(f,g,h)$, the alternating maximization procedure is only intended to provide a numerical approximation of $\widehat{U}$, and it may not attain the exact supremum. In practice, we start with random initializations, and iterations are terminated when the relative change of the objective value falls below a prescribed tolerance level, e.g. $10^{-3}$. Empirically, the algorithm is simple to implement and exhibits rapid convergence.
Note that our theoretical justifications are stated for the exact supremum, yet the present asymptotic theory does not include the effect of algorithmic approximation.
\end{rmk}

The computation of the bootstrap statistic $\widehat U^b$ follows the same alternating maximization strategy, and we defer the computation details to the supplement. Finally, we conclude this section by presenting an end-to-end algorithm summarizing the full testing procedure.

\begin{algorithm}[h!]
\caption{Proposed testing procedure}\label{Alg:full_test}

\begin{algorithmic}[1]
\Require $\cd = \{(X_{1i},Y_{1i},X_{2i},Y_{2i})\}_{i=1}^{n} \in \br^{2p+2q}$, split number $L$, significance level $\alpha$
\Ensure statistical decision

\vspace{3mm}
\Procedure{Proposed Testing Procedure}{}
\State Partition the index set into $L$ disjoint groups $\{\ci_1,\cdots,\ci_L\}$
\Comment{Sample splitting}

\State Sample generator noises $\cz = \{(Z_{1i},Z_{2i})\}_{i=1}^{n}$ independent of $\cd$
\For{$\ell=1,\cdots,L$}
    \State Estimate $\widehat{G}_1^{\pell}$ using $\{(X_{1i},Y_{1i})\}_{i\notin\ci_\ell}$ and estimate $\widehat{G}_2^{\pell}$ using $\{(X_{2i},Y_{2i})\}_{i\notin\ci_\ell}$
    \vspace{1mm}
    \For{$i\in\ci_\ell$}        
        \State Generate $\widehat{Y}_{1i} = \widehat{G}_2^{\pell}(X_{1i},Z_{1i})$ and $\widehat{Y}_{2i} = \widehat{G}_1^{\pell}(X_{2i},Z_{2i})$
        \Comment{Cross-generating}
    \EndFor
\EndFor
\State Apply Algorithm~\ref{Alg:stat} to approximate $\widehat{U}$ given by Equation~\eqref{Equ:U}

\For{$\beta=1,\cdots,B$}
    \State Sample multipliers $\varepsilon^{(\beta)} = (\varepsilon_1^{(\beta)},\cdots,\varepsilon_n^{(\beta)}) \stsim{iid} \cn(0,1)$ independent of $\cd,\cz$
    \State Approximate $\widehat{U}^{b(\beta)}$ given by Equation~\eqref{Equ:U_boot} using $\varepsilon^{(\beta)}$
\EndFor
\State Set $\widehat{\gamma}_{1-\alpha}$ as the $(1-\alpha)$ sample quantile of $\{\sqrt{n} \widehat{U}^{b(\beta)}\}_{\beta=1}^{B}$
\Comment{Critical value}
\vspace{1mm}

\State Reject the null hypothesis if $\sqrt{n} \widehat{U} > \widehat{\gamma}_{1-\alpha}$
\Comment{Statistical decision}
\EndProcedure
\end{algorithmic}
\end{algorithm}

\begin{rmk}
The sample-splitting and cross-generation mechanism together help avoid overfitting and preserve the conditional independence of $\{(X_{1i},Y_{1i},\widehat{Y}_{1i},X_{2i},Y_{2i},\widehat{Y}_{2i})\}_{i\in\ci_\ell}$ across $i\in\ci_\ell$, given $\cd^{\pmell}$ for each $\ell$. This conditional independence property facilitates the control of the approximation error of $\widehat{U}$ through symmetrization and Rademacher complexity bounds.
Here, the split number $L$ is specified by the user. Typically, a larger value of $L$ increases the effective training sample size used for estimating $\widehat{G}_1^{\pell}$ and $\widehat{G}_2^{\pell}$, though at the cost of increased computational burden. Therefore, the choice of $L$ reflects a trade-off between the approximation accuracy of $\widehat{U}$ and computational efficiency. In this article, we set $L=2$ in numerical experiments while establishing the theory for a generic $L\ge2$.
\end{rmk}

\section{Theoretical Results}\label{Sec:Theory}

In this section, we establish the asymptotic properties of the proposed test statistic and its bootstrap calibration. Specifically, Section~\ref{Sec:Theory_Convergence} first establishes the null limiting distribution of the oracle statistic and then derives the approximation rate between the feasible statistic and its oracle counterpart.
Furthermore, Section~\ref{Sec:Theory_Validity} establishes the Type-I error control and characterizes the local alternatives against which the proposed test achieves asymptotic power one.

\subsection{Asymptotic Theory of the Oracle and Feasible Statistics}\label{Sec:Theory_Convergence}
We first study the oracle statistics $U^{\ast}$ and $U^{b\ast}$, both of which are suprema of empirical processes indexed by $\Phi$. To control the complexity of the function class $\Phi$, we impose the entropy condition below.

\begin{assumpt}\label{Assumpt:entropy}
We assume that there exists deterministic constants $\tau>0, a>0$ and a constant $v\geq1$ that solely depends on $p,q$, such that $a>u_{K,K'}^{1/2}\exp(v/2)$, and
    \BEqn
    & & \sup\limits_{P\in\cp(\br^{p+q},\cb(\br^{p+q}))} \log N(\ch\cup\{0\},\|\cdot\|_{L_2(P)},u_{K,K'}\varepsilon)
    \le \tau \log^v\lrp{\frac{a}{u_{K,K'}\varepsilon}}, 
    ~\forall~ 0<\varepsilon<1, \\
    & & \sup\limits_{P'\in\cp(\br^{2p},\cb(\br^{2p}))} \log N(\ch'\cup\{0\},\|\cdot\|_{L_2(P')},u_{K,K'}\varepsilon)
    \le \tau \log^v\lrp{\frac{a}{u_{K,K'}\varepsilon}}, 
    ~\forall~ 0<\varepsilon<1, 
    \EEqn
    where $\cp(\br^{p+q},\cb(\br^{p+q}))$ and $\cp(\br^{2p},\cb(\br^{2p}))$ denote the set of all probability measures on $(\br^{p+q},\cb(\br^{p+q}))$ and $(\br^{2p},\cb(\br^{2p}))$ respectively, $N(\ch, \|\cdot\|_{L_2(P)}, \varepsilon)$ denotes the $\varepsilon$-covering number of $\ch$ with respect to $L_2(P)$, and $u_{K,K'}$ is given by Assumption~\ref{Assumpt:kernel}\ref{Assumpt:kernel_bound}. 
\end{assumpt}
\begin{rmk}
Assumption~\ref{Assumpt:entropy} controls the complexity of the RKHS classes through uniform entropy bounds, and can be satisfied by Gaussian RKHSs on compact subsets of Euclidean spaces. The logarithmic-power covering entropy condition used here accommodates rich infinite-dimensional function classes beyond standard finite-dimensional or VC-type classes; see \cite{gine1999laws,belkin2018approximation,mendelson2003entropy} for related conditions. 
\end{rmk}

Under Assumption~\ref{Assumpt:entropy}, the function class $\Phi$ satisfies the entropy condition required for Donsker-type empirical process convergence. Then, we can leverage classical weak convergence and multiplier central limit theorems for empirical processes to establish the asymptotic behaviors of $U^{\ast}$ and $U^{b\ast}$.
To facilitate the statement, we use $\|\bg\|_{\Phi} := \sup_{\phi_{f,g,h}\in\Phi} |\bg(\phi_{f,g,h})|$ to denote the supremum norm of a process $\{\bg(\phi_{f,g,h})\}_{\phi_{f,g,h}\in\Phi}$ indexed by $\Phi$, and use $\leadsto^p_{\varepsilon}$ to denote conditional weak convergence in probability with respect to the bootstrap multipliers $\varepsilon$; see Section 2.2.3 of \cite{kosorok2008introduction}.

\begin{theorem}\label{Thm:Uast_null}
Under Assumption~\ref{Assumpt:kernel}--\ref{Assumpt:X1X2} and \ref{Assumpt:entropy}, it holds that
\begin{enumerate}[label=(\roman*)]
    \item $\sqrt{n} U^{\ast} \stra{d} \|\bg\|_{\Phi}$ under the null hypothesis.
    \item Given $\cd,\cz^{c\ast}$, $\sqrt{n} U^{b\ast} \leadsto^p_{\varepsilon} \|\bg\|_{\Phi}$ under both the null and alternative hypotheses.
\end{enumerate}
where $\bg$ is a tight Gaussian process indexed by $\Phi$ with mean zero and the covariance function
\BEqn
& & \Cov\lrp{\bg(\phi_{f_1,g_1,h_1}), \bg(\phi_{f_2,g_2,h_2})} \\
&=& \bbe{\phi_{f_1,g_1,h_1}(X_1,Y_1,Y_1^{c\ast},X_2,Y_2,Y_2^{c\ast}) \,
          \phi_{f_2,g_2,h_2}(X_1,Y_1,Y_1^{c\ast},X_2,Y_2,Y_2^{c\ast})} \\
& & - \bbe{\phi_{f_1,g_1,h_1}(X_1,Y_1,Y_1^{c\ast},X_2,Y_2,Y_2^{c\ast})} \,
      \bbe{\phi_{f_2,g_2,h_2}(X_1,Y_1,Y_1^{c\ast},X_2,Y_2,Y_2^{c\ast})}.
\EEqn 
\end{theorem}

\begin{rmk}
By Theorem~\ref{Thm:Validity}, the oracle statistic $U^{\ast}$ is centered only under the null because the population discrepancy vanishes exclusively under conditional distribution equality. Hence, the limiting theory of $\sqrt{n} U^{\ast}$ is solely valid under the null. In contrast, the empirical process associated with the bootstrap statistic $U^{b\ast}$ is explicitly centered through the multiplier construction (cf. Remark~\ref{Rmk:U_boot}). Consequently, its asymptotic behavior follows directly from the conditional multiplier central limit theorem for Donsker classes (see Theorem 2.9.6 of \cite{van1996weak} and Theorem 10.4 of \cite{kosorok2008introduction}), and remains valid under both the null and the alternative.    
\end{rmk}

Theorem~\ref{Thm:Uast_null} presents the asymptotic behaviors of $U^{\ast}$ and $U^{b\ast}$ with the rescaling rate $\sqrt{n}$. For both statistics, the limiting distribution corresponds to the supremum of a tight Gaussian process whose covariance function depends on the unknown law of $(X_1,Y_1,Y_1^{c\ast},X_2,Y_2,Y_2^{c\ast})$. Consequently, both limiting distributions are non-pivotal. 

The feasible statistics $\widehat{U}$ and $\widehat{U}^{b}$ are constructed by replacing the oracle cross-generated samples in $U^{\ast}$ and $U^{b\ast}$ with counterparts from the estimated conditional generators. To control the resulting plug-in error, we impose the following estimation accuracy condition on the cross-fitted generators.

\begin{assumpt}\label{Assumpt:double_robustness}
For each $\ell=1,\ldots,L$, let $(X_1,Y_1^{c\ast},X_2,Y_2^{c\ast})$ denote generic held-out observations independent of $\cd^{(-\ell)}$, and let $\widehat{Y}_1^{\pell} = \widehat{G}_2^{(\ell)}(X_1, Z_1)$, $\widehat{Y}_2^{\pell} = \widehat{G}_1^{(\ell)}(X_2, Z_2)$, where $Z_1$ and $Z_2$ are auxiliary input noises independent of $\cd$.

We assume that there exist constants $0<k_1,k_2<\frac{1}{2}$ and $M>0$ such that
\begin{eqnarray}
& & \max\limits_{1\le \ell\le L}
    \bigp{\bbe{ \|K\bigp{(X_1, Y_1^{c\ast}), \cdot} - K\bigp{(X_1, \widehat{Y}_1^{\pell}), \cdot}\|_{\bh}^2 }}^{1/2} 
    \le M n^{-k_1}, \label{Equ:double_robustness_f} \\
& & \max\limits_{1\le \ell\le L}
    \bigp{\bbe{ \|K\bigp{(X_2, Y_2^{c\ast}), \cdot} - K\bigp{(X_2, \widehat{Y}_2^{\pell}), \cdot}\|_{\bh}^2 }}^{1/2}
    \le M n^{-k_2}. \label{Equ:double_robustness_g}
\end{eqnarray}
\end{assumpt}

Importantly, Assumption~\ref{Assumpt:double_robustness} is formulated as a coupled generator approximation condition on the joint law of $Y^{c\ast}$ and $\widehat{Y}$, rather than solely on the discrepancy between their marginal conditional distributions. Although $(Y_1^{c\ast},Y_2^{c\ast})$ and
$(\widehat{Y}_1^{\pell},\widehat{Y}_2^{\pell})$ can be represented through auxiliary variables $(Z_1^{c\ast},Z_2^{c\ast})$ and $(Z_1,Z_2)$,  Assumption~\ref{Assumpt:double_robustness} is formulated directly at the response level, and imposes no independence, equality, or distributional requirements on the auxiliary variables $(Z_1^{c\ast},Z_2^{c\ast})$ and $(Z_1,Z_2)$.


\begin{rmk}
Recent works on conditional diffusion models have established minimax-optimal estimation rates under several probability metrics, including total variation and Wasserstein distances under smoothness and manifold assumptions; see \cite{tang2024conditional}. For mixture-density networks, existing theoretical guarantees are typically formulated in terms of conditional density or conditional distribution consistency under additional regularity and optimization assumptions; see \cite{rothfuss2019conditional}. By Jensen's inequality, Assumption~\ref{Assumpt:double_robustness} implies the weaker conditional kernel-mean embedding error bound 
\begin{equation*}
    \be_X[\|\mu_{\widehat P(\cdot|X)} -\mu_{P(\cdot|X)}\|_{\bh}^2 ] 
\le \be[ \|K((X,Y^{c\ast}),\cdot) -K((X,\widehat Y),\cdot)\|_{\bh}^2 ],
\end{equation*}
where $\mu_{P(\cdot|X)}$ and $\mu_{\widehat P(\cdot|X)}$ denote the conditional kernel mean embeddings of the oracle and estimated conditional distributions, respectively. However, the reverse inequality does not generally hold.
Establishing Assumption~\ref{Assumpt:double_robustness} from such distributional guarantees for specific learning algorithms generally requires additional coupling arguments between the oracle and estimated generators, which is beyond the scope of this article.    
\end{rmk}

The quantity $|\widehat{U}-U^{\ast}|$ measures the plug-in error induced by replacing the oracle conditional generators with their estimated counterparts. Since both $\widehat{U}$ and $U^{\ast}$ involve suprema over infinite-dimensional RKHS classes, establishing the approximation rate requires uniform control of the associated empirical process indexed by $\Phi$. To this end, we combine symmetrization and Rademacher complexity arguments with a telescoping decomposition of 
\begin{equation*}
    \phi_{f,g,h}(X_1,Y_1,\widehat{Y}_1,X_2,Y_2,\widehat{Y}_2)
  - \phi_{f,g,h}(X_1,Y_1,Y_1^{c\ast},X_2,Y_2,Y_2^{c\ast})
\end{equation*}
to derive explicit approximation rates under both the null and alternative hypotheses.

\begin{theorem}\label{Thm:Double-robustness}
Suppose that Assumption~\ref{Assumpt:kernel}--\ref{Assumpt:X1X2} and Assumption~\ref{Assumpt:entropy}--\ref{Assumpt:double_robustness} hold.
\begin{enumerate}[label=(\roman*)]
    \item Under the null, it holds that $\sqrt{n}|\widehat{U}-U^{\ast}| = O_p\lrp{n^{-(k_1+k_2-\frac{1}{2})}\log(n)}$.
    
    \item Under the alternative, it holds that $\sqrt{n}|\widehat{U}-U^{\ast}| = O_p\lrp{n^{-\min\{k_1,k_2\}+\frac{1}{2}}\log^{1/2}(n)}$.
\end{enumerate}
\end{theorem}

Theorem~\ref{Thm:Double-robustness} establishes the approximation rate between the feasible statistic $\widehat{U}$ and its oracle counterpart $U^\ast$, and the discrepancy is measured at the $\sqrt{n}$ scale to align with the limiting theory in Theorem~\ref{Thm:Uast_null}. The cross-fitting mechanism is essential for establishing Theorem~\ref{Thm:Double-robustness}. Indeed, the conditional generators $(\widehat G_1^{\pell},\widehat G_2^{\pell})$ are trained on $\cd^{\pmell}$ and are therefore independent of the held-out observations in $\cd^{\pell}$. Conditioning on $\cd^{\pmell}$, the samples $\{(X_{1i},Y_{1i},\widehat{Y}_{1i},X_{2i},Y_{2i},\widehat{Y}_{2i})\}_{i\in\ci_\ell}$ remain conditionally iid, which enables the asymptotic analysis through symmetrization and Rademacher complexity arguments.

\begin{rmk}\label{rmk:double_ro}
Under the null hypothesis, the leading first-order plug-in terms vanish because $Y_1 =^d Y_1^{c\ast} \mid X_1$ and $Y_2 =^d Y_2^{c\ast} \mid X_2$. Consequently, the approximation error depends only on the product of the two generator estimation errors, which yields the faster convergence rate of $n^{-(k_1+k_2-\frac12)}$. In particular, the plug-in error is asymptotically negligible whenever $k_1+k_2>\frac{1}{2}$. This condition allows one conditional generator to be estimated at a slower rate, provided that the other generator is estimated sufficiently accurately. This resembles the double-robustness property established in Theorem 3 of \cite{shi2021double} and Theorem 1 of \cite{zhang2024doubly}. In contrast, under the alternative hypothesis, the oracle discrepancy is no longer centered, and perturbations from either conditional generator contribute at the first order. As a result, the approximation error is dominated by the slower generator estimation rate, leading to the weaker convergence rate of $n^{-\min\{k_1,k_2\}+\frac{1}{2}}\log^{1/2}(n)$. 
\end{rmk}

Unlike the feasible statistic $\widehat{U}$, the bootstrap statistic $\widehat{U}^{b}$ remains asymptotically equivalent to its oracle counterpart under both the null and alternative hypotheses.

\begin{prop}\label{Prop:Double-robustness-boot}
Under Assumption~\ref{Assumpt:kernel}--\ref{Assumpt:X1X2} and Assumption~\ref{Assumpt:entropy}--\ref{Assumpt:double_robustness}, it holds that 
\begin{equation*}
    \sqrt{n} |\widehat{U}^{b} - U^{b\ast}| = O_p\lrp{n^{-\min\{k_1,k_2\}/2} \log^v(n)},
\end{equation*}
where $k_1,k_2$ are given by Assumption~\ref{Assumpt:double_robustness} and $v$ is given by Assumption~\ref{Assumpt:entropy}.
\end{prop}

Proposition \ref{Prop:Double-robustness-boot} establishes the approximation rate between the feasible bootstrap statistic $\widehat{U}^b$ and its oracle counterpart $U^{b\ast}$. The stochastic order is unconditional, with respect to the joint randomness in $(\cd,\cz,\cz^{\ast},\varepsilon)$.
Unlike Theorem~\ref{Thm:Double-robustness}, the approximation rate in Proposition~\ref{Prop:Double-robustness-boot} does not distinguish between the null and alternative hypotheses. The improved behavior arises because the multiplier bootstrap process is conditionally centered; see Remark~\ref{Rmk:U_boot}. Consequently, the leading deterministic discrepancy terms vanish after multiplier weighting under both the null and alternative hypotheses, and the resulting approximation error is asymptotically negligible.

\subsection{Validity and Power of the Proposed Test}\label{Sec:Theory_Validity}

Next, we establish the validity and power properties of the proposed test. The key step is to show that the conditional distribution of the bootstrap statistic $\widehat{U}^{b}$ consistently approximates the null limiting distribution of $\widehat{U}$. To this end, we impose the following non-degeneracy condition.

\begin{assumpt}\label{Assumpt:continuity}
We assume that there exist $f_0,g_0\in\ch$ and $h_0\in\ch'$ such that
\begin{equation*}
    \Var\bigp{\phi_{f_0,g_0,h_0}(X_1,Y_1,Y_1^{c\ast},X_2,Y_2,Y_2^{c\ast})} > 0.
\end{equation*}
\end{assumpt}

\begin{rmk}
Assumption~\ref{Assumpt:continuity} is necessary to justify that the limiting Gaussian seminorm $\|\bg\|_{\Phi}$ in Theorem~\ref{Thm:Uast_null} admits a continuous distribution. It is a mild non-degeneracy condition that excludes the pathological case in which the limiting Gaussian process is identically zero. 
Unlike the anti-concentration conditions employed in Gaussian approximation results for high-dimensional maxima (e.g., \cite{chernozhukov2012gaussian,chernozhukov2015comparison}), our analysis relies on the classical weak convergence theory for Donsker empirical processes and the conditional multiplier central limit theorem. Since we do not require quantitative Kolmogorov-distance bounds or explicit anti-concentration inequalities for Gaussian maxima, a substantially weaker non-degeneracy condition is sufficient.
\end{rmk}

Under the null hypothesis, Theorem~\ref{Thm:Double-robustness} quantifies the approximation error between $\widehat{U}$ and $U^{\ast}$, while Proposition~\ref{Prop:Double-robustness-boot} establishes the corresponding approximation result for $\widehat{U}^b$ and $U^{b\ast}$. In addition, Theorem~\ref{Thm:Uast_null} characterizes the asymptotic behaviors of $U^{\ast}$ and $U^{b\ast}$. Combining these results yields the bootstrap validity theorem below.

\begin{theorem}\label{Thm:Bootstrap_null}
Suppose that Assumption~\ref{Assumpt:kernel}--\ref{Assumpt:X1X2} and Assumption~\ref{Assumpt:entropy},\ref{Assumpt:continuity} hold, and Assumption \ref{Assumpt:double_robustness} is satisfied with $k_1+k_2>\frac{1}{2}$. Then under the null,
\begin{equation*}
    \sup\limits_{x\in\br} \lrabs{ \blrp{\sqrt{n}\widehat{U} \le x \mid H_0} - \blrp{\sqrt{n}\widehat{U}^{b} \le x \mid \cd, \cz} } = o_p(1),
\end{equation*}
where $\cd$ and $\cz$ are defined in Section~\ref{Sec:Method_TestStat}.
\end{theorem}

\begin{rmk}
Theorem~\ref{Thm:Bootstrap_null} establishes the asymptotic validity of the proposed multiplier bootstrap procedure under the null hypothesis.
The proof proceeds in three steps. First, Theorem~\ref{Thm:Double-robustness} and Proposition~\ref{Prop:Double-robustness-boot} show that the feasible statistics $(\widehat{U},\widehat{U}^{b})$ are asymptotically equivalent to their oracle counterparts $(U^{\ast},U^{b\ast})$ whenever $k_1+k_2>\frac{1}{2}$. Second, Theorem~\ref{Thm:Uast_null} establishes that $U^{\ast}$ converges weakly to the supremum of a tight Gaussian process and that the multiplier bootstrap consistently reproduces the same limit conditionally on the observed data. Finally, Assumption~\ref{Assumpt:continuity} guarantees the continuity of the limiting distribution, which upgrades the weak convergence to the uniform approximation result in Theorem~\ref{Thm:Bootstrap_null}.
\end{rmk}

As a direct consequence of Theorem~\ref{Thm:Bootstrap_null}, the multiplier bootstrap procedure yields asymptotically valid critical values. The following corollary establishes the asymptotic Type-I error control of the proposed test.

\begin{cor}\label{Cor:Bootstrap_size}
Given the conditions stated in Theorem~\ref{Thm:Bootstrap_null}, it holds under the null that 
\begin{equation*}
    \blrp{\sqrt{n}\widehat{U} > \gamma_{1-\alpha} \mid H_0} = \alpha + o_p(1),
\end{equation*}
where $\gamma_{1-\alpha}$ denotes the conditional $(1-\alpha)$ quantile of $\sqrt{n} \widehat{U}^{b}$ given $(\cd,\cz)$. 
\end{cor}

Recall that $\sup_{\phi_{f,g,h}\in\Phi} \lrabs{\bbe{\phi_{f,g,h}(X_1, Y_1, Y_1^{c\ast}, X_2, Y_2, Y_2^{c\ast})}}$ characterizes the population-level conditional discrepancy under the alternative hypothesis. The following theorem establishes an explicit signal strength condition under which the proposed test achieves asymptotic power one against the local alternatives.

\begin{theorem}\label{Thm:Bootstrap_power}
Suppose that Assumption~\ref{Assumpt:kernel}--\ref{Assumpt:X1X2} and Assumption~\ref{Assumpt:entropy}--\ref{Assumpt:continuity} hold. Additionally, if  
\begin{equation}\label{Equ:PowerCondition}
    \sup\limits_{\phi_{f,g,h}\in\Phi} \lrabs{\bbe{\phi_{f,g,h}(X_1,Y_1,Y_1^{c\ast},X_2,Y_2,Y_2^{c\ast})}}
  = \omega\lrp{n^{-\min\{k_1,k_2\}} \log^{1/2}(n)},
\end{equation}
then it holds under the alternative that $\bp(\sqrt{n}\widehat{U} \geq \gamma_{1-\alpha} \mid H_1) \rightarrow 1$, where $\gamma_{1-\alpha}$ denotes the conditional $(1-\alpha)$ quantile of $\sqrt{n} \widehat{U}^{b}$ given $(\cd,\cz)$.
\end{theorem}

\begin{rmk}    
Theorem~\ref{Thm:Bootstrap_power} shows that the proposed test is consistent whenever the population discrepancy signal dominates the plug-in approximation error. Specifically, Theorem~\ref{Thm:Double-robustness} implies that $|\widehat{U}-U^{\ast}| = O_p\bigp{n^{-\min\{k_1,k_2\}}\log^{1/2}(n)}$, while the LHS of \eqref{Equ:PowerCondition} measures the population-level conditional discrepancy under the alternative hypothesis. Therefore, condition~\eqref{Equ:PowerCondition} guarantees that the signal asymptotically dominates the plug-in error.
\end{rmk}

We emphasize that the signal strength in \eqref{Equ:PowerCondition} is allowed to decay with the sample size. The detectable rate depends explicitly on the approximation rates $k_1$ and $k_2$ of the estimated conditional generators. More accurate estimation of the conditional generators yields a smaller plug-in error, thereby allowing the proposed test to detect weaker local alternatives.

\section{Simulation Studies} \label{Sec:Simulation}

To evaluate the finite-sample performance of the proposed max-type test, we conduct simulation studies using three models: conditional mean shift (Model A), conditional variance shift (Model B), and conditional covariance shift (Model C). For Models A and B, we consider two response settings: univariate responses ($D_Y=1$) and multivariate responses ($D_Y>1$). For Model C, we consider only the multivariate response setting. As comparisons, we include the procedure of \cite{hu2024two} with four implementations: linear logistic regression (HL-LL), kernel logistic regression (HL-KLR), likelihood ratio test (HL-LR), and likelihood ratio with sample splitting (HL-LRS). For the max-type test, we report results for both the oracle generator (Oracle) and two practical implementations based on a mixture density network (MDN, \cite{bishop1994mixture}) and a conditional diffusion model (CDM). All experiments are repeated $500$ times. For each alternative, we report both the raw rejection rate ($H_1$) and the size-adjusted power ($H_1$ Adj), where the latter is computed by replacing the nominal $5\%$ level with the empirical $5\%$ quantile of the null $p$-values, thereby correcting for any finite-sample size distortion. Unless otherwise specified, each group contains $n_1 = n_2 = 1000$ observations ($N = 2000$ in total), the bootstrap size is set to $B = 100$, and all kernel bandwidths are set by the median heuristic as described in the supplement. The unadjusted empirical rejection rates corresponding to all size-adjusted results are also provided in the supplement.



To reduce the Monte Carlo variability induced by a single cross-generated draw, we generate $m$ independent response pairs $\{(\widehat{Y}_{1i}^{(j)},\widehat{Y}_{2i}^{(j)})\}_{j=1}^{m}$ for each observation pair $(X_{1i},X_{2i})$. We then replace the single-draw discrepancy contribution in $\widehat{U}$ by its Monte Carlo average across the $m$ generated samples and compute
\begin{equation*}
    \widehat{U}^{m}
  = \sup\limits_{\phi_{f,g,h} \in \Phi}
    \lrabs{\,\frac{1}{n} \sum\limits_{i=1}^{n} 
           \,\frac{1}{m} \sum\limits_{j=1}^{m}
    \phi_{f,g,h}(X_{1i},Y_{1i},\widehat{Y}_{1i}^{(j)},X_{2i},Y_{2i},\widehat{Y}_{2i}^{(j)})\,}.
\end{equation*}
All numerical results in this section use $m = 20$.
For $m>1$, both the test statistic and the multiplier bootstrap are computed by the same alternating maximization as in Section~\ref{Sec:Method_Algorithm}, with two modifications. First, the Gaussian multipliers are attached at the observation level: a single $\varepsilon_i-\bar\varepsilon$ is assigned to each observation pair $i$ and shared across its $m$ cross-generated draws, so that the bootstrap statistic $\widehat{U}^{b,m}$ is obtained from $\widehat{U}^{m}$ by inserting the centered multiplier $(\varepsilon_i-\bar\varepsilon)$ outside the inner average $\frac{1}{m}\sum_{j=1}^{m}$. Second, at the kernel level the $f$- and $g$-blocks of Proposition~\ref{Prop:Update_U} are formed on the $nm$ stacked generated samples, so that $\Omega_1$ and $\Omega_2$ become $nm\times nm$ kernel matrices indexed by the pairs $(i,j)$ while retaining the same four-term difference structure, whereas the interaction block $\Omega_3$ remains $n\times n$ because $h(X_{1i},X_{2i})$ does not depend on $j$. Since the two centered factors in $\phi_{f,g,h}$ are evaluated at the same draw $j$ before being multiplied, the $m$ draws enter through a within-observation average of paired products and cannot be collapsed into a single pre-averaged response. The remaining steps of Algorithm~\ref{Alg:stat} are unchanged.
The theoretical results in Section~\ref{Sec:Theory} are stated for the single-draw case $m=1$. The proofs for the fixed $m$ case are expected to be similar to those for $m=1$, so are omitted.

\subsection{Model A (Conditional Mean Shift)}

This model is adapted from Model A (Gaussian, linear) of \cite{hu2024two}. Assume the data $\{(x_{\ell i}, y_{\ell i})\}$ are generated from the linear model
\[
y_{\ell i} = \alpha_\ell + \beta^\top x_{\ell i} + \varepsilon_{\ell i},
\qquad i = 1,\ldots,n_\ell,\quad \ell = 1,2,
\]
where $x_{1i} \stackrel{\text{i.i.d.}}{\sim} \mathcal{N}(0, I_p)$, $x_{2i} \stackrel{\text{i.i.d.}}{\sim} \mathcal{N}(\mu, I_p)$ with $\{x_{1i}\}$ independent of $\{x_{2i}\}$, $\varepsilon_{1i}, \varepsilon_{2i} \stackrel{\text{i.i.d.}}{\sim} \mathcal{N}(0,1)$ independent of the covariates, and $\beta_j \in \{-1,+1\}$ with independent random signs. The mean-shift vector is $\mu = (1,1,-1,-1,0,\ldots,0)^\top$. Under $H_0$ we set $\alpha_1=\alpha_2=0$, while under $H_1$ we set $\alpha_1=0$ and vary $\alpha_2 \in \{0.1,0.2,0.3,0.5,0.8,1.0\}$. We consider $(p, D_y) \in \{(5,1), (100,1), (100,3)\}$; for $D_y=3$, the coefficient matrix $\boldsymbol{\beta} \in \mathbb{R}^{p \times D_y}$ has independent random sign entries $\beta_{ij} \in \{-1,+1\}$ and the intercept shift is applied coordinatewise. Since \cite{hu2024two} is designed for scalar responses, HL methods are only included for $D_y=1$, and the $D_y=3$ MDN implementation uses diagonal covariance matrices.

From Table~\ref{tab:modelA}, the max-type test controls the Type I error near the nominal $5\%$ level across all settings. In the low-dimensional scalar setting ($p=5$, $D_y=1$), the HL methods attain higher size-adjusted power than max-type at small to moderate $\alpha_2$, with HL-LR and HL-LRS reaching above $55\%$ already at $\alpha_2=0.1$, while the max-type Oracle and CDM achieve comparable power to HL-KLR and HL-LL at moderate signal levels. In the high-dimensional scalar setting ($p=100$, $D_y=1$), HL-KLR exhibits substantial size inflation ($14.2\%$ under $H_0$) and loses most of its apparent power advantage after size adjustment, while the max-type MDN and CDM track the Oracle closely and remain comparable to HL-LL, HL-LR, and HL-LRS at larger signal levels. For the multivariate setting ($p=100$, $D_y=3$), the max-type test attains substantially higher power than the $D_y=1$ version at small to moderate $\alpha_2$: the Oracle achieves $23.8\%$ versus $7.6\%$ at $\alpha_2=0.3$, and $64.8\%$ versus $38.0\%$ at $\alpha_2=0.5$, as the $D_y=3$ test jointly exploits all response coordinates through the kernel. At large $\alpha_2$ ($\geq 0.8$), the two versions converge to similar power levels. Among the practical implementations, MDN attains higher power than CDM at moderate $\alpha_2$ ($55.6\%$ versus $53.4\%$ at $\alpha_2=0.5$), while CDM slightly surpasses MDN at larger signal levels ($86.4\%$ versus $85.8\%$ at $\alpha_2=0.8$), with both implementations benefiting substantially from the move to $D_y=3$. Overall, while the HL methods attain higher power than the max-type test under the low-dimensional linear mean shift ($p=5$, $D_y=1$), the max-type test maintains reliable size control across all settings and demonstrates a clear advantage in the multivariate response setting ($D_y=3$), where jointly exploiting all response coordinates yields substantially higher power.

\begin{table}[!ht]
\centering
\caption{Empirical size and size-adjusted power (\%) under Model A.}
\label{tab:modelA}
\scalebox{0.8}{
\resizebox{\textwidth}{!}{
\begin{tabular}{ccc|ccccccc}
        \multirow{2}{*}{$(p, D_y)$} & & \multirow{2}{*}{$\alpha_2$} & \multicolumn{3}{c}{max-type} & HL-KLR & HL-LL & HL-LR & HL-LRS \\  
         & & & Oracle & MDN & CDM & Practical & Practical & Practical & Practical \\ \hline

\multirow{7}{*}{$(5,\ 1)$}
& $H_0$ & $0.0$ & 6.0 & 4.2 & 7.8 & 5.0 & 5.6 & 5.8 & 4.6 \\\cline{2-10}
& \multirow{6}{*}{$H_1$} & $0.1$ & 4.6 & 6.8 & 5.6 & 12.6 & 19.2 & 55.6 & 35.6 \\\cline{3-10}
& & $0.2$ & 7.6 & 9.2 & 11.2 & 32.8 & 43.8 & 99.0 & 87.4 \\\cline{3-10}
& & $0.3$ & 25.2 & 16.8 & 21.4 & 62.4 & 69.6 & 100 & 99.6 \\\cline{3-10}
& & $0.5$ & 99.4 & 72.2 & 65.4 & 97.2 & 97.0 & 100 & 100 \\\cline{3-10}
& & $0.8$ & 100 & 96.0 & 90.6 & 100 & 100 & 100 & 100 \\\cline{3-10}
& & $1.0$ & 100 & 100 & 97.8 & 100 & 100 & 100 & 100 \\\cline{1-10}

\multirow{7}{*}{$(100,\ 1)$}
& $H_0$ & $0.0$ & 5.2 & 4.4 & 4.8 & 14.2 & 5.0 & 8.0 & 8.2 \\\cline{2-10}
& \multirow{6}{*}{$H_1$} & $0.1$ & 5.2 & 4.0 & 4.0 & 4.8 & 11.0 & 51.8 & 27.4 \\\cline{3-10}
& & $0.2$ & 4.6 & 7.0 & 6.0 & 5.0 & 32.8 & 97.6 & 81.8 \\\cline{3-10}
& & $0.3$ & 7.6 & 9.4 & 7.0 & 5.8 & 60.6 & 100 & 99.6 \\\cline{3-10}
& & $0.5$ & 38.0 & 35.8 & 23.8 & 7.4 & 93.6 & 100 & 100 \\\cline{3-10}
& & $0.8$ & 83.2 & 84.4 & 77.0 & 12.0 & 100 & 100 & 100 \\\cline{3-10}
& & $1.0$ & 98.8 & 94.2 & 94.0 & 19.4 & 100 & 100 & 100 \\\cline{1-10}

\multirow{7}{*}{$(100,\ 3)$}
& $H_0$ & $0.0$ & 6.4 & 5.0 & 3.0 & -- & -- & -- & -- \\\cline{2-10}
& \multirow{6}{*}{$H_1$} & $0.1$ & 5.0 & 6.6 & 7.0 & -- & -- & -- & -- \\\cline{3-10}
& & $0.2$ & 9.2 & 13.0 & 10.4 & -- & -- & -- & -- \\\cline{3-10}
& & $0.3$ & 23.8 & 23.0 & 18.8 & -- & -- & -- & -- \\\cline{3-10}
& & $0.5$ & 64.8 & 55.6 & 53.8 & -- & -- & -- & -- \\\cline{3-10}
& & $0.8$ & 83.2 & 85.8 & 86.4 & -- & -- & -- & -- \\\cline{3-10}
& & $1.0$ & 92.8 & 91.4 & 91.2 & -- & -- & -- & -- \\\cline{1-10}
\end{tabular}}}
\end{table}

\subsection{Model B (Conditional Variance Shift)}

Assume the data are generated from
\[
Y_{\ell i} = \beta^\top X_{\ell i} + 0.4(\alpha_\ell + 1)\varepsilon_{\ell i}, \qquad \ell = 1, 2,
\]
where $\varepsilon_{1i}, \varepsilon_{2i} \stackrel{\text{i.i.d.}}{\sim} \mathcal{N}(0,1)$ independent of the covariates and $\beta = \frac{1}{\sqrt{p}}(1,\ldots,1)^\top$. Under $H_0$ we set $\alpha_1=\alpha_2=0$, while under $H_1$ we set $\alpha_1=0$ and vary $\alpha_2 \in \{1,2,3,4\}$. We consider $(p, D_y) \in \{(100, 1), (5, 1), (100, 3)\}$ under both same- and different-distribution covariate settings, where the different-distribution case sets $X_{2i} \stackrel{\text{i.i.d.}}{\sim} \mathcal{N}(\mu, I_p)$ with $\mu_j = 0.4$ for $j \leq \lfloor p/2 \rfloor$ and $\mu_j = -0.4$ otherwise.

From Table~\ref{tab:modelB}, a key finding is that the size behavior of the HL methods varies substantially across settings, requiring careful interpretation of their power figures. In the same-distribution setting ($p=100$, $D_y=1$), HL-KLR controls size well ($4.8\%$) and attains high size-adjusted power ($69.6\%$ at $\alpha_2=1$, reaching $100\%$ at $\alpha_2 \geq 2$), while HL-LL, HL-LR, and HL-LRS exhibit trivial power, failing entirely to detect the conditional variance shift; this is expected because the specific linear and likelihood-ratio implementations used here have limited flexibility to capture scale differences when the conditional mean is correctly specified. In the different-distribution setting with $p=100$, however, HL-KLR suffers from severe size inflation ($39.8\%$ under $H_0$) and HL-LL inflates even further to $53.2\%$; after size adjustment, both methods collapse to near-zero power. In the low-dimensional different-distribution setting ($p=5$), HL-KLR swings to the opposite extreme, becoming overly conservative ($2.2\%$ under $H_0$), while HL-LL, HL-LR, and HL-LRS again have trivial power. In contrast, the max-type test maintains correct size across all three settings and achieves consistent power, with the Oracle and CDM attaining $55.8\%$ and $60.4\%$ size-adjusted power at $\alpha_2=1$ under $p=5$, and MDN and CDM closely tracking the Oracle throughout. For the multivariate setting ($p=100$, $D_y=3$, different distribution), the max-type Oracle and CDM both achieve $100\%$ power already at $\alpha_2=1$, compared to $38.6\%$ and $49.2\%$ for $D_y=1$, and MDN improves from $31.4\%$ to $79.8\%$, confirming that jointly modeling the multivariate conditional distribution provides a clear power advantage when the variance shift affects all response dimensions simultaneously.

\begin{table}[!ht]
    \centering
    \caption{Empirical size and size-adjusted power (\%) under Model B.}
    \label{tab:modelB}
    \scalebox{.85}{
    \resizebox{\textwidth}{!}{
    \begin{tabular}{ccccl|ccccccc}
    \multirow{2}{*}{Covariate} & \multirow{2}{*}{$p$} & \multirow{2}{*}{$D_y$} & & \multirow{2}{*}{$\alpha_2$} & \multicolumn{3}{c}{max-type} & HL-KLR & HL-LL & HL-LR & HL-LRS \\
    & & & & & Oracle & MDN & CDM & Practical & Practical & Practical & Practical \\ \hline
    \multirow{5}{*}{Same} & \multirow{5}{*}{$100$} & \multirow{5}{*}{$1$}
    & $H_0$ & $0$ & 7.2 & 3.2 & 3.8 & 4.8 & 6.2 & 7.2 & 7.6 \\\cline{4-12}
    & & & \multirow{4}{*}{$H_1$} & $1$ & 38.6 & 20.2 & 27.4 & 69.6 & 4.8 & 4.0 & 4.2 \\\cline{5-12}
    & & & & $2$ & 100 & 100 & 99.8 & 100 & 5.4 & 3.0 & 3.4 \\\cline{5-12}
    & & & & $3$ & 100 & 100 & 99.8 & 100 & 5.8 & 2.8 & 3.2 \\\cline{5-12}
    & & & & $4$ & 100 & 100 & 100 & 100 & 6.4 & 3.0 & 3.2 \\\cline{1-12}
    \multirow{15}{*}{Different} & \multirow{5}{*}{$5$} & \multirow{5}{*}{$1$}
    & $H_0$ & $0$ & 8.2 & 4.2 & 4.4 & 2.2 & 6.0 & 6.8 & 5.6 \\\cline{4-12}
    & & & \multirow{4}{*}{$H_1$} & $1$ & 55.8 & 30.2 & 60.4 & 99.8 & 5.6 & 5.0 & 5.6 \\\cline{5-12}
    & & & & $2$ & 100 & 100 & 96.6 & 100 & 6.6 & 5.2 & 5.4 \\\cline{5-12}
    & & & & $3$ & 100 & 100 & 99.6 & 100 & 8.2 & 5.2 & 5.0 \\\cline{5-12}
    & & & & $4$ & 100 & 100 & 100 & 100 & 7.8 & 5.4 & 4.8 \\\cline{2-12}
    & \multirow{5}{*}{$100$} & \multirow{5}{*}{$1$}
    & $H_0$ & $0$ & 7.2 & 4.0 & 5.0 & 39.8 & 53.2 & 5.6 & 6.4 \\\cline{4-12}
    & & & \multirow{4}{*}{$H_1$} & $1$ & 38.6 & 31.4 & 49.2 & 8.0 & 3.6 & 4.8 & 5.0 \\\cline{5-12}
    & & & & $2$ & 100 & 99.6 & 100 & 25.8 & 2.6 & 4.0 & 4.2 \\\cline{5-12}
    & & & & $3$ & 100 & 100 & 100 & 54.6 & 2.6 & 3.8 & 4.2 \\\cline{5-12}
    & & & & $4$ & 100 & 100 & 100 & 80.2 & 3.0 & 3.8 & 4.2 \\\cline{2-12}
    & \multirow{5}{*}{$100$} & \multirow{5}{*}{$3$}
    & $H_0$ & $0$ & 6.0 & 7.2 & 4.4 & -- & -- & -- & -- \\\cline{4-12}
    & & & \multirow{4}{*}{$H_1$} & $1$ & 100 & 79.8 & 100 & -- & -- & -- & -- \\\cline{5-12}
    & & & & $2$ & 100 & 100 & 100 & -- & -- & -- & -- \\\cline{5-12}
    & & & & $3$ & 100 & 100 & 100 & -- & -- & -- & -- \\\cline{5-12}
    & & & & $4$ & 100 & 100 & 100 & -- & -- & -- & -- \\\cline{1-12}
    \end{tabular}}}
\end{table}

\subsection{Model C (Conditional Covariance Shift) with multivariate response $Y$ ($D_Y = 10$)}
We consider a conditional covariance shift model where the conditional mean depends on $X$ and is identical across groups, but the conditional covariance differs. The covariates are generated as $x_{1i}, x_{2i} \stackrel{\text{i.i.d.}}{\sim} \mathcal{N}(0, I_p)$ with $\{X_{1i}\}$ independent of $\{X_{2i}\}$, and the coefficient matrix is $\boldsymbol{\beta} = \frac{0.5}{\sqrt{p}}\mathbf{1}_{p \times d_y}$. Conditional on the covariates, the responses follow
\[
Y_{1i} \mid X_{1i} \sim \mathcal{N}(s_1, I_{d_y}),
\qquad
Y_{2i} \mid X_{2i} \sim \mathcal{N}(s_2, \Sigma_\rho),
\]
where $s_\ell = \boldsymbol{\beta}^\top X_{\ell i}$, $d_y = 10$, and $\Sigma_\rho$ is the equicorrelation matrix with off-diagonal entries $\rho$. Under $H_0$ we set $\rho = 0$, while under $H_1$ we vary $\rho \in \{0.3, 0.5, 0.7, 0.9\}$.

From Table~\ref{tab:modelC}, both the Oracle and CDM control size near the nominal level under $H_0$ and have low power at small $\rho$, with rejection rates near the null level at $\rho=0.3$ and moderate power of $24.5\%$ and $27.3\%$ respectively at $\rho=0.5$. Power increases sharply as $\rho$ grows, with both implementations reaching $99.9\%$ at $\rho=0.7$ and $100\%$ at $\rho=0.9$. The CDM implementation attains results close to the Oracle across all $\rho$ levels, confirming that the diffusion model is effective at capturing high-dimensional covariance structure in this challenging setting.

\begin{table}[!ht]
\centering
\caption{Empirical rejection rates (\%) and size-adjusted power under Model C, based on $2000$ independent trials.}
\label{tab:modelC}
\scalebox{.9}{
\begin{tabular}{cl|ccccc}
    \multicolumn{2}{c|}{\multirow{2}{*}{max-type}} & \multicolumn{5}{c}{$\rho$} \\
    \multicolumn{2}{c|}{} & $0.0$ & $0.3$ & $0.5$ & $0.7$ & $0.9$ \\ \hline
    \multirow{2}{*}{Oracle} & Raw     & 6.1 & 6.5  & 24.5 & 99.9 & 100 \\
                            & Size-adjusted  &  -  & 4.1  & 18.0 & 99.9 & 100 \\ \hline
    \multirow{2}{*}{CDM}    & Raw     & 5.1 & 7.3  & 27.3 & 99.9 & 100 \\
                            & Size-adjusted  &  -  & 6.0  & 24.1 & 99.9 & 100 \\
\end{tabular}}
\end{table}

\section{Real Data Experiments} \label{Sec:RealData}

We conduct two real-data experiments on the cropped and aligned \emph{UTK-Face} dataset \citep{Zhang_2017_CVPR}. The first experiment evaluates the proposed test with a scalar response under controlled covariate degradation, comparing our method against the procedure of \cite{hu2024two}. The second experiment examines the multivariate-response setting, where the signal lies in the dependence structure between response components rather than in their marginals.
\subsection{Scalar Response with Covariate Degradation}

We first evaluate the proposed method on a real-data example with scalar responses and progressively degraded covariate information. The analysis is based on the UTK-Face dataset, which contains $23{,}705$ face images with ages ranging from $1$ to $116$. To reduce sparsity at very young and very old ages, we restrict the analysis to individuals aged between $20$ and $59$, resulting in $16{,}425$ samples. Each image is a cropped and aligned RGB image of size $(224,224,3)$. We extract image features using a fixed pretrained ResNet-18 network, which maps each image to a $512$-dimensional feature vector $X$. The response $Y$ is the individual's age, linearly normalized to $[0,1]$.

\begin{figure}[!ht]
\footnotesize
\centering
\begin{subfigure}[t]{0.18\textwidth}
\centering
\includegraphics[width=\textwidth]{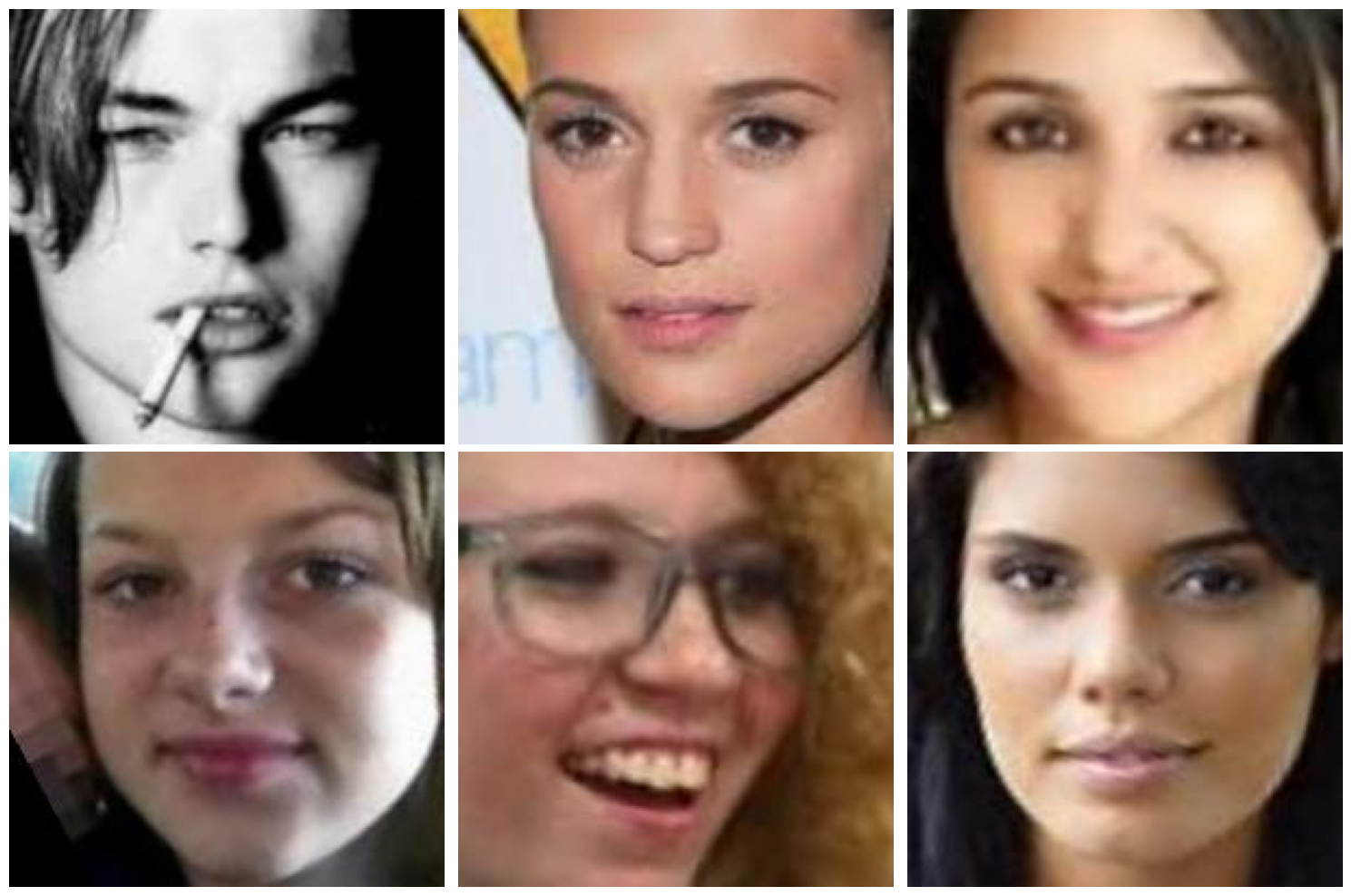}
\caption{$X_1$: Full image}
\end{subfigure}
\hfill
\begin{subfigure}[t]{0.18\textwidth}
\centering
\includegraphics[width=\textwidth]{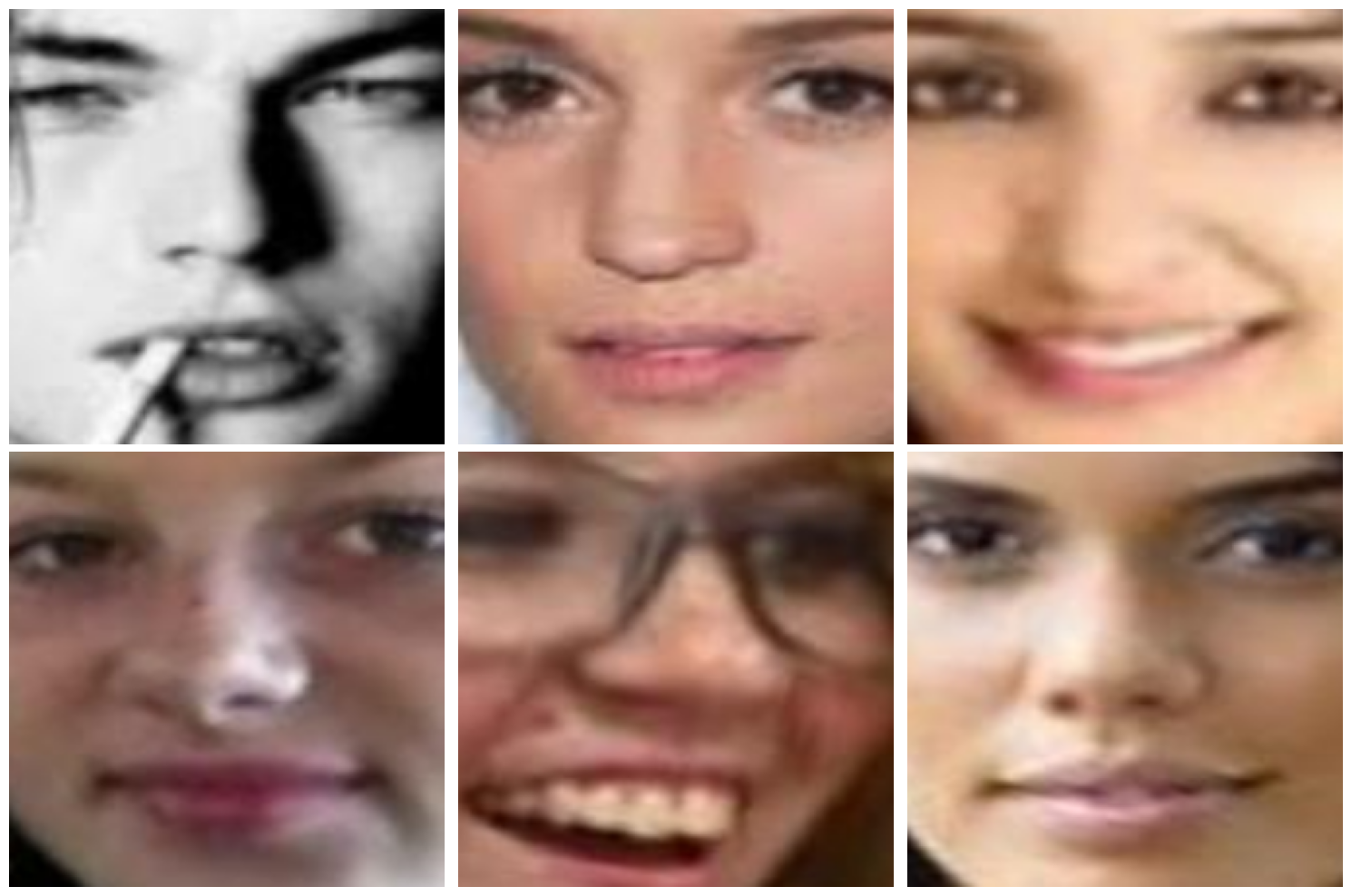}
\caption{$X_2$: Face-preserving crop}
\end{subfigure}
\hfill
\begin{subfigure}[t]{0.18\textwidth}
\centering
\includegraphics[width=\textwidth]{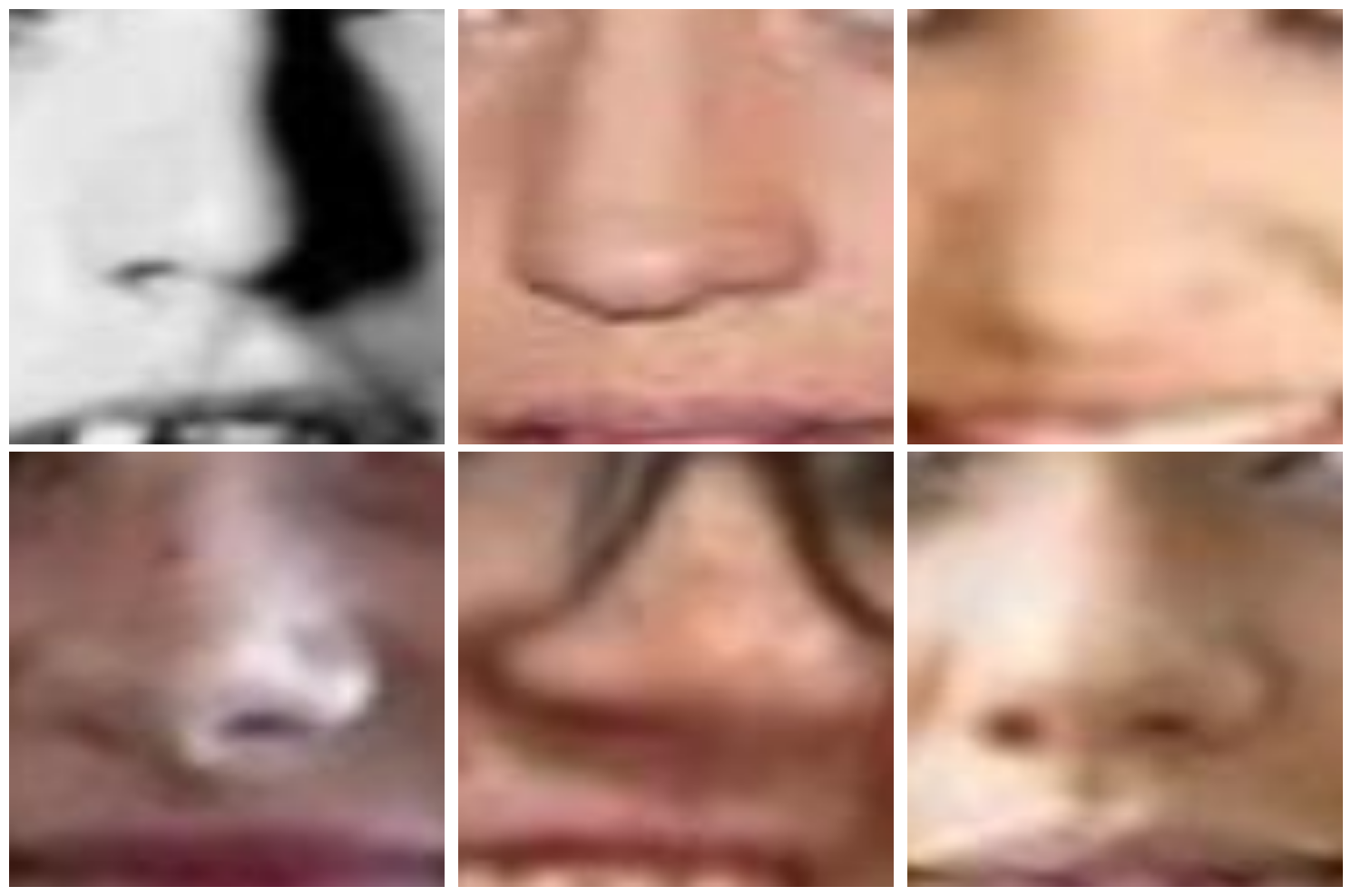}
\caption{$X_3$: Localized crop}
\end{subfigure}
\hfill
\begin{subfigure}[t]{0.18\textwidth}
\centering
\includegraphics[width=\textwidth]{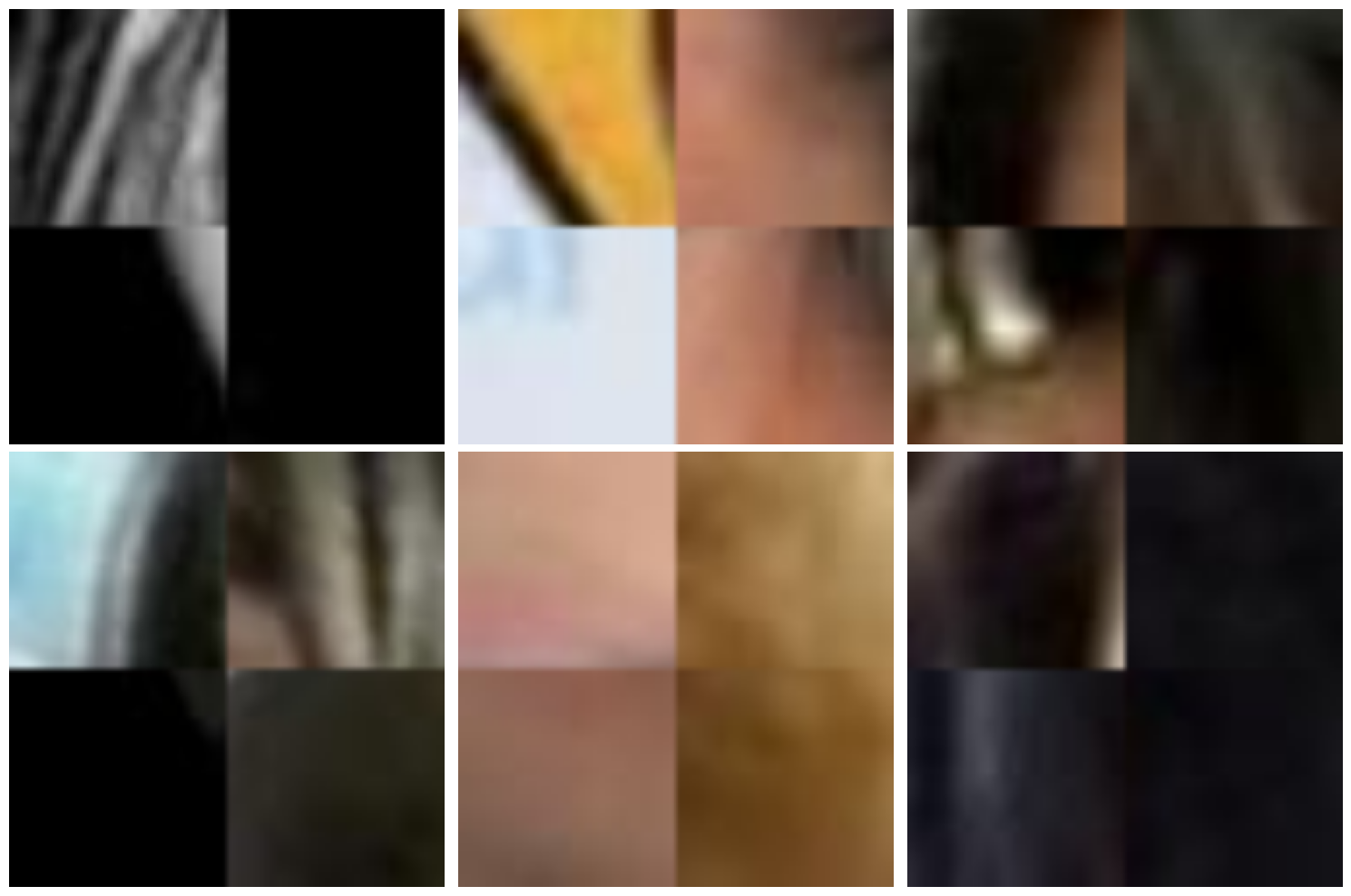}
\caption{$X_4$: Corner-only composition}
\end{subfigure}
\hfill
\begin{subfigure}[t]{0.18\textwidth}
\centering
\includegraphics[width=\textwidth]{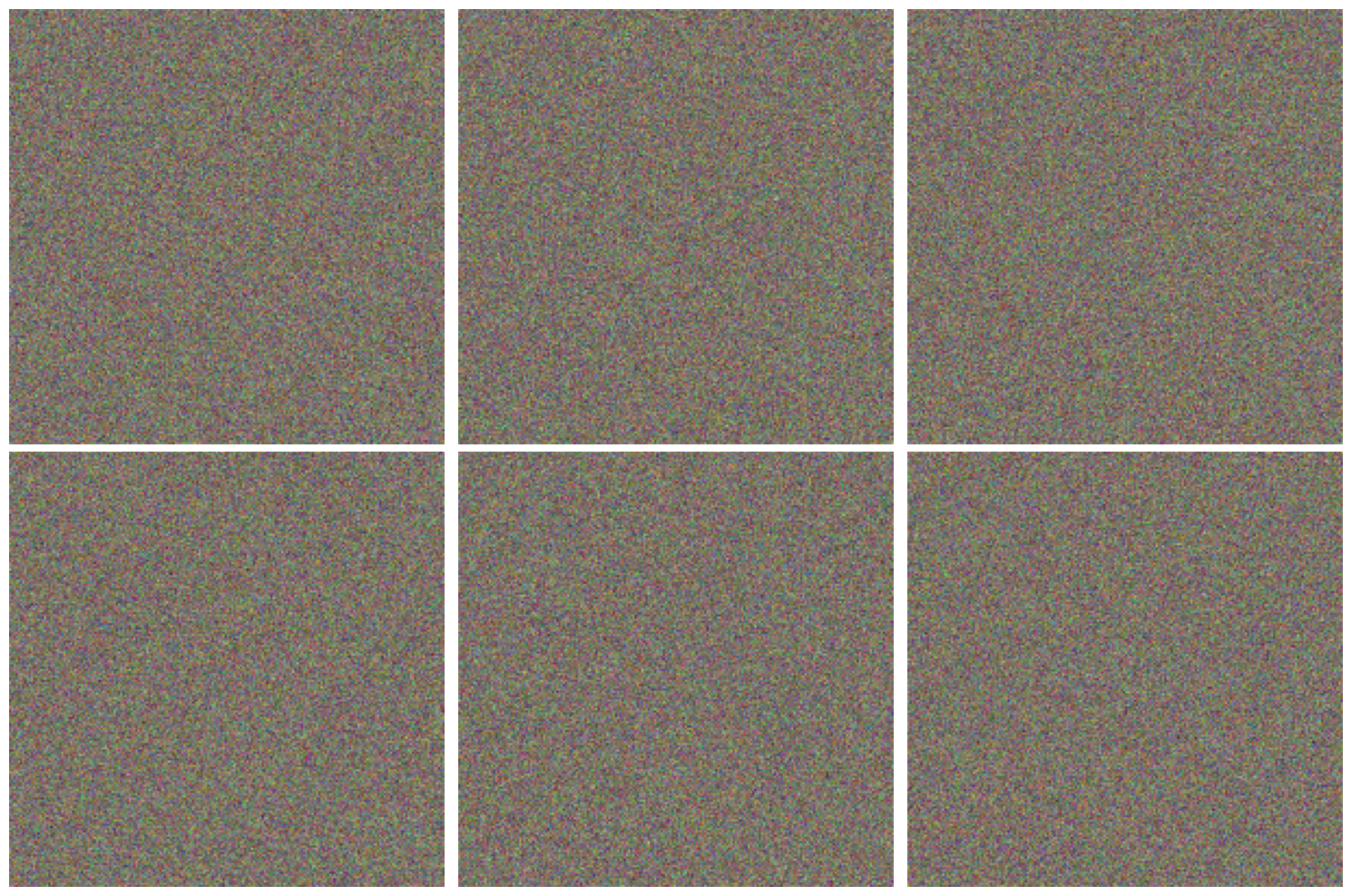}
\caption{$X_5$: Noise input}
\end{subfigure}
\caption{Representative samples of the five input variants, illustrating the progressive degradation of facial semantic content from left to right.}
\label{fig:masked_inputs}
\end{figure}

To create controlled changes in the covariate representation while keeping the response fixed, we construct five input variants by applying masking and cropping operations before feature extraction, as shown in Figure~\ref{fig:masked_inputs}. The first input, $X_1$, is the full image and contains complete facial information. The second input, $X_2$, is a face-preserving central crop that retains most salient facial features, including the eyes, nose, and mouth. The third input, $X_3$, is a localized crop that focuses on a smaller central facial region and removes a substantial amount of global facial structure. The fourth input, $X_4$, is a corner-only composition formed by concatenating the four corner patches of the original image, thereby removing nearly all semantic facial content while preserving some low-level texture information. The fifth input, $X_5$, is a purely noisy image with independently generated pixel values and serves as a semantic-free covariate.

Let $X_k$ denote the ResNet-18 features extracted from the $k$-th input variant. Since all transformations are applied only to the image input, the response remains unchanged across variants, so $Y_1=\cdots=Y_5$. We consider four testing scenarios:
\begin{align*}
\text{Case 1:} &\quad (X_1, Y_1) \ \text{vs.}\ (X_2, Y_2), \qquad
\text{Case 2:} \quad (X_1, Y_1) \ \text{vs.}\ (X_3, Y_3), \\
\text{Case 3:} &\quad (X_1, Y_1) \ \text{vs.}\ (X_4, Y_4), \qquad
\text{Case 4:} \quad (X_1, Y_1) \ \text{vs.}\ (X_5, Y_5).
\end{align*}
Case~1 is designed to be close to the null hypothesis, since both covariates retain rich facial information. Cases~3 and~4 correspond to clear alternatives, where the covariates contain little or no age-relevant semantic information. Case~2 represents an intermediate setting with partial semantic degradation. For each scenario, we form two independent samples of size $n_1=n_2=2000$ from disjoint sets of original images, and repeat the experiment $50$ times using independent random splits.

We compare the proposed max-type test with the MDN implementation against the procedure of \cite{hu2024two} under five implementations: HL-KLR, HL-LL, HL-LR, HL-LRS, and HL-NN, where HL-NN denotes the neural-network logistic-regression version. Although HL-NN was excluded from the simulation studies because of its substantially higher computational cost, we include it in this real-data experiment to verify whether its empirical performance is consistent with the other HL variants.

\begin{figure}[!ht]
\centering
\includegraphics[width=0.65\textwidth]{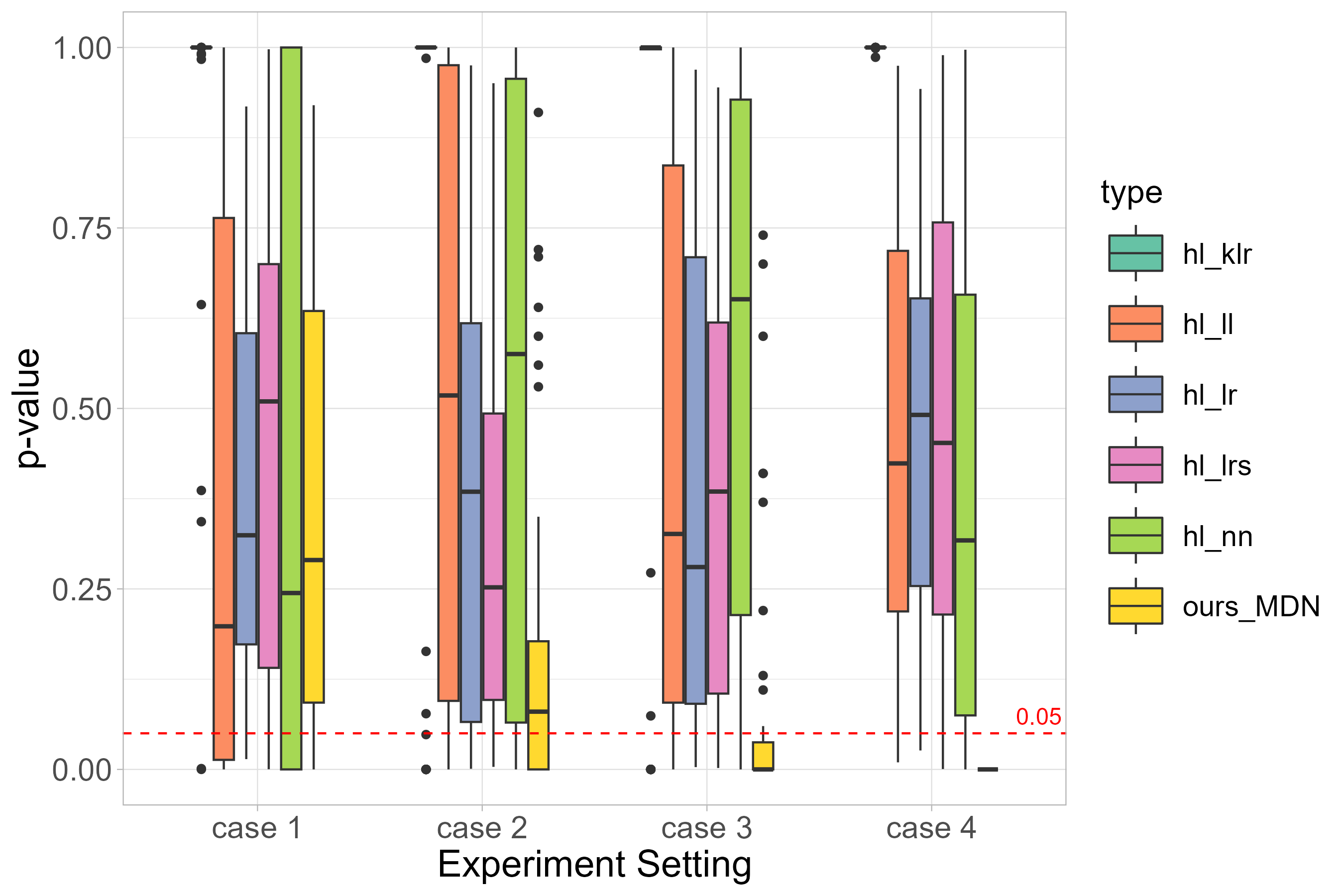}
\caption{Empirical distribution of $p$-values across four testing scenarios. Case~1 corresponds to the near-null setting; Cases~2--4 represent increasing levels of covariate degradation. The dashed red line denotes $\alpha=0.05$.}
\label{fig:hl_ours_boxplot}
\end{figure}

The empirical distributions of the resulting $p$-values are shown in Figure~\ref{fig:hl_ours_boxplot}. The HL methods exhibit limited power across all four cases. HL-KLR produces $p$-values concentrated near $1$, while HL-LL, HL-LR, HL-LRS, and HL-NN yield $p$-values that remain broadly spread over $[0,1]$, even in the most extreme alternatives in Cases~3 and~4. This suggests little ability to detect the distributional change induced by severe covariate degradation. In contrast, the proposed max-type MDN test shows a clear monotone response as the covariate information deteriorates. In Case~1, the $p$-values are broadly distributed over $[0,1]$, consistent with type-I error control under the near-null setting. In Case~2, more than half of the $p$-values remain above $0.05$, reflecting the moderate strength of the signal. In Case~3, more than $75\%$ of the $p$-values fall below $0.05$, indicating strong evidence against the null. In Case~4, all $p$-values are close to zero, showing near-certain detection when the covariate is replaced by pure noise.

\subsection{Multivariate Response with Dependence Structure Shift}

We next evaluate the proposed method in a multivariate-response setting where the signal lies primarily in the dependence structure between response components. We use the same UTK-Face dataset, age restriction, and ResNet-18 feature extraction as in the preceding subsection, but now consider a two-dimensional response
\begin{align*}
Y &= \bigl(Y^{(1)},Y^{(2)}\bigr), \quad\mbox{with}\quad
Y^{(1)} = \text{age} \in [0,1] \quad \mbox{and}\quad
Y^{(2)} = \text{gender} \in \{0,1\},
\end{align*}
where age is linearly rescaled from $[20,59]$ to $[0,1]$, and gender is encoded as $0$ for male and $1$ for female. We use only the unmasked images so that the distribution of $X$ is comparable across all groups.

\begin{figure}[!ht]
\centering
\begin{subfigure}[t]{0.32\textwidth}
\centering
\includegraphics[width=\textwidth]{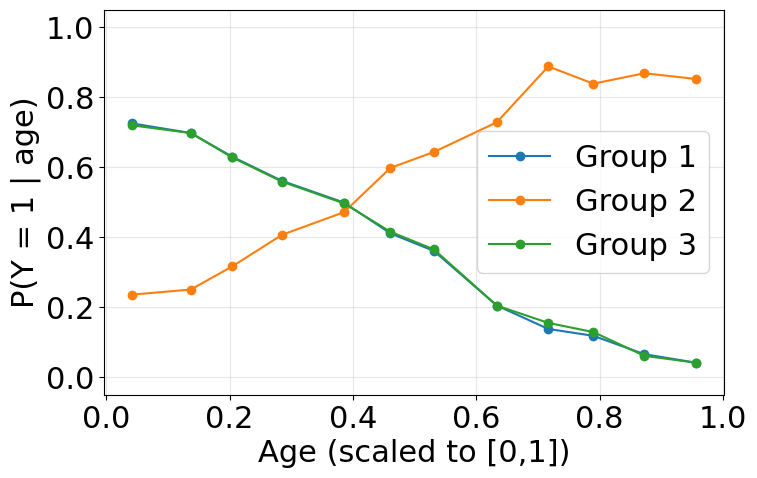}
\caption{Conditional trend}
\label{fig:conditional_trend}
\end{subfigure}
\hfill
\begin{subfigure}[t]{0.32\textwidth}
\centering
\includegraphics[width=\textwidth]{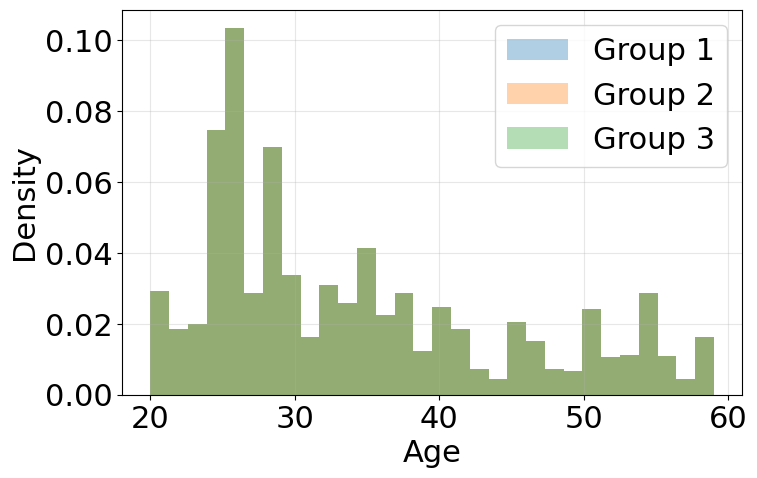}
\caption{Marginal distribution of age}
\label{fig:marginal_age}
\end{subfigure}
\hfill
\begin{subfigure}[t]{0.32\textwidth}
\centering
\includegraphics[width=\textwidth]{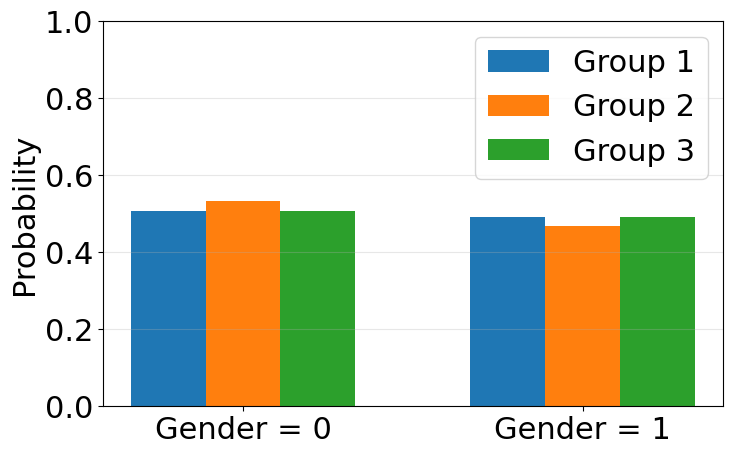}
\caption{Marginal distribution of gender}
\label{fig:marginal_gender}
\end{subfigure}
\caption{Left: empirical conditional trend $P(Y^{(2)}=1 \mid Y^{(1)}=\text{age})$ for the three constructed groups. Groups~1 and~3 share a decreasing trend, while Group~2 has an increasing trend. Center and right: marginal distributions of age and gender across the three groups, showing that the groups are well matched in their marginals.}
\label{fig:group_construction}
\end{figure}

We construct three groups from disjoint subsets of UTK-Face images by controlling the conditional trend $P(Y^{(2)}=1 \mid Y^{(1)}=\text{age})$ at each age level while keeping the marginal age distribution comparable across groups. Groups~1 and~3 are designed to have the same decreasing age--gender trend, with older individuals more likely to be male, whereas Group~2 is designed to have the opposite increasing trend. Figure~\ref{fig:group_construction} displays the empirical conditional trends together with the marginal distributions of age and gender across the three groups. The groups are well matched in their marginal distributions, but differ in their conditional dependence structure.

We consider two comparisons: Group~1 versus Group~3 as the null setting, where the groups share the same age--gender trend, and Group~1 versus Group~2 as the alternative setting, where the groups have opposite trends. Thus, the primary signal distinguishing the two distributions comes from the dependence structure between $Y^{(1)}$ and $Y^{(2)}$, rather than from their marginal distributions.

To model $P(Y \mid X)$, we use the sequential factorization
\begin{align*}
P(Y \mid X)
\,=\, P\bigl(Y^{(1)} \mid X\bigr) \cdot 
P\bigl(Y^{(2)} \mid Y^{(1)}, X\bigr).
\end{align*}
The age component $P(Y^{(1)} \mid X)$ is modeled by an MDN, and the gender component $P(Y^{(2)} \mid Y^{(1)}, X)$ is modeled by a binary classifier that takes both the generated age and $X$ as inputs. This sequential construction is important for capturing the age--gender dependence that distinguishes the groups; modeling gender as a function of $X$ alone would not directly represent the changing gender probability across age levels. The age model is trained first by maximizing the conditional likelihood, after which the gender classifier is trained using both observed and MDN-sampled ages to reduce the gap between training and generation.

\begin{figure}[!ht]
\centering
\includegraphics[width=0.5\textwidth]{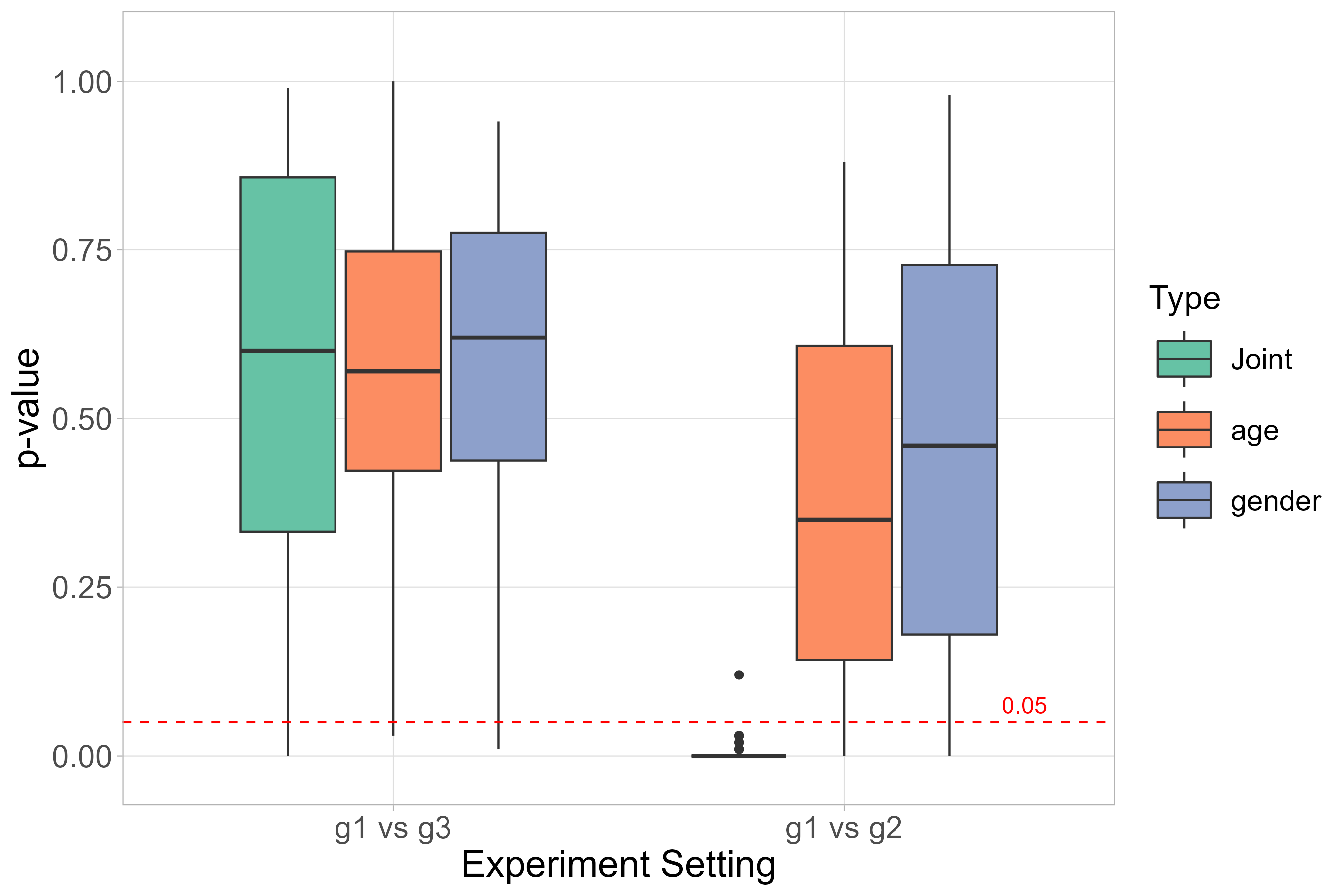}
\caption{Empirical distribution of $p$-values under the null setting (Group~1 vs.\ Group~3) and the alternative setting (Group~1 vs.\ Group~2). The dashed red line denotes $\alpha=0.05$.}
\label{fig:multivariate_boxplot_results}
\end{figure}

Figure~\ref{fig:multivariate_boxplot_results} shows the empirical distributions of the resulting $p$-values. In the null setting, Group~1 versus Group~3, the joint $p$-values are broadly distributed over $[0,1]$, indicating satisfactory type-I error control. In the alternative setting, Group~1 versus Group~2, the marginal test for gender produces $p$-values that remain broadly spread over $[0,1]$, showing essentially no power. The marginal test for age shows only a moderate downward shift, with median $p$-value around $0.3$, reflecting weak sensitivity to the group difference. In contrast, the joint test produces $p$-values concentrated near zero, with nearly all repetitions falling below the $0.05$ threshold. This contrast highlights the advantage of jointly testing the full multivariate conditional distribution: when the main signal lies in the dependence structure between response components rather than in their marginals, marginal tests can lose substantial power, whereas the proposed joint max-type test successfully detects the difference.

\section{Discussion and Future Work}\label{Sec:Discussion}

This paper shows that conditional generative models can serve as useful inference tools for two-sample testing of conditional distributions, not merely as devices for prediction or simulation. The proposed framework is designed for settings where classical approaches based on conditional density ratios or local smoothing become difficult, particularly with high-dimensional covariates and multivariate responses. The theory shows that valid calibration is still possible when the conditional laws are learned from data, and the experiments illustrate that the method is sensitive to changes beyond conditional means, including variance, covariance, and dependence-structure shifts.

Several directions remain for future work. First, it would be useful to develop more primitive and model-specific conditions under which modern conditional generators satisfy the high-level approximation assumptions used in our theory, particularly for diffusion-based generators and other neural conditional density estimators. Second, the finite-sample performance of the procedure can depend on implementation choices such as kernel bandwidths, generator architecture, the number of sample splits, the number of generated samples, bootstrap size, and the optimization accuracy of the alternating maximization algorithm. A systematic study of these choices could lead to more adaptive and computationally efficient implementations. Third, while the present work focuses on global testing, future work could develop diagnostic or localization tools to identify which regions of the covariate space or which components of a multivariate response are responsible for rejection. Finally, the cross-generation principle may be useful beyond the two-sample setting considered here, including multi-sample conditional homogeneity testing, conditional distribution monitoring under temporal or distributional drift, fairness auditing, simulator validation, and testing problems involving structured responses such as images, text, or networks.

\section*{Acknowledgments}
 This paper is a substantially revised version of Chapter 5, “A Two-Sample Conditional Distribution Test via Generative Adversarial Networks,” from Hanjia Gao’s PhD thesis at the University of Illinois Urbana-Champaign in 2024. The authors would like to thank Yi Zhang for helpful comments on an early draft.

\vspace{10em}
\begin{center}    
\section*{Supplement to ``Testing Equality of Conditional Distributions via Generative Models"}
\end{center}

The supplementary material is organized as follows.
Appendix~\ref{Appendix:Additional_simulation_results} contains additional simulation results, including the unadjusted empirical rejection rates corresponding to the size-adjusted results reported in the simulation studies.
Appendix~\ref{Appendix:Implementation} provides implementation details of the proposed testing procedure used in the simulation studies and real data applications. 
Appendix~\ref{Appendix:Counter} presents two counterexamples illustrating the necessity of the proposed test statistic and several key assumptions. 
Appendix~\ref{Appendix:Proof} contains the proofs of all main results in the article.
For completeness, Appendix~\ref{Appendix:EmpTheory} reviews several empirical process results used throughout the theoretical analysis. 
Finally, Appendix~\ref{Appendix:AuxDoubleRobustness} and Appendix~\ref{Appendix:AuxBoot} collect the auxiliary lemmas and their detailed proofs.
\vspace{5pt}

\appendix
\section{Additional simulation results} \label{Appendix:Additional_simulation_results}

Tables~\ref{tab:modelA_full} and~\ref{tab:modelB_full} report the unadjusted empirical rejection rates ($H_1$) corresponding to the size-adjusted results presented in Tables~\ref{tab:modelA} and~\ref{tab:modelB} of the main text. The size-adjusted power ($H_1$ Adj) is obtained by replacing the nominal $5\%$ level with the empirical $5\%$ quantile of the null $p$-values, and is the primary metric used for comparison in the main text. The unadjusted rates are included here for completeness and to allow assessment of the degree of size distortion for each method.

\begin{table}[H]
\centering
\caption{Empirical rejection rates and size-adjusted power (\%) under Model A. The HL methods are not included for $D_y=3$ as \cite{hu2024two} is designed for scalar responses. All results are based on $500$ independent trials.}
\label{tab:modelA_full}
\scalebox{.8}{
\resizebox{\textwidth}{!}{
\begin{tabular}{ccl|ccccccc}
        \multirow{2}{*}{$(p, D_y)$} & \multirow{2}{*}{$\alpha_2$} & & \multicolumn{3}{c}{max-type} & HL-KLR & HL-LL & HL-LR & HL-LRS \\  
         & & & Oracle & MDN & CDM & Practical & Practical & Practical & Practical \\ \hline

\multirow{15}{*}{$(5,\ 1)$}
& $0.0$ & $H_0$ & 6 & 4.2 & 7.8 & 5 & 5.6 & 5.8 & 4.6 \\\cline{2-10}
& \multirow{2}{*}{$0.1$} & $H_1$ & 5.8 & 6.2 & 9.4 & 12 & 20.2 & 58 & 34 \\ 
& & $H_1$ Adj & 4.6 & 6.8 & 5.6 & 12.6 & 19.2 & 55.6 & 35.6 \\\cline{2-10}
& \multirow{2}{*}{$0.2$} & $H_1$ & 9.2 & 8 & 15.6 & 32 & 44.4 & 99.2 & 86.6 \\ 
& & $H_1$ Adj & 7.6 & 9.2 & 11.2 & 32.8 & 43.8 & 99 & 87.4 \\\cline{2-10}
& \multirow{2}{*}{$0.3$} & $H_1$ & 29.4 & 14.2 & 29.6 & 61.8 & 69.6 & 100 & 99.4 \\ 
& & $H_1$ Adj & 25.2 & 16.8 & 21.4 & 62.4 & 69.6 & 100 & 99.6 \\\cline{2-10}
& \multirow{2}{*}{$0.5$} & $H_1$ & 99.4 & 70.2 & 73.6 & 96.8 & 97.2 & 100 & 100 \\ 
& & $H_1$ Adj & 99.4 & 72.2 & 65.4 & 97.2 & 97 & 100 & 100 \\\cline{2-10}
& \multirow{2}{*}{$0.8$} & $H_1$ & 100 & 95.6 & 92 & 100 & 100 & 100 & 100 \\ 
& & $H_1$ Adj & 100 & 96 & 90.6 & 100 & 100 & 100 & 100 \\\cline{2-10}
& \multirow{2}{*}{$1.0$} & $H_1$ & 100 & 99.8 & 99 & 100 & 100 & 100 & 100 \\ 
& & $H_1$ Adj & 100 & 100 & 97.8 & 100 & 100 & 100 & 100 \\\cline{1-10}

\multirow{15}{*}{$(100,\ 1)$}
& $0.0$ & $H_0$ & 5.2 & 4.4 & 4.8 & 14.2 & 5 & 8 & 8.2 \\\cline{2-10}
& \multirow{2}{*}{$0.1$} & $H_1$ & 6.2 & 4 & 2.4 & 14.2 & 10.8 & 58 & 32.4 \\ 
& & $H_1$ Adj & 5.2 & 4 & 4 & 4.8 & 11 & 51.8 & 27.4 \\\cline{2-10}
& \multirow{2}{*}{$0.2$} & $H_1$ & 6 & 7 & 4.6 & 14.6 & 32 & 99 & 85 \\ 
& & $H_1$ Adj & 4.6 & 7 & 6 & 5 & 32.8 & 97.6 & 81.8 \\\cline{2-10}
& \multirow{2}{*}{$0.3$} & $H_1$ & 10.8 & 9.4 & 5.6 & 16.4 & 60 & 100 & 99.8 \\ 
& & $H_1$ Adj & 7.6 & 9.4 & 7 & 5.8 & 60.6 & 100 & 99.6 \\\cline{2-10}
& \multirow{2}{*}{$0.5$} & $H_1$ & 39.6 & 35.8 & 21.8 & 18.8 & 93.2 & 100 & 100 \\ 
& & $H_1$ Adj & 38 & 35.8 & 23.8 & 7.4 & 93.6 & 100 & 100 \\\cline{2-10}
& \multirow{2}{*}{$0.8$} & $H_1$ & 84.8 & 84.4 & 75 & 27.6 & 100 & 100 & 100 \\ 
& & $H_1$ Adj & 83.2 & 84.4 & 77 & 12 & 100 & 100 & 100 \\\cline{2-10}
& \multirow{2}{*}{$1.0$} & $H_1$ & 99.2 & 94.2 & 93.8 & 33 & 100 & 100 & 100 \\ 
& & $H_1$ Adj & 98.8 & 94.2 & 94 & 19.4 & 100 & 100 & 100 \\\cline{1-10}

\multirow{15}{*}{$(100,\ 3)$}
& $0.0$ & $H_0$ & 6.4 & 5 & 3 & -- & -- & -- & -- \\\cline{2-10}
& \multirow{2}{*}{$0.1$} & $H_1$ & 7.4 & 6.6 & 5.2 & -- & -- & -- & -- \\ 
& & $H_1$ Adj & 5 & 6.6 & 7 & -- & -- & -- & -- \\\cline{2-10}
& \multirow{2}{*}{$0.2$} & $H_1$ & 12.4 & 13 & 8.2 & -- & -- & -- & -- \\ 
& & $H_1$ Adj & 9.2 & 13 & 10.4 & -- & -- & -- & -- \\\cline{2-10}
& \multirow{2}{*}{$0.3$} & $H_1$ & 29.8 & 23 & 17.8 & -- & -- & -- & -- \\ 
& & $H_1$ Adj & 23.8 & 23 & 18.8 & -- & -- & -- & -- \\\cline{2-10}
& \multirow{2}{*}{$0.5$} & $H_1$ & 68.6 & 55.6 & 53.4 & -- & -- & -- & -- \\ 
& & $H_1$ Adj & 64.8 & 55.6 & 53.8 & -- & -- & -- & -- \\\cline{2-10}
& \multirow{2}{*}{$0.8$} & $H_1$ & 84.8 & 85.8 & 86.4 & -- & -- & -- & -- \\ 
& & $H_1$ Adj & 83.2 & 85.8 & 86.4 & -- & -- & -- & -- \\\cline{2-10}
& \multirow{2}{*}{$1.0$} & $H_1$ & 92.8 & 91.4 & 91.2 & -- & -- & -- & -- \\ 
& & $H_1$ Adj & 92.8 & 91.4 & 91.2 & -- & -- & -- & -- \\\cline{1-10}
\end{tabular}}}
\end{table}

\begin{table}[H]
    \centering
    \caption{Empirical rejection rates (\%) and size-adjusted power under Model B. All results are based on $500$ independent trials. ``Same'' and ``Different'' refer to whether $X_1$ and $X_2$ share the same marginal distribution. The HL methods are not included for $D_y=3$ as \cite{hu2024two} is designed for scalar responses.}
    \label{tab:modelB_full}
    \scalebox{.9}{
    \resizebox{\textwidth}{!}{
    \begin{tabular}{ccccl|ccccccc}
    \multirow{2}{*}{Covariate} & \multirow{2}{*}{$p$} & \multirow{2}{*}{$D_y$} & \multirow{2}{*}{$\alpha_2$} & & \multicolumn{3}{c}{max-type} & HL-KLR & HL-LL & HL-LR & HL-LRS \\
    & & & & & Oracle & MDN & CDM & Practical & Practical & Practical & Practical \\ \hline

    \multirow{9}{*}{Same} & \multirow{9}{*}{$100$} & \multirow{9}{*}{$1$}
    & $0.0$ & $H_0$ & 7.2 & 3.2 & 3.8 & 4.8 & 6.2 & 7.2 & 7.6 \\\cline{4-12}
    & & & \multirow{2}{*}{$1$} & $H_1$ & 46.4 & 15 & 22.8 & 68.8 & 6.4 & 6.4 & 7.8 \\
    & & & & $H_1$ Adj & 38.6 & 20.2 & 27.4 & 69.6 & 4.8 & 4 & 4.2 \\\cline{4-12}
    & & & \multirow{2}{*}{$2$} & $H_1$ & 100 & 100 & 99.8 & 100 & 7.2 & 4.4 & 6.8 \\
    & & & & $H_1$ Adj & 100 & 100 & 99.8 & 100 & 5.4 & 3 & 3.4 \\\cline{4-12}
    & & & \multirow{2}{*}{$3$} & $H_1$ & 100 & 100 & 99.8 & 100 & 8.4 & 4.4 & 6.2 \\
    & & & & $H_1$ Adj & 100 & 100 & 99.8 & 100 & 5.8 & 2.8 & 3.2 \\\cline{4-12}
    & & & \multirow{2}{*}{$4$} & $H_1$ & 100 & 100 & 100 & 100 & 9.4 & 4 & 5.4 \\
    & & & & $H_1$ Adj & 100 & 100 & 100 & 100 & 6.4 & 3 & 3.2 \\\cline{1-12}

    \multirow{27}{*}{Different} & \multirow{9}{*}{$5$} & \multirow{9}{*}{$1$}
    & $0.0$ & $H_0$ & 8.2 & 4.2 & 4.4 & 2.2 & 6 & 6.8 & 5.6 \\\cline{4-12}
    & & & \multirow{2}{*}{$1$} & $H_1$ & 63.2 & 30.2 & 57 & 99.6 & 6.8 & 6.6 & 5.8 \\
    & & & & $H_1$ Adj & 55.8 & 30.2 & 60.4 & 99.8 & 5.6 & 5 & 5.6 \\\cline{4-12}
    & & & \multirow{2}{*}{$2$} & $H_1$ & 100 & 100 & 96.4 & 100 & 7.8 & 7 & 5.8 \\
    & & & & $H_1$ Adj & 100 & 100 & 96.6 & 100 & 6.6 & 5.2 & 5.4 \\\cline{4-12}
    & & & \multirow{2}{*}{$3$} & $H_1$ & 100 & 100 & 99.6 & 100 & 8.8 & 7 & 5.4 \\
    & & & & $H_1$ Adj & 100 & 100 & 99.6 & 100 & 8.2 & 5.2 & 5 \\\cline{4-12}
    & & & \multirow{2}{*}{$4$} & $H_1$ & 100 & 100 & 100 & 100 & 9.2 & 6.4 & 5.4 \\
    & & & & $H_1$ Adj & 100 & 100 & 100 & 100 & 7.8 & 5.4 & 4.8 \\\cline{2-12}
    
    & \multirow{9}{*}{$100$} & \multirow{9}{*}{$1$}
    & $0.0$ & $H_0$ & 7.2 & 4 & 5 & 39.8 & 53.2 & 5.6 & 6.4 \\\cline{4-12}
    & & & \multirow{2}{*}{$1$} & $H_1$ & 46.4 & 30 & 49.2 & 50.8 & 47.2 & 5.2 & 6.4 \\
    & & & & $H_1$ Adj & 38.6 & 31.4 & 49.2 & 8 & 3.6 & 4.8 & 5 \\\cline{4-12}
    & & & \multirow{2}{*}{$2$} & $H_1$ & 100 & 99.6 & 100 & 72.4 & 46.6 & 4.2 & 6.2 \\
    & & & & $H_1$ Adj & 100 & 99.6 & 100 & 25.8 & 2.6 & 4 & 4.2 \\\cline{4-12}
    & & & \multirow{2}{*}{$3$} & $H_1$ & 100 & 100 & 100 & 90.4 & 46.6 & 4.2 & 5.6 \\
    & & & & $H_1$ Adj & 100 & 100 & 100 & 54.6 & 2.6 & 3.8 & 4.2 \\\cline{4-12}
    & & & \multirow{2}{*}{$4$} & $H_1$ & 100 & 100 & 100 & 98.2 & 46.6 & 4 & 5.2 \\
    & & & & $H_1$ Adj & 100 & 100 & 100 & 80.2 & 3 & 3.8 & 4.2 \\\cline{2-12}

    & \multirow{9}{*}{$100$} & \multirow{9}{*}{$3$}
    & $0.0$ & $H_0$ & 6 & 7.2 & 4.4 & -- & -- & -- & -- \\\cline{4-12}
    & & & \multirow{2}{*}{$1$} & $H_1$ & 100 & 84 & 100 & -- & -- & -- & -- \\
    & & & & $H_1$ Adj & 100 & 79.8 & 100 & -- & -- & -- & -- \\\cline{4-12}
    & & & \multirow{2}{*}{$2$} & $H_1$ & 100 & 100 & 100 & -- & -- & -- & -- \\
    & & & & $H_1$ Adj & 100 & 100 & 100 & -- & -- & -- & -- \\\cline{4-12}
    & & & \multirow{2}{*}{$3$} & $H_1$ & 100 & 100 & 100 & -- & -- & -- & -- \\
    & & & & $H_1$ Adj & 100 & 100 & 100 & -- & -- & -- & -- \\\cline{4-12}
    & & & \multirow{2}{*}{$4$} & $H_1$ & 100 & 100 & 100 & -- & -- & -- & -- \\
    & & & & $H_1$ Adj & 100 & 100 & 100 & -- & -- & -- & -- \\\cline{1-12}

    \end{tabular}}}
\end{table}

\section{Implementation Details of Numerical Studies} \label{Appendix:Implementation}

\subsection{Computation of the Bootstrap Test Statistic}\label{Appendix:CompBootTestStat}

We extend the discussions in Section~\ref{Sec:Method_TestStat} to approximate $\widehat{U}^{b}$ for any fixed multipliers $\{\varepsilon_i\}_{i=1}^{n}$ by the alternating maximization algorithm. The blockwise updating formulas are summarized in Proposition~\ref{Prop:Update_U_boot}.

\begin{prop}\label{Prop:Update_U_boot}
Let $\alpha_f^{b(r-1)}, \beta_g^{b(r-1)}, \gamma_h^{b(r-1)}$ denote the updates at the $(r-1)$-th iteration to approximate $\widehat{U}^{b}$, and let $e=(e_1,\cdots,e_n)^{\top}$ with $e_i = \varepsilon_i - \bar\varepsilon$. We have that
\BEqn
    \alpha_f^{b(r)}
&=& \argmax\limits_{\alpha_f^b:~f\in\ch} 
    \lrabs{\frac{1}{n}
           \sum\limits_{i=1}^{n} 
           \alpha_{f,i}^b \beta_{g,i}^{b(r-1)} \gamma_{h,i}^{b(r-1)} e_i}
  = \frac{\Omega_1 p_f^{b(r)}}
         {\sqrt{p_f^{b(r)\top} \Omega_1 p_f^{b(r)}}}, \quad
    p_f^{b(r)} = e \odot \beta_g^{b(r-1)} \odot \gamma_h^{b(r-1)}, \\
    \beta_g^{b(r)}
&=& \argmax\limits_{\beta_g^b:~g\in\ch} 
    \lrabs{\frac{1}{n}
           \sum\limits_{i=1}^{n} 
           \alpha_{f,i}^{b(r)} \beta_{g,i}^b \gamma_{h,i}^{b(r-1)} e_i} 
 = \frac{\Omega_2 p_g^{b(r)}}
        {\sqrt{p_g^{b(r)\top} \Omega_2 p_g^{b(r)}}}, \quad
    p_g^{b(r)} = e \odot \alpha_f^{b(r)} \odot \gamma_h^{b(r-1)}, \\
    \gamma_h^{b(r)}
&=& \argmax\limits_{\gamma_h^b:~h\in\ch'} 
    \lrabs{\frac{1}{n}
           \sum\limits_{i=1}^{n} 
           \alpha_{f,i}^{b(r)} \beta_{g,i}^{b(r)} \gamma_{h,i}^b e_i}
 = \frac{\Omega_3 p_h^{b(r)}}
        {\sqrt{p_h^{b(r)\top} \Omega_3 p_h^{b(r)}}}, \quad
   p_h^{b(r)} = e \odot \alpha_f^{b(r)} \odot \beta_g^{b(r)},
\EEqn
where $\Omega_1,\Omega_2,\Omega_3$ are given by Proposition~\ref{Prop:Update_U}. Again, if the denominator in any of the above updates is zero, then we set the updated vector equal to the previous iterate.
\end{prop}

Compared with Proposition~\ref{Prop:Update_U}, the bootstrap updates differ from those in Proposition~\ref{Prop:Update_U} only through the additional multiplier weights contained in the vectors $p_f^{b(r)}, p_g^{b(r)}, p_h^{b(r)}$, thus we present the updating formula without proof.

The full alternating maximization algorithm to approximate $\widehat{U}^{b}$ is given by Algorithm~\ref{Alg:boot}. 

\begin{algorithm}[h!]
\caption{Alternating maximization for approximating $\widehat{U}^{b}$}\label{Alg:boot}

\begin{algorithmic}[1]
\Require $\{(X_{1i},Y_{1i},\widehat{Y}_{1i},X_{2i},Y_{2i},\widehat{Y}_{2i})\}_{i=1}^{n} \in \br^{2p+4q}$ \\
\hspace{1.5em} $\{\varepsilon_i\}_{i=1}^{n} \stsim{iid} \cn(0,1)$ independent of $\{(X_{1i},Y_{1i},\widehat{Y}_{1i},X_{2i},Y_{2i},\widehat{Y}_{2i})\}_{i=1}^{n}$ \\
\hspace{1.5em} tolerance level $\delta>0$, maximum iteration number $R$
\Ensure the bootstrap test statistic $\widehat{U}^{b}$

\vspace{3mm}
\Procedure{Bootstrap Test Statistic $\widehat{U}^{b}$}{}
\State Initialize $r=0$
\State Initialize $\alpha_f^{b(0)},\beta_g^{b(0)},\gamma_h^{b(0)} \in \br^{n}$
\State Compute $\Omega_1, \Omega_2, \Omega_3 \in \br^{n\times n}$ given by Proposition~\ref{Prop:Update_U}
\State Compute $e=(e_1,\cdots,e_n)^{\top}\in\br^n$ with $e_i = \varepsilon_i-\bar\varepsilon$
\State Initialize $\widehat{U}^{b(-1)} = -\infty$ and $\widehat{U}^{b(0)} = \lrabs{\frac{1}{n} \sum\limits_{i=1}^{n} \alpha_{f,i}^{b(0)} \beta_{g,i}^{b(0)} \gamma_{h,i}^{b(0)} e_i}$.
\While{$|\widehat{U}^{b(r)} - \widehat{U}^{b(r-1)}| > \delta$ and $r\le R$}
    \State $r \leftarrow r+1$    
    \State Compute $\alpha_f^{b(r)}, \beta_g^{b(r)}, \gamma_h^{b(r)}$ by updating formulas in Proposition~\ref{Prop:Update_U_boot}
    \Comment{Blockwise update}
    
    
    
    \State Evaluate $\widehat{U}^{b(r)} = \lrabs{ \frac{1}{n} \sum\limits_{i=1}^{n} \alpha_{f,i}^{b(r)} \beta_{g,i}^{b(r)} \gamma_{h,i}^{b(r)} e_i}$
    \Comment{Check for early stopping}
\EndWhile

\State Approximate $\widehat{U}^{b} = \lrabs{ \frac{1}{n} \sum\limits_{i=1}^{n} \alpha_{f,i}^{b(r)} \beta_{g,i}^{b(r)} \gamma_{h,i}^{b(r)} e_i}$
\EndProcedure
\end{algorithmic}
\end{algorithm}

\subsection{Overview of Generative Models}\label{Appendix:MDN_CDM_overview}

We first provide implementation details for the conditional generative models used to estimate the oracle conditional generators. These models are used purely as flexible conditional distribution estimators, while the asymptotic theory in the main text only relies on the high-level approximation conditions in Assumption \ref{Assumpt:double_robustness}.

Let $p_{Y|X}(\cdot|x)$ denote the conditional density of $Y$ given $X=x$. MDN parameterizes $p_{Y|X}(\cdot|x)$ as a finite mixture
\begin{equation*}
    p_{Y|X}(\cdot|x)
  = \sum_{k=1}^K \pi_k(x) p_k(\cdot|x),
\end{equation*}
where $\pi_k(x)\ge 0$ are mixture weights with $\sum_{k=1}^K \pi_k(x)=1$, and $p_k(\cdot|x)$ are parametric component densities, typically chosen as multivariate Gaussian distributions $\cn(\mu_k(x), \Sigma_k(x))$. The parameters $\{\pi_k(x), \mu_k(x), \Sigma_k(x)\}_{k=1}^{K}$ are learned via a neural network by minimizing the negative log-likelihood
\begin{equation*}
    \cl(\theta)
  = - \sum_{i=1}^{n}
    \log\Big( \sum_{k=1}^K
              \pi_k(x_i)\,
              \phi(y_i;\mu_k(x_i),\Sigma_k(x_i)) \Big),
\end{equation*}
where $\phi(\cdot;\mu,\Sigma)$ denotes the Gaussian density. After training, conditional samples can be generated by first sampling a mixture component according to $\{\pi_k(x)\}_{k=1}^K$ and then drawing from the corresponding Gaussian distribution.

In contrast, CDM approximates the conditional distribution implicitly through a sequence of noise-perturbed variables and a denoising mechanism. Let $t\in\{0,1,\cdots,T-1\}$ and $\{\alpha_t\}_{t=1}^{T}$ be a noise schedule. For each $y_i$, define
\begin{equation*}
    y_{it} = \sqrt{\alpha_t} y_i + \sqrt{1-\alpha_t} \varepsilon_i, 
    \qquad
    \varepsilon_i \sim \cn(0, I_q).
\end{equation*}
Let $\varepsilon_{\theta}$ be a neural network taking $(y_{it},t,x_i)$ as input. The network is trained by minimizing
\begin{equation*}
    \cl(\theta)
  = \sum_{i=1}^{n}
    \|\varepsilon_i - \varepsilon_{\theta}(y_{it},t,x_i)\|_2^2.
\end{equation*}
After training, conditional samples are generated through the corresponding reverse denoising process.

We omit architecture-specific choices such as network depth, optimizer selection, and tuning procedures, since these are implementation-dependent and not essential to the theoretical development.

\subsection{Numerical Simulation}

\subsubsection{Mixture Density Network (MDN)}

\paragraph{Architecture}
The MDN follows the formulation in Appendix~\ref{Appendix:MDN_CDM_overview}, with $K = 5$ Gaussian components and diagonal covariance matrices. The network adopts a trunk-and-head architecture: the trunk maps the covariate $X \in \mathbb{R}^{d_x}$ to a shared $W$-dimensional hidden representation through $L_{\mathrm{trunk}}$ fully connected layers with LeakyReLU activations, and three separate head networks then map this shared representation to $\pi$, $\mu$, and $\sigma$ respectively, each consisting of $L_{\mathrm{head}}$ fully connected layers with LeakyReLU activations and a final output projection. A softmax activation is applied to the $\pi$ outputs to ensure valid mixture weights, and component standard deviations are clipped to $[10^{-4}, 10]$ for numerical stability. Inputs $X$ and $Y$ are standardized to zero mean and unit variance before training. The Adam optimizer \citep{kingma2014adam} is used with a \texttt{ReduceLROnPlateau} scheduler and early stopping; hyperparameters are selected via Optuna \citep{akiba2019optuna} by minimizing the validation negative log-likelihood, and the chosen configurations are reported in Table~\ref{tab:mdn_hyperparams}.

\begin{table}[h!]
    \centering
    \scalebox{.95}{
    \resizebox{\textwidth}{!}{
    \begin{tabular}{lcccccccc}
        \toprule
        & \multicolumn{3}{c}{Model A} & \multicolumn{3}{c}{Model B} & Model B \\
        \cmidrule(lr){2-4} \cmidrule(lr){5-7}
        Hyperparameter & $d_x=5$ & $d_x=100$ & $d_x=100$ & $d_x=5$ & $d_x=100$ & $d_x=100$ & $d_x=100$ \\
        & $d_y=1$ & $d_y=1$ & $d_y=3$ & $d_y=1$ & $d_y=1$ & $d_y=3$ & $d_y=1$ (same) \\
        \midrule
        Mixture components $K$           & 5     & 5     & 5     & 5     & 5     & 5     & 5     \\
        Trunk width $W$                  & 128   & 256   & 256   & 32    & 16    & 16    & 16    \\
        Trunk layers $L_{\mathrm{trunk}}$& 1     & 1     & 1     & 0     & 1     & 1     & 1     \\
        Trunk LeakyReLU slope            & 0.342 & 0.454 & 0.500 & 0.384 & 0.499 & 0.500 & 0.499 \\
        Trunk dropout                    & 0.0   & 0.0   & 0.0   & 0.0   & 0.0   & 0.0   & 0.0   \\
        Head width                       & 64    & 128   & 64    & 32    & 256   & 128   & 256   \\
        Head layers $L_{\mathrm{head}}$  & 1     & 1     & 1     & 0     & 1     & 1     & 1     \\
        Head LeakyReLU ($\pi$)           & 0.318 & 0.481 & 0.241 & 0.260 & 0.391 & 0.420 & 0.391 \\
        Head LeakyReLU ($\mu$)           & 0.496 & 0.492 & 0.500 & 0.395 & 0.454 & 0.453 & 0.454 \\
        Head LeakyReLU ($\sigma$)        & 0.186 & 0.032 & 0.051 & 0.486 & 0.013 & 0.030 & 0.013 \\
        Learning rate                    & $1.00\times10^{-3}$ & $1.00\times10^{-3}$ & $1.00\times10^{-3}$ & $1.00\times10^{-3}$ & $1.00\times10^{-3}$ & $1.00\times10^{-3}$ & $1.00\times10^{-3}$ \\
        Weight decay                     & $8.44\times10^{-5}$ & $6.84\times10^{-3}$ & $2.81\times10^{-3}$ & $1.29\times10^{-6}$ & $0$ & $0$ & $0$ \\
        LR patience $p_{\mathrm{lr}}$    & 8     & 3     & 5     & 3     & 3     & 4     & 3     \\
        LR decay $\gamma$                & 0.330 & 0.110 & 0.189 & 0.275 & 0.471 & 0.443 & 0.471 \\
        Early stopping $p_{\mathrm{stop}}$& 30   & 30    & 30    & 20    & 20    & 20    & 20    \\
        Mini-batch size $B$              & 128   & 512   & 64    & 64    & 64    & 64    & 128   \\
        \bottomrule
    \end{tabular}}}
    \caption{Hyperparameters of the MDN selected by Optuna for each simulation setting.} \label{tab:mdn_hyperparams}
\end{table}

\subsubsection{Conditional Diffusion Model (CDM)}

\paragraph{Architecture}
The CDM follows the formulation in Appendix~\ref{Appendix:MDN_CDM_overview}, with a cosine noise schedule over $T$ discrete steps. The noise prediction network $\epsilon_\theta$ consists of a covariate encoder that maps $X \in \mathbb{R}^{d_x}$ to a $d_{\mathrm{emb}}$-dimensional embedding through a trunk MLP with batch normalization and LeakyReLU activations, and a diffusion head that takes the concatenation of the noisy response $Y_t$, a $d_{\mathrm{time}}$-dimensional sinusoidal time embedding, and the covariate embedding as input, outputting the predicted noise through a head MLP with LeakyReLU activations and optional dropout. For high-dimensional covariate settings ($d_x = 100$), FiLM conditioning \citep{perez2018film} is additionally applied in the head MLP, where the covariate embedding modulates the hidden features via learned affine transformations at each layer. The training procedure, input standardization, optimizer, and hyperparameter selection follow the same protocol as the MDN described above, with hyperparameters tuned by minimizing the validation noise-prediction loss; the chosen configurations are reported in Table~\ref{tab:cdm_hyperparams}.

\begin{table}[h!]
    \centering
    \scalebox{.95}{
    \resizebox{\textwidth}{!}{
    \begin{tabular}{lcccccccc}
        \toprule
        & \multicolumn{3}{c}{Model A} & \multicolumn{3}{c}{Model B} & Model B & Model C \\
        \cmidrule(lr){2-4} \cmidrule(lr){5-7}
        Hyperparameter & $d_x=5$ & $d_x=100$ & $d_x=100$ & $d_x=5$ & $d_x=100$ & $d_x=100$ & $d_x=100$ & $d_x=100$ \\
        & $d_y=1$ & $d_y=1$ & $d_y=3$ & $d_y=1$ & $d_y=1$ & $d_y=3$ & $d_y=1$ (same) & $d_y=10$ \\
        \midrule
        Diffusion steps $T$              & 150   & 150   & 150   & 150   & 150   & 150   & 150   & 200   \\
        Hidden dim $W$                   & 256   & 256   & 64    & 256   & 128   & 64    & 128   & 512   \\
        Trunk hidden layers              & 2     & 3     & 1     & 3     & 2     & 3     & 3     & 6     \\
        Head hidden layers               & 2     & 1     & 2     & 1     & 1     & 1     & 2     & 3     \\
        Time emb dim $d_{\mathrm{time}}$ & 128   & 128   & 128   & 32    & 64    & 32    & 64    & 256   \\
        Covariate emb dim $d_{\mathrm{emb}}$ & 64 & 16  & 128   & 128   & 64    & 64    & 64    & 256   \\
        LeakyReLU slope                  & 0.262 & 0.208 & 0.274 & 0.012 & 0.293 & 0.268 & 0.300 & 0.252 \\
        Dropout probability              & 0.0   & 0.1   & 0.0   & 0.1   & 0.1   & 0.1   & 0.1   & 0.202 \\
        Learning rate                    & $1.98\times10^{-3}$ & $2.97\times10^{-3}$ & $1.08\times10^{-3}$ & $7.20\times10^{-4}$ & $1.69\times10^{-3}$ & $2.49\times10^{-3}$ & $1.49\times10^{-3}$ & $1.82\times10^{-4}$ \\
        Weight decay                     & $6.67\times10^{-7}$ & $1.75\times10^{-6}$ & $4.92\times10^{-3}$ & $1.29\times10^{-6}$ & $1.19\times10^{-3}$ & $2.02\times10^{-3}$ & $7.88\times10^{-7}$ & $7.54\times10^{-4}$ \\
        LR patience $p_{\mathrm{lr}}$    & 3     & 6     & 7     & 3     & 3     & 4     & 4     & 2     \\
        LR decay $\gamma$                & 0.307 & 0.256 & 0.345 & 0.275 & 0.471 & 0.443 & 0.445 & 0.455 \\
        Early stopping $p_{\mathrm{stop}}$ & 20  & 25    & 20    & 20    & 20    & 20    & 20    & 25    \\
        Mini-batch size $B$              & 64    & 128   & 64    & 64    & 64    & 64    & 128   & 256   \\
        \bottomrule
    \end{tabular}}}
    \caption{Hyperparameters of the CDM selected by Optuna for each simulation setting.} \label{tab:cdm_hyperparams}
\end{table}

\subsubsection{Test statistic and bootstrap}

For all simulation settings, the Gaussian kernel is used for both $K$ and $K'$, with bandwidths set by the median heuristic: $\gamma_x$ is computed as the median pairwise Euclidean distance of the pooled covariates $(X_1, X_2)$, $\gamma_y$ is computed from the pooled observed responses $(Y_1, Y_2)$, and $\gamma_{x_1}$, $\gamma_{x_2}$ are computed separately from $X_1$ and $X_2$ respectively for the interaction kernel $K'$. The alternating maximization algorithm is run until convergence
to approximate $\widehat{U}$, and the bootstrap critical value is obtained from $B = 100$ Gaussian multiplier samples at significance level $\alpha = 0.05$.

\subsection{Real Data Experiments}

\subsubsection{Experiment I: UTK-Face (Scalar Response)}

\paragraph{Generator}
The conditional distribution $P_{Y \mid X}$ is modeled by an MDN with $K = 20$ Gaussian components and diagonal covariance matrices, following the same trunk-and-head architecture and training protocol as in the simulation studies. The trunk maps the 512-dimensional ResNet-18 feature vector $X$ to a 32-dimensional hidden representation through 1 fully connected layer with LeakyReLU activation (slope 0.043), and each head network ($\pi$, $\mu$, $\sigma$) consists of 1 hidden layer of width 32 with LeakyReLU activations (slopes 0.472, 0.237, 0.354, respectively). The same MDN architecture and tuning protocol are used across all four cases, with the conditional generators fitted separately for each comparison, since the response $Y$ (age) and its relationship to the image features are identical across cases. The remaining hyperparameters, selected via Optuna by minimizing the validation negative log-likelihood, are summarized in Table~\ref{tab:mdn_exp1}.

\begin{table}[h!]
    \centering
    \scalebox{.9}{
    \begin{tabular}{lc}
        \toprule
        Hyperparameter & Value \\
        \midrule
        Mixture components $K$            & 20 \\
        Trunk width $W$                   & 32 \\
        Trunk layers $L_{\mathrm{trunk}}$ & 1 \\
        Trunk LeakyReLU slope             & 0.043 \\
        Head width                        & 32 \\
        Head layers $L_{\mathrm{head}}$   & 1 \\
        Head LeakyReLU ($\pi$)            & 0.472 \\
        Head LeakyReLU ($\mu$)            & 0.237 \\
        Head LeakyReLU ($\sigma$)         & 0.354 \\
        Learning rate                     & $1.00\times10^{-3}$ \\
        Weight decay                      & $1.28\times10^{-7}$ \\
        LR patience $p_{\mathrm{lr}}$     & 4 \\
        LR decay $\gamma$                 & 0.654 \\
        Early stopping $p_{\mathrm{stop}}$& 30 \\
        Mini-batch size                   & 64 \\
        \bottomrule
    \end{tabular}}
    \caption{Hyperparameters of the MDN for Experiment~I (all cases)} \label{tab:mdn_exp1}
\end{table}

\paragraph{Test statistic and bootstrap}
The Gaussian kernel is used for both $K$ and $K'$, with bandwidths set by the median heuristic applied to the pooled samples. The alternating maximization algorithm is run until convergence 
and the bootstrap critical value is obtained from $B = 100$ Gaussian multiplier samples at significance level $\alpha = 0.05$, with $m = 20$ noise draws per test observation.

\subsubsection{Experiment II: UTK-Face (Multivariate Response)}

The joint conditional distribution $P_{Y^{(1)}, Y^{(2)} \mid X}$ is modeled via the sequential factorization
\begin{equation*}
    P(Y^{(1)}, Y^{(2)} \mid X) = P(Y^{(1)} \mid X) \cdot P(Y^{(2)} \mid Y^{(1)}, X),
\end{equation*}
where $P(Y^{(1)} \mid X)$ is an MDN and $P(Y^{(2)} \mid Y^{(1)}, X)$ is a binary classifier. For the marginal age and marginal gender tests, only the relevant component of the generator is used. The three testing scenarios and their corresponding generator configurations are summarized in Table~\ref{tab:exp2_generators}.

For the joint test, the age MDN is trained first by maximizing the conditional log-likelihood, after which the gender classifier is trained using the fixed age MDN. During training of the joint gender classifier, the age input is a weighted combination of observed and MDN-sampled ages with weights $\lambda_{\mathrm{true}} = 0.7$ and $\lambda_{\mathrm{sample}} = 0.3$, reducing the gap between training and generation; at inference, the age is first sampled from the MDN and then passed together with $X$ into the classifier to produce a Bernoulli sample of $Y^{(2)}$. The same generator settings are used for both the null (Group~1 vs.\ Group~3) and alternative (Group~1 vs.\ Group~2) comparisons.

\begin{table}[h!]
    \centering
    \scalebox{.85}{
    \begin{tabular}{lccc}
        \toprule
        & Marginal age & Marginal gender & Joint \\
        \midrule
        Response & $Y^{(1)}$ & $Y^{(2)}$ & $(Y^{(1)}, Y^{(2)})^\top$ \\
        Generator & Age MDN & Gender classifier & Age MDN + Gender classifier \\
        \midrule
        \multicolumn{4}{l}{\textit{Age MDN ($P_{Y^{(1)} \mid X}$)}} \\
        Mixture components $K$ & 20 & -- & 20 \\
        Trunk width & 32 & -- & 32 \\
        Trunk layers & 1 & -- & 1 \\
        Trunk LeakyReLU & 0.234 & -- & 0.234 \\
        Head width & 16 & -- & 16 \\
        Head layers & 1 & -- & 1 \\
        Head LeakyReLU ($\pi / \mu / \sigma$) & 0.281 / 0.216 / 0.399 & -- & 0.281 / 0.216 / 0.399 \\
        $\sigma$ clip range & $[10^{-4}, 0.6]$ & -- & $[10^{-4}, 0.6]$ \\
        Learning rate & $1.21\times10^{-3}$ & -- & $1.21\times10^{-3}$ \\
        Weight decay & $1.00\times10^{-3}$ & -- & $1.00\times10^{-3}$ \\
        \midrule
        \multicolumn{4}{l}{\textit{Gender classifier ($P_{Y^{(2)} \mid X}$ or $P_{Y^{(2)} \mid Y^{(1)}, X}$)}} \\
        Input & -- & $X$ & $(X, \widehat{Y}^{(1)})$ \\
        Hidden layers & -- & 3 & 3 \\
        Hidden width & -- & 128 & 128 \\
        LeakyReLU slope & -- & 0.188 & 0.188 \\
        Dropout & -- & 0.3 & 0.3 \\
        $\lambda_{\mathrm{true}} / \lambda_{\mathrm{sample}}$ & -- & -- & 0.7 / 0.3 \\
        Learning rate & -- & $6.75\times10^{-4}$ & $6.75\times10^{-4}$ \\
        Weight decay & -- & $9.01\times10^{-3}$ & $9.01\times10^{-3}$ \\
        \midrule
        \multicolumn{4}{l}{\textit{Shared training settings}} \\
        Optimizer & \multicolumn{3}{c}{Adam} \\
        LR patience $p_{\mathrm{lr}}$ & \multicolumn{3}{c}{8} \\
        LR decay $\gamma$ & \multicolumn{3}{c}{0.649} \\
        Early stopping $p_{\mathrm{stop}}$ & \multicolumn{3}{c}{19} \\
        Mini-batch size & \multicolumn{3}{c}{32} \\
        \bottomrule
    \end{tabular}}
    \caption{Generator configurations for the three testing scenarios in Experiment~II.} \label{tab:exp2_generators}
\end{table}

\paragraph{Test statistic and bootstrap.}
The test statistic and bootstrap procedure follow the same settings as Experiment~I: Gaussian kernel with median heuristic bandwidths, 
$B = 100$ Gaussian multiplier samples, $m = 20$ noise draws, and significance level $\alpha = 0.05$.

\section{Counterexamples}\label{Appendix:Counter}

\begin{example}[Necessity of the interaction term]\label{Example:interaction}
We first illustrate why the interaction term $h$ is necessary. Consider $X_1,X_2\in\{0,1\}$ with joint probability $\bp(X_1=0, X_2=0) = 0.4$, $\bp(X_1=0, X_2=1) = 0.3$, $\bp(X_1=1, X_2=0) = 0.2$ and $\bp(X_1=1, X_2=1) = 0.1$. Then $X_1$ and $X_2$ are dependent, while Assumption \ref{Assumpt:X1X2} is satisfied. Let $\varepsilon\in(0,\frac{1}{2})$, we construct $Y_1,Y_2\in\{0,1\}$ with conditional distributions:
\begin{equation*}
    \bp(Y_1=1 \mid X_1=x) 
  = \left\{
    \begin{array}{ll}
        \frac{1}{2}+\varepsilon, & x=0, \\
        \frac{1}{2}-\varepsilon, & x=1.
    \end{array}
    \right.
  \qquad
    \bp(Y_2=1 \mid X_2=x) 
  = \left\{
    \begin{array}{ll}
        \frac{1}{2}-\varepsilon, & x=0, \\
        \frac{1}{2}+\varepsilon, & x=1.
    \end{array}
    \right.
\end{equation*}
Then the null hypothesis fails since $\varepsilon > 0$. Let $\psi$ be a bounded continuous function with $\psi(0) \neq \psi(1)$, and consider the bounded continuous kernel $K((x,y),(x',y')) = 1 + \psi(y)\psi(y')$. For every fixed $x$, $K_x(y,y') = 1 + \psi(y)\psi(y')$ is characteristic on $\{0,1\}$, since its kernel mean embedding satisfies
\begin{equation*}
    \mu_Y(\cdot) 
  = \be[K_x(Y,\cdot)] 
  = 1 + \be[\psi(Y)] \psi(\cdot),
\end{equation*}
and $\be\{\psi(Y)\}=P(Y=1)\psi(1)+\{1-P(Y=1)\}\psi(0)$ uniquely determines $P(Y=1)$. 
Since the kernel $K((x,y),(x',y')) = 1 + \psi(y)\psi(y')$ does not depend on $x$, every function in the associated RKHS takes the form $f(x,y) = a+b\psi(y)$ for some constants $a,b\in\br$, thus $f(x,y)-f(x,y')$ does not depend on $x$. Define $\Delta_f = f(x,1)-f(x,0)$ and $\Delta_g = g(x,1)-g(x,0)$.
\BEqn
    A_f(x)
&=& \be[f(X_1,Y_1) - f(X_1,Y_1^{c\ast}) \mid X_1=x] \\
&=& \Delta_f \bigp{\bp(Y_1=1|X_1=x) - \bp(Y_2=1|X_2=x)} \\
&=& 2\varepsilon \Delta_f \bigp{\bone\{x=0\} - \bone\{x=1\}},
\EEqn
and similarly, 
\begin{equation*}
    B_g(x)
  = \be[g(X_2,Y_2) - g(X_2,Y_2^{c\ast}) \mid X_2=x]
  = - 2\varepsilon \Delta_g \bigp{\bone\{x=0\} - \bone\{x=1\}}.
\end{equation*}
If $h\equiv1$, the reduced population discrepancy involves
\BEqn
& & \bbe{A_f(X_1) B_g(X_2)} \\
&=& -4\varepsilon^2 \Delta_f \Delta_g 
    \bbe{\bigp{\bone\{X_1=0\} - \bone\{X_1=1\}} \bigp{\bone\{X_2=0\} - \bone\{X_2=1\}}} \\
&=& -4\varepsilon^2 \Delta_f \Delta_g 
    \bigp{\bp(X_1=0,X_2=0) + \bp(X_1=1,X_2=1) - \bp(X_1=0,X_2=1) - \bp(X_1=1,X_2=0)} \\
&=& 0.
\EEqn
Hence the reduced discrepancy with $h\equiv1$ vanishes for all $f,g\in\ch$, even though the conditional distributions are different. In contrast, since $\bh'$ is dense in $L_2(P_{X_1,X_2})$, it can approximate the function $h(x_1,x_2) = \bigp{\bone\{x_1=0\} - \bone\{x_1=1\}} \bigp{\bone\{x_2=0\} - \bone\{x_2=1\}}$, and therefore,
\begin{equation*}
    \bbe{A_f(X_1)B_g(X_2)h(X_1,X_2)} 
  = -4\varepsilon^2 \Delta_f \Delta_g \bbe{h^2(X_1,X_2)} 
  = -4\varepsilon^2 \Delta_f \Delta_g,
\end{equation*}
which is nonzero whenever $\Delta_f\Delta_g\neq0$. Thus, the interaction function $h$ allows the discrepancy measure to adapt to the dependence structure of $(X_1,X_2)$ and prevents the cancellation phenomenon that may occur when using only the constant interaction $h\equiv1$.
\end{example}

\begin{example}[Necessity of the overlap condition]\label{Example:overlap}
In this example, we demonstrate the necessity of Assumption \ref{Assumpt:X1X2}. Consider $X_1\sim\mathrm{Bernoulli}(1/2)$ and $X_2=1-X_1$, then $P_{X_1}=P_{X_2}$, but the two conditioning variables never take the same value simultaneously, so Assumption \ref{Assumpt:X1X2} is violated. We construct $Y_1,Y_2\in\{0,1\}$ such that
\begin{equation*}
    P_{Y_1|X_1=0}\neq P_{Y_2|X_2=0},
    \qquad
    P_{Y_1|X_1=1}=P_{Y_2|X_2=1}.
\end{equation*}
Then the null hypothesis fails. 

For any $f,g\in\ch$ and any interaction function $h$, we have
\BEqn
& & \bbe{(f(X_1,Y_1) - f(X_1,Y_1^{c\ast})) (g(X_2,Y_2) - g(X_2,Y_2^{c\ast})) h(X_1,X_2)} \\
&=& \frac{1}{2} \bbe{f(0,Y_1)-f(0,Y_1^{c\ast})} \bbe{g(1,Y_2) - g(1,Y_2^{c\ast})} h(0,1) \\
& & + \frac{1}{2} \bbe{f(1,Y_1)-f(1,Y_1^{c\ast})} \bbe{g(0,Y_2) - g(0,Y_2^{c\ast})} h(1,0).
\EEqn
Since $P_{Y_1|X_1=1} = P_{Y_2|X_2=1}$, we have that $\bbe{f(1,Y_1)-f(1,Y_1^{c\ast})} = \bbe{g(1,Y_2) - g(1,Y_2^{c\ast})} = 0$ for any $f,g\in\ch$. Therefore, $\bbe{(f(X_1,Y_1) - f(X_1,Y_1^{c\ast})) (g(X_2,Y_2) - g(X_2,Y_2^{c\ast})) h(X_1,X_2)} = 0$ for any $f,g\in\ch$ and $h\in\bh'$. Thus, even with a nontrivial interaction function $h$, the discrepancy fails to detect the difference between the conditional distributions when the overlap condition is violated. This counterexample illustrates why Assumption \ref{Assumpt:X1X2} is necessary.
\end{example}

\section{Proof of Main Theorems}\label{Appendix:Proof}

To facilitate the analysis, we follow Section 2.1 of \cite{kosorok2008introduction} to introduce some common notation of the empirical processes that we will use throughout the proof.

In our setup, given the random sample $\{D_i^{\ast}:=(X_{1i},Y_{1i},Y_{1i}^{c\ast},X_{2i},Y_{2i},Y_{2i}^{c\ast})\}_{i=1}^{n}$ of independent draws from the probability measure $P$ on the sample space $\br^{2p+4q}$, we define the empirical measure as 
\begin{equation*}
    \bp_n = \frac{1}{n} \sum\limits_{i=1}^{n} \delta_{D_i^{\ast}},
\end{equation*}
where $\delta_x$ is the measure that assigns mass one at a single point $x$ and zero elsewhere. For any measurable function $\phi_{f,g,h}\in\Phi:~\br^{2p+4q}\rightarrow\br$, we denote 
\BEqn
    \bp_n \phi_{f,g,h}
&=& \frac{1}{n} \sum\limits_{i=1}^{n} 
    \phi_{f,g,h}(X_{1i},Y_{1i},Y_{1i}^{c\ast},X_{2i},Y_{2i},Y_{2i}^{c\ast}) \\
&=& \frac{1}{n} \sum\limits_{i=1}^{n} 
    \bigp{f(X_{1i}, Y_{1i}) - f(X_{1i}, Y_{1i}^{c\ast})}
    \bigp{g(X_{2i}, Y_{2i}) - g(X_{2i}, Y_{2i}^{c\ast})}
    h(X_{1i},X_{2i}).
\EEqn
Then for the class $\Phi=\{\phi_{f,g,h}:~f,g\in\ch, h\in\ch'\}$ of measurable functions, we can define an empirical process $\{\bp_n\phi_{f,g,h}: \phi_{f,g,h}\in\Phi\}$ and the random measure $\bg_n = \sqrt{n}(\bp_n - P)$. Furthermore, we use $\|\cdot\|_\Phi$ to denote the supremum map over $\Phi$, i.e. $\|\bg_n\|_\Phi = \sup\limits_{\phi_{f,g,h}\in\Phi} |\bg_n(\phi_{f,g,h})|$. Additionally, we use $\cp(\cs, \cb(\cs))$ to denote the set of all probability measures on the space $\cs$.

In the numerical studies in Sections~\ref{Sec:Simulation}--\ref{Sec:RealData}, we draw $m$ independent cross-generated samples for each observation pair and consider
\begin{equation*}
    \widehat{U}^{m}
  = \sup\limits_{\phi_{f,g,h} \in \Phi}
    \lrabs{\,\frac{1}{n} \sum\limits_{i=1}^{n} 
           \,\frac{1}{m} \sum\limits_{j=1}^{m}
    \phi_{f,g,h}(X_{1i},Y_{1i},\widehat{Y}_{1i}^{(j)},X_{2i},Y_{2i},\widehat{Y}_{2i}^{(j)})\,},
\end{equation*}
whose oracle version is given by
\begin{equation*}
    U^{m\ast}
  = \sup\limits_{\phi_{f,g,h} \in \Phi}
    \lrabs{\,\frac{1}{n} \sum\limits_{i=1}^{n} 
           \,\frac{1}{m} \sum\limits_{j=1}^{m}
    \phi_{f,g,h}(X_{1i},Y_{1i},Y_{1i}^{c\ast(j)},X_{2i},Y_{2i},Y_{2i}^{c\ast(j)})\,}.
\end{equation*}
Since $\{(X_{1i},Y_{1i},Y_{1i}^{c\ast(1)},\cdots,Y_{1i}^{c\ast(m)},X_{2i},Y_{2i},Y_{2i}^{c\ast(1)},\cdots,Y_{2i}^{c\ast(m)})\}_{i=1}^{n}$ can be viewed as an random sample from a common distribution on $\br^{2p+2(m+1)q}$, the corresponding oracle statistic $U^{m\ast}$ remains an empirical process indexed by the same function class $\Phi$. Consequently, the empirical process and bootstrap arguments developed below extend to any fixed $m\ge1$ with only minor modifications, while the approximation analysis between the feasible and oracle statistics can be adapted analogously. To avoid unnecessary notational complexity, we present the proofs for the case $m=1$ only.

\subsection{Proof of Theorem \ref{Thm:Validity}}
\begin{Proof}
We first show that, under the null, $\sup\limits_{\phi_{f,g,h}\in\Phi} \lrabs{\bbe{\phi_{f,g,h}(X_1,Y_1,Y_1^{c \ast},X_2,Y_2,Y_2^{c \ast})}} = 0$. It suffices to prove that the expectation is zero for each $\phi_{f,g,h}\in\Phi$.

By the definition of the oracle generator $G_1^{\ast}$ and $G_2^{\ast}$, we have that 
\begin{equation*}
    Y_1 = G_1^{\ast}(X_1,Z_1^{\ast}), \quad
    Y_1^{c\ast} = G_2^{\ast}(X_1,Z_1^{c \ast}), \quad
    Y_2 = G_2^{\ast}(X_2,Z_2^{\ast}), \quad
    Y_2^{c\ast} = G_1^{\ast}(X_2,Z_2^{c \ast}),
\end{equation*}
where $G_1^{\ast}, G_2^{\ast}$ are deterministic functions independent of $X_1, X_2$, and $Z_1^{\ast}, Z_2^{\ast}, Z_1^{c \ast}, Z_2^{c \ast}$ are some independent random vectors that are all independent of $(X_1, X_2)$. Under the null, we have $G_1^{\ast}(X_1, Z_1^{\ast}) =^d G_2^{\ast}(X_1,Z_1^{c \ast}) \mid X_1$, it follows that
\BEqn
& & \blre{f(X_1,Y_1) - f(X_1,Y_1^{c\ast}) \mid X_1, X_2} \\
&=& \blre{f(X_1,G_1^{\ast}(X_1,Z_1^{\ast})) - f(X_1,G_2^{\ast}(X_1,Z_1^{c \ast})) \mid X_1, X_2} \\
&=& \blre{f(X_1,G_1^{\ast}(X_1,Z_1^{\ast})) - f(X_1,G_2^{\ast}(X_1,Z_1^{c \ast})) \mid X_1} \\
&=& 0.
\EEqn

Similarly, since $G_1^{\ast}(X_2, Z_2^{c \ast}) =^d G_2^{\ast}(X_2, Z_2^{\ast}) \mid X_2$ holds under the null, we also have 
\begin{equation*}
    \blre{g(X_2,Y_2) - g(X_2,Y_2^{c\ast}) \mid X_1, X_2} 
  = 0.
\end{equation*}
It follows from simple calculations that for any $\phi_{f,g,h}\in\Phi$, we have that
\BEqn
& & \bbe{\phi_{f,g,h}(X_1,Y_1,Y_1^{c\ast},X_2,Y_2,Y_2^{c\ast})} \\
&=& \blre{
    \blre{\bigp{f(X_1,Y_1) - f(X_1,Y_1^{c\ast})}
          \bigp{g(X_2,Y_2) - g(X_2,Y_2^{c\ast})}
          h(X_1,X_2) \mid X_1,X_2}} \\
&=& \blre{ \blre{f(X_1,Y_1) - f(X_1,Y_1^{c\ast}) \mid X_1, X_2}
           \cdot \blre{g(X_2,Y_2) - g(X_2,Y_2^{c\ast}) \mid X_1, X_2}
           \cdot h(X_1,X_2)} \\
&=& 0,
\EEqn
where the second equality follows from the conditional independence between $(Y_1,Y_1^{c\ast})$ and $(Y_2,Y_2^{c\ast})$ given $(X_1,X_2)$.

Note that the equality holds for arbitrary $\phi_{f,g,h}\in\Phi$, we conclude that under the null,
\begin{equation*}
    \sup\limits_{\phi_{f,g,h}\in\Phi}
    \lrabs{\bbe{\phi_{f,g,h}(X_1, Y_1, Y_1^{c \ast}, X_2, Y_2, Y_2^{c \ast})}} 
  = 0.
\end{equation*}

It remains to show that $\sup\limits_{\phi_{f,g,h}\in\Phi} \lrabs{\bbe{\phi_{f,g,h}(X_1, Y_1, Y_1^{c\ast}, X_2, Y_2, Y_2^{c\ast})}} = 0$ implies the null hypothesis. In this case, it holds for any fixed $f,g\in\ch$ and $h\in\ch'$ that 
\BEqn
& & \bbe{\phi_{f,g,h}(X_1,Y_1,Y_1^{c\ast},X_2,Y_2,Y_2^{c\ast})} \\
&=& \blre{ \blre{f(X_1,Y_1) - f(X_1,Y_1^{c\ast}) \mid X_1, X_2}
           \cdot \blre{g(X_2,Y_2) - g(X_2,Y_2^{c\ast}) \mid X_1, X_2}
           \cdot h(X_1,X_2)} \\
&=& 0.
\EEqn
We define the functional 
\begin{equation*}
    \Psi_{f,g}(X_1,X_2)
  = \blre{f(X_1,Y_1) - f(X_1,Y_1^{c\ast}) \mid X_1, X_2}
    \cdot \blre{g(X_2,Y_2) - g(X_2,Y_2^{c\ast}) \mid X_1, X_2}.
\end{equation*}
Note that $\blre{\Psi_{f,g}(X_1,X_2)h(X_1,X_2)} = 0$ for any $f,g\in\ch$ and $h\in\ch'$, and the equality also holds for any $h\in\bh'$. Under Assumption \ref{Assumpt:kernel}\ref{Assumpt:kernel_L2_dense}, $\bh'$ is dense in $L_2(P_{X_1,X_2})$. Since $\Psi_{f,g} \in L_2(P_{X_1,X_2})$, then we claim that $\Psi_{f,g}(X_1,X_2) = 0$ with respect to $P_{X_1,X_2}$ almost surely for any $f,g\in\ch$.

Recall that
\begin{equation*}
    \blre{f(X_1,Y_1) - f(X_1,Y_1^{c\ast}) \mid X_1, X_2}
  = \blre{f(X_1,Y_1) - f(X_1,Y_1^{c\ast}) \mid X_1},
\end{equation*}
and
\begin{equation*}
    \blre{g(X_2,Y_2) - g(X_2,Y_2^{c\ast}) \mid X_1, X_2}
  = \blre{g(X_2,Y_2) - g(X_2,Y_2^{c\ast}) \mid X_2},
\end{equation*}
we define
\BEqn
    A_f(x) &=& \blre{f(X_1,Y_1) - f(X_1,Y_1^{c\ast}) \mid X_1 = x}, \\
    B_g(x) &=& \blre{g(X_2,Y_2) - g(X_2,Y_2^{c\ast}) \mid X_2 = x},
\EEqn
and it follows that
\begin{equation*}
    \Psi_{f,g}(X_1,X_2) = A_f(X_1) B_g(X_2).
\end{equation*}

Note that 
\begin{equation*}
    \blre{f(X_1,Y_1) - f(X_1,Y_1^{c\ast}) \mid X_1 = x} 
  = \int_{\cy} f(x,y) d\lrp{\bp_{Y_1|X_1}(\cdot|x) - \bp_{Y_2|X_2}(\cdot|x)}(y).
\end{equation*}
Under Assumption \ref{Assumpt:kernel}\ref{Assumpt:kernel_characteristic}, it follows from similar arguments used in \cite{gretton2012kernel} that $\sup\limits_{f\in\ch} |A_f(x)| = 0$ if and only if $P_{Y_1|X_1}(\cdot|x) = P_{Y_2|X_2}(\cdot|x)$. Let 
\begin{equation*}
    \ca = \lrcp{x\in\cx:~ P_{Y_1|X_1}(\cdot|x) \neq P_{Y_2|X_2}(\cdot|x)},
\end{equation*}
then we have that $\ca = \lrcp{x\in\cx: \sup\limits_{f\in\ch} |A_f(x)| > 0}$. Similarly, we also have $\ca = \lrcp{x\in\cx: \sup\limits_{g\in\ch} |B_g(x)| > 0}$. 

Then we prove by contradiction. Suppose $\bp(X_1 \in \ca) > 0$ or $\bp(X_2 \in \ca) > 0$, then under Assumption \ref{Assumpt:X1X2}\ref{Assumpt:mutual_abs_continuity}, we have that $\bp(X_1 \in \ca) > 0$ and $\bp(X_2 \in \ca) > 0$. Furthermore, under Assumption \ref{Assumpt:X1X2}\ref{Assumpt:joint_overlap}, it holds that 
\begin{equation*}
    \blrp{X_1\in\ca, ~X_2\in\ca} > 0.
\end{equation*}
On the other hand, since $\cx\times\cy\subset\br^{p+q}$ is a separable metric space, and the kernel $K$ is continuous under Assumption~\ref{Assumpt:kernel}, then the unit ball $\ch$ in the associated RKHS is separable. Let $\{f_m\}_{m\geq1}, \{g_m\}_{m\geq1}$ be the countable dense subsets of $\ch$. By the continuity of the linear functionals $f \mapsto A_f(x)$ and $g \mapsto B_g(x)$ on $\ch$ and the separability, we have
\begin{equation*}
    \sup_{f\in\ch} |A_f(x)|
  = \sup_{m\ge1} |A_{f_m}(x)|,
  \qquad
    \sup_{g\in\ch} |B_g(x)|
  = \sup_{m\ge1}|B_{g_m}(x)|.  
\end{equation*}
then it holds that 
\begin{equation*}
    \ca
  = \bigcup\limits_{m=1}^{\infty} 
    \lrcp{x\in\cx:~ |A_{f_m}(x)| > 0}
  = \bigcup\limits_{m=1}^{\infty} 
    \lrcp{x\in\cx:~ |B_{g_m}(x)| > 0}.
\end{equation*}
Therefore, $\blrp{X_1\in\ca, ~X_2\in\ca} > 0$ implies that 
\begin{equation*}
    \blrp{|A_{f_{m_1}}(X_1)| > 0, ~|B_{g_{m_2}}(X_2)| > 0} > 0
\end{equation*}
for some $m_1,m_2\geq1$, which further implies that $\lrabs{A_{f_{m_1}}(X_1) B_{g_{m_2}}(X_2)} > 0$ with positive probability. Consequently, $\sup_{f,g\in\ch} \lrabs{A_f(X_1) B_g(X_2)} > 0$ on an event with nonzero probability, which contradicts the previously established fact that $A_f(X_1)B_g(X_2) = 0$ with respect to $P_{X_1,X_2}$ almost surely for any $f,g\in\ch$. Therefore, we conclude that $\bp(X_1 \in \ca) = \bp(X_2 \in \ca) = 0$, and hence $P_{Y_1|X_1}(\cdot|x) = P_{Y_2|X_2}(\cdot|x)$ for both $P_{X_1}$- and $P_{X_2}$-almost every $x$, which completes the proof.
\end{Proof}

\subsection{Proof of Proposition \ref{Prop:Update_U}}
\begin{Proof}
Suppose $\alpha_f^{(r-1)}, \beta_g^{(r-1)}$ and $\gamma_h^{(r-1)}$ denote the updates of the $(r-1)$-th iteration. We first fix $\beta_g^{(r-1)}$ and $\gamma_h^{(r-1)}$ and find $\alpha_f^{(r)}$. 

By the reproducing property of $f\in\ch$ in Equation \eqref{Equ:kernel_trick}, we have
\begin{equation*}
    f(X_{1i},Y_{1i}) - f(X_{1i},\widehat{Y}_{1i})
  = \lrag{f, K((X_{1i},Y_{1i}),\cdot) - K((X_{1i},\widehat{Y}_{1i}),\cdot)}_{\bh}.
\end{equation*}
With $p_f^{(r)} = \beta_g^{(r-1)} \odot \gamma_h^{(r-1)}$, we further have
\BEqn
& & \lrabs{\frac{1}{n} \sum\limits_{i=1}^{n} 
           \alpha_{f,i}^{(r)} \beta_{g,i}^{(r-1)} \gamma_{h,i}^{(r-1)} } \\
&=& \sup\limits_{f\in\ch}
    \lrabs{\frac{1}{n} \sum\limits_{i=1}^{n} p_{f,i}^{(r)}
    \bigp{f(X_{1i},Y_{1i}) - f(X_{1i},\widehat{Y}_{1i})}} \\
&=& \sup\limits_{f\in\ch}
    \lrabs{\lrag{f, 
    ~\frac{1}{n} \sum\limits_{i=1}^{n} p_{f,i}^{(r)} 
    \bigp{K((X_{1i},Y_{1i}),\cdot) - K((X_{1i},\widehat{Y}_{1i}),\cdot)}}_{\bh}} \\
&=& \lrnorm{\frac{1}{n} \sum\limits_{i=1}^{n} p_{f,i}^{(r)} 
    \bigp{K((X_{1i},Y_{1i}),\cdot) - K((X_{1i},\widehat{Y}_{1i}),\cdot)}}_{\bh},
\EEqn
where the supremum is attained at
\begin{equation*}
    f^{(r)}(\cdot) 
  = \frac{ \frac{1}{n} \sum\limits_{i=1}^{n} p_{f,i}^{(r)} 
    \bigp{K((X_{1i},Y_{1i}),\cdot) - K((X_{1i},\widehat{Y}_{1i}),\cdot)} }
    { \lrnorm{\frac{1}{n} \sum\limits_{i=1}^{n} p_{f,i}^{(r)} 
    \bigp{K((X_{1i},Y_{1i}),\cdot) - K((X_{1i},\widehat{Y}_{1i}),\cdot)}}_{\bh} }.
\end{equation*}
By using the property of RKHS, we further have that
\BEqn
& & \lrnorm{\sum\limits_{i=1}^{n} p_{f,i}^{(r)} 
    \bigp{K((X_{1i},Y_{1i}),\cdot) - K((X_{1i},\widehat{Y}_{1i}),\cdot)}}_{\bh}^2 \\
&=& \lrag{
    \sum\limits_{i=1}^{n} p_{f,i}^{(r)} 
    \bigp{K((X_{1i},Y_{1i}),\cdot) - K((X_{1i},\widehat{Y}_{1i}),\cdot)},
    \sum\limits_{k=1}^{n} p_{f,k}^{(r)} 
    \bigp{K((X_{1k},Y_{1k}),\cdot) - K((X_{1k},\widehat{Y}_{1k}),\cdot)} }_{\bh} \\
&=& \sum\limits_{i,k=1}^{n} 
    p_{f,i}^{(r)} p_{f,k}^{(r)} \Omega_{1,ik} \\
&=& p_f^{(r)\top} \Omega_1 p_f^{(r)},
\EEqn
where $\Omega_1$ is given by Proposition~\ref{Prop:Update_U}.

It follows from direct calculations that 
\BEqn
& & f^{(r)}(X_{1k},Y_{1k})
 =  \frac{ \sum\limits_{i=1}^{n} p_{f,i}^{(r)}
    \bigp{K((X_{1i},Y_{1i}),(X_{1k},Y_{1k})) - K((X_{1i},\widehat{Y}_{1i}),(X_{1k},Y_{1k}))} }{\sqrt{p_f^{(r)\top} \Omega_1 p_f^{(r)}}}, \\
& & f^{(r)}(X_{1k},\widehat{Y}_{1k})
 =  \frac{ \sum\limits_{i=1}^{n} p_{f,i}^{(r)}
    \bigp{K((X_{1i},Y_{1i}),(X_{1k},\widehat{Y}_{1k})) - K((X_{1i},\widehat{Y}_{1i}),(X_{1k},\widehat{Y}_{1k}))} }{\sqrt{p_f^{(r)\top} \Omega_1 p_f^{(r)}}}.
\EEqn
Then we have that 
\begin{equation*}
    \alpha_{f,k}^{(r)}
  = f^{(r)}(X_{1k},Y_{1k}) - f^{(r)}(X_{1k},\widehat{Y}_{1k})
  = \frac{ \sum\nolimits_{i=1}^{n}
    p_{f,i}^{(r)} \Omega_{1,ik} }{ \sqrt{p_f^{(r)\top} \Omega_1 p_f^{(r)}} }
  = \frac{\bigp{\Omega_1 p_f^{(r)}}_{k}}{\sqrt{p_f^{(r)\top} \Omega_1 p_f^{(r)}}}.
\end{equation*}
By translating the result in the vector form, we update $\alpha_f$ in the $r$-th iteration by
\begin{equation*}
    \alpha_{f}^{(r)}
 =  \frac{\Omega_1 p_f^{(r)}}{\sqrt{p_f^{(r)\top} \Omega_1 p_f^{(r)}}}.
\end{equation*}
The update formulas for $\beta_g^{(r)}$ and $\gamma_h^{(r)}$ can be justified in a similar way, for which we spare the details and complete the proof.
\end{Proof}


\subsection{Proof of Theorem \ref{Thm:Uast_null}}
\begin{Proof}
Under Assumption~\ref{Assumpt:kernel}--\ref{Assumpt:X1X2}, it follows from Theorem~\ref{Thm:Validity} that $\|P(\phi_{f,g,h})\|_{\Phi} = 0$ and $\|(\bp_n-P)(\phi_{f,g,h})\|_{\Phi} = U^{\ast}$. Then the desired results directly follow from Lemma~\ref{Lemma:Donsker_Uast}--\ref{Lemma:Donsker_Ubast}.
\end{Proof}

\subsection{Proof of Theorem \ref{Thm:Double-robustness}}
\begin{Proof}
Recall that we have shown in Lemma \ref{Lemma:decomp} that 
\begin{equation*}
    |\widehat{U}-U^{\ast}| 
\le \max\limits_{1\le\ell\le L} U_1^{\pell} 
  + \max\limits_{1\le\ell\le L} U_2^{\pell} 
  + \max\limits_{1\le\ell\le L} U_3^{\pell},
\end{equation*}
To establish the stochastic boundedness of $|\widehat{U}-U^{\ast}|$, it suffices to prove the stochastic boundedness of $\max\limits_{1\le\ell\le L} U_i^{\pell}$ for $i=1,2,3$ respectively. Under Assumption \ref{Assumpt:kernel}--\ref{Assumpt:X1X2} and Assumption \ref{Assumpt:entropy}--\ref{Assumpt:double_robustness}, by Lemma \ref{Lemma:Rate_U12} and Lemma \ref{Lemma:Rate_U3}, we have that
\BEqn
& & \max\limits_{1\le\ell\le L} U_1^{\pell} = O_p\lrp{n^{-k_1}\log^{1/2}(n)}, \\
& & \max\limits_{1\le\ell\le L} U_2^{\pell} = O_p\lrp{n^{-k_2}\log^{1/2}(n)}, \\
& & \max\limits_{1\le\ell\le L} U_3^{\pell} = O_p\lrp{n^{-(k_1+k_2)}\log(n)}.
\EEqn
It follows that 
\begin{equation*}
    |\widehat{U}-U^{\ast}|
  = O_p\lrp{n^{-k_1}\log^{1/2}(n) + n^{-k_2}\log^{1/2}(n) + n^{-(k_1+k_2)}\log(n)}
  = O_p\lrp{n^{-\min\{k_1,k_2\}}\log^{1/2}(n)},
\end{equation*}
and consequently, $\sqrt{n}|\widehat{U}-U^{\ast}| = O_p\lrp{n^{-\min\{k_1,k_2\}+\frac{1}{2}}\log^{1/2}(n)}$

Under the null, it follows from Lemma \ref{Lemma:Rate_U12} that
\begin{equation*}
    \max\limits_{1\le\ell\le L} U_1^{\pell} = O_p\lrp{n^{-(k_1+\frac{1}{2})}\log^{(v+1)/2}(n)}, \quad
    \max\limits_{1\le\ell\le L} U_2^{\pell} = O_p\lrp{n^{-(k_2+\frac{1}{2})}\log^{(v+1)/2}(n)},
\end{equation*}
and $\max\limits_{1\le\ell\le L} U_3^{\pell} = O_p\lrp{n^{-(k_1+k_2)}\log(n)}$, then we have that 
\begin{equation*}
    \sqrt{n}|\widehat{U}-U^{\ast}|
  = O_p\lrp{  n^{-k_1}\log^{(v+1)/2}(n)
            + n^{-k_2}\log^{(v+1)/2}(n) 
            + n^{-(k_1+k_2-\frac{1}{2})}\log(n)}.
\end{equation*}
Under Assumption \ref{Assumpt:double_robustness}, we have $0<k_1,k_2<\frac{1}{2}$, then we have that
\begin{equation*}
    \frac{n^{-(k_1+k_2-\frac{1}{2})}\log(n)}{n^{-k_1}\log^{(v+1)/2}(n)}
  = \frac{n^{\frac{1}{2}-k_2}}{\log^{(v-1)/2}(n)},
  \rightarrow \infty,
\end{equation*}
and similarly, $\frac{n^{-(k_1+k_2-\frac{1}{2})}\log(n)}{n^{-k_2}\log^{(v+1)/2}(n)} \rightarrow \infty$. Thus, $n^{-(k_1+k_2-\frac{1}{2})}\log(n)$ dominates $n^{-k_1}\log^{(v+1)/2}(n) + n^{-k_2}\log^{(v+1)/2}(n)$ asymptotically, and we conclude that under the null,
\begin{equation*}
    \sqrt{n}|\widehat{U}-U^{\ast}|
  = O_p\lrp{n^{-(k_1+k_2-\frac{1}{2})}\log(n)},
\end{equation*}
which completes the proof.
\end{Proof}

\subsection{Proof of Proposition \ref{Prop:Double-robustness-boot}}
\begin{Proof}
By Lemma \ref{Lemma:decomp-boot}, we have that
\begin{equation*}
    |\widehat{U}^b - U^{b\ast}| 
\le \max\limits_{1\le\ell\le L} R_1^{\pell} 
  + \max\limits_{1\le\ell\le L} R_2^{\pell} 
  + \max\limits_{1\le\ell\le L} R_3^{\pell}
  + \max\limits_{1\le\ell\le L} U_1^{\pell} \cdot |\bar\varepsilon| 
  + \max\limits_{1\le\ell\le L} U_2^{\pell} \cdot |\bar\varepsilon| 
  + \max\limits_{1\le\ell\le L} U_3^{\pell} \cdot |\bar\varepsilon|.
\end{equation*}
Together with the established rates in Lemma \ref{Lemma:Rate_R123} and Lemma \ref{Lemma:Rate_R456}, we conclude that
\BEqn
& & |\widehat{U}^b - U^{b\ast}| \\
&=& O_p\lrp{ n^{-(k_1+1)/2} \log^v(n) }
 +  O_p\lrp{ n^{-(k_2+1)/2} \log^v(n) }
 +  O_p\lrp{ n^{-(k_1+k_2+1)/2} \log^v(n)} \\
& & + O_p\lrp{ n^{-(k_1+\frac{1}{2})} \log^{1/2}(n) } 
 +  O_p\lrp{ n^{-(k_2+\frac{1}{2})} \log^{1/2}(n) }
 +  O_p\lrp{ n^{-(k_1+k_2+\frac{1}{2})} \log(n) }.
\EEqn
Asymptotically, $n^{-(\min\{k_1,k_2\}+1)/2} \log^v(n)$ dominates all the individual terms, that is,
\begin{equation*}
    |\widehat{U}^b - U^{b\ast}| = O_p\lrp{n^{-(\min\{k_1,k_2\}+1)/2} \log^v(n)}.
\end{equation*}
Multiplying by $\sqrt{n}$, we conclude that
\begin{equation*}
    \sqrt{n} |\widehat{U}^b - U^{b\ast}| 
  = O_p\lrp{n^{-\min\{k_1,k_2\}/2} \log^v(n)}.
\end{equation*}
\end{Proof}

\subsection{Proof of Theorem \ref{Thm:Bootstrap_null}}
\begin{Proof}
It follows from the triangle inequality for the supremum norm that
\BEqn
& & \sup\limits_{x\in\br}
    \lrabs{ \blrp{\sqrt{n} \widehat{U} \le x \mid H_0} 
    - \blrp{\sqrt{n} \widehat{U}^{b} \le x \mid \cd, \cz} } \\
&\le& \sup\limits_{x\in\br}
    \lrabs{\blrp{\sqrt{n} \widehat{U} \le x \mid H_0} 
    - \blrp{\sqrt{n} U^{\ast} \le x \mid H_0}} 
    + \sup\limits_{x\in\br}
    \lrabs{\blrp{\sqrt{n} U^{\ast} \le x \mid H_0}
    - \blrp{\lrnorm{\bg(\phi_{f,g,h})}_{\Phi} \le x}} \\
& & + \sup\limits_{x\in\br}
    \lrabs{\blrp{\lrnorm{\bg(\phi_{f,g,h})}_{\Phi} \le x}
    - \blrp{\sqrt{n} U^{b\ast} \le x \mid \cd, \cz^{c \ast}}} \\
& & + \sup\limits_{x\in\br}
    \lrabs{\blrp{\sqrt{n} U^{b\ast} \le x \mid \cd, \cz^{c \ast}}
    - \blrp{\sqrt{n} \widehat{U}^{b} \le x \mid \cd, \cz}}.
\EEqn
Under the null, it follows from Theorem \ref{Thm:Validity} that 
\begin{equation*}
    \|P(\phi_{f,g,h})\|_{\Phi}
  = \sup\limits_{\phi_{f,g,h}\in\Phi} 
    \lrabs{\bbe{\phi_{f,g,h}(X_1,Y_1,Y_1^{c\ast},X_2,Y_2,Y_2^{c\ast})}}
  = 0.
\end{equation*}
Thus, it holds under the null that, 
\begin{equation*}
    U^{\ast}
  = \|\bp_n(\phi_{f,g,h})\|_{\Phi}
  = \|(\bp_n-P)(\phi_{f,g,h})\|_{\Phi}.
\end{equation*}
Consequently,
\BEqn
& & \sup\limits_{x\in\br}
    \lrabs{\blrp{\sqrt{n} U^{\ast} \le x \mid H_0}
    - \blrp{\lrnorm{\bg(\phi_{f,g,h})}_{\Phi} \le x}} \\
&=& \sup\limits_{x\in\br}
    \lrabs{\blrp{\sqrt{n} \|(\bp_n-P)(\phi_{f,g,h})\|_{\Phi} \le x \mid H_0}
    - \blrp{\lrnorm{\bg(\phi_{f,g,h})}_{\Phi} \le x}}.
\EEqn
By putting together the results we have established in Lemma \ref{Lemma:S2}, Lemma \ref{Lemma:S3} and Lemma \ref{Lemma:S4}, we conclude that
\begin{equation*}
    \sup\limits_{x\in\br}
    \lrabs{ \blrp{\sqrt{n} \widehat{U} \le x \mid H_0} 
    - \blrp{\sqrt{n} \widehat{U}^{b} \le x \mid \cd, \cz} }
  = o_p(1),
\end{equation*}
which completes the proof.
\end{Proof}

\subsection{Proof of Corollary \ref{Cor:Bootstrap_size}}
\begin{Proof}
For simplicity, define 
\begin{equation*}
    B_n(x) := \blrp{\sqrt{n} \widehat{U}^{b} \le x \mid \cd, \cz}.
\end{equation*}
In light of Theorem~\ref{Thm:Bootstrap_null}, it suffices to justify $\blrp{\sqrt{n} \widehat{U}^{b} > \gamma_{1-\alpha} \mid \cd, \cz} = \alpha + o_p(1)$, or equivalently, $B_n(\gamma_{1-\alpha}) = 1 - \alpha + o_p(1)$. Note that $\gamma_{1-\alpha}$ is the conditional $(1-\alpha)$ quantile of $\sqrt{n} \widehat{U}^{b}$ given $(\cd,\cz)$, it follows from the definition of quantile that 
\begin{equation*}
    \gamma_{1-\alpha}
  = \inf\{x\in\br:~B_n(x) \geq 1-\alpha\}.
\end{equation*}
Thus, $B_n(\gamma_{1-\alpha}) \geq 1-\alpha$ and $B_n(\gamma_{1-\alpha} - \eta) < 1 - \alpha$ for any $\eta > 0$. Moreover, we have
\begin{equation*}
    B_n(\gamma_{1-\alpha}) 
\le B_n(\gamma_{1-\alpha} - \eta)
  + \sup\limits_{x\in\br} \lrabs{B_n(x) - B_n(x-\eta)}.
\end{equation*}
By Lemma~\ref{Lemma:S3} and Lemma~\ref{Lemma:S4}, it holds under the stated assumptions that
\begin{equation*}
    \sup\limits_{x\in\br} 
    \lrabs{B_n(x) - \blrp{\|\bg\|_{\Phi} \le x}}
  = o_p(1),
\end{equation*}
implying that
\begin{equation}\label{Equ:Size_aux}
    B_n(\gamma_{1-\alpha}) 
\le B_n(\gamma_{1-\alpha} - \eta)
  + \sup\limits_{x\in\br}
    \lrabs{\blrp{\|\bg\|_{\Phi} \le x} - \blrp{\|\bg\|_{\Phi} \le x-\eta}}
  + o_p(1).
\end{equation}
By Lemma~\ref{Lemma:ContinuousCDF}, the distribution function of $\|\bg\|_{\Phi}$ is continuous on $\br$, and hence uniformly continuous. Therefore, for any $\delta>0$, we can choose a fixed $\eta>0$ such that
\begin{equation*}
    \sup\limits_{x\in\br}
    \lrabs{\blrp{\|\bg\|_{\Phi} \le x} - \blrp{\|\bg\|_{\Phi} \le x-\eta}}
  < \delta,
\end{equation*}
Together with $B_n(\gamma_{1-\alpha}-\eta)<1-\alpha$, Equation~\eqref{Equ:Size_aux} yields $B_n(\gamma_{1-\alpha}) \le 1 - \alpha + \delta + o_p(1)$. Since $\delta>0$ is arbitrary, we have 
\begin{equation*}
    B_n(\gamma_{1-\alpha}) \le 1 - \alpha + o_p(1).
\end{equation*}
Combined with the previously stated $B_n(\gamma_{1-\alpha}) \geq 1-\alpha$, we conclude that $B_n(\gamma_{1-\alpha}) = 1 - \alpha + o_p(1)$, which completes the proof.
\end{Proof}

\subsection{Proof of Theorem \ref{Thm:Bootstrap_power}}
\begin{Proof}
For each $\phi_{f,g,h}\in\Phi$, define
\begin{equation*}
    \widehat\bp_n(\phi_{f,g,h})
  = \frac{1}{n}\sum\limits_{i=1}^{n} 
    \phi_{f,g,h}(X_{1i},Y_{1i},\widehat{Y}_{1i},X_{2i},Y_{2i},\widehat{Y}_{2i}),
\end{equation*}
By the triangle inequality of the supremum norm $\|\cdot\|_{\Phi}$, we have that
\BEqn
    \sqrt{n} \lrnorm{P(\phi_{f,g,h})}_{\Phi} - \sqrt{n} \lrnorm{\widehat\bp_n(\phi_{f,g,h})}_{\Phi} 
&\le& \sqrt{n} \lrnorm{(\widehat\bp_n - P)(\phi_{f,g,h})}_{\Phi} \\
&\le& \sqrt{n} \lrnorm{(\widehat\bp_n - \bp_n)(\phi_{f,g,h})}_{\Phi} + \sqrt{n} \lrnorm{(\bp_n - P)(\phi_{f,g,h})}_{\Phi},
\EEqn
which implies that
\begin{equation}\label{Equ:Power_aux_1}
    \sqrt{n} \lrnorm{\widehat\bp_n(\phi_{f,g,h})}_{\Phi}
\geq \sqrt{n} \lrnorm{P(\phi_{f,g,h})}_{\Phi} 
   - \sqrt{n} \lrnorm{(\widehat\bp_n - \bp_n)(\phi_{f,g,h})}_{\Phi} 
   - \sqrt{n} \lrnorm{(\bp_n - P)(\phi_{f,g,h})}_{\Phi}.
\end{equation}
By the telescoping decomposition arguments used in Lemma \ref{Lemma:decomp}, we have that
\BEqn
& & \lrnorm{(\widehat\bp_n - \bp_n)(\phi_{f,g,h})}_{\Phi} \\
&=& \sup\limits_{\phi_{f,g,h}\in\Phi} 
    \lrabs{\frac{1}{n} \sum\limits_{i=1}^{n} 
    \lrp{\phi_{f,g,h}(X_{1i},Y_{1i},\widehat{Y}_{1i},X_{2i},Y_{2i},\widehat{Y}_{2i})
    - \phi_{f,g,h}(X_{1i},Y_{1i},Y_{1i}^{c\ast},X_{2i},Y_{2i},Y_{2i}^{c\ast})}} \\
&\le& \max\limits_{1\le\ell\le L} U_1^{\pell} 
  + \max\limits_{1\le\ell\le L} U_2^{\pell} 
  + \max\limits_{1\le\ell\le L} U_3^{\pell}.
\EEqn
It follows from Lemma \ref{Lemma:Rate_U12} and Lemma \ref{Lemma:Rate_U3} that 
\BEqn
    \lrnorm{(\widehat\bp_n - \bp_n)(\phi_{f,g,h})}_{\Phi} 
&=& O_p(n^{-k_1}\log^{1/2}(n) + n^{-k_2}\log^{1/2}(n) + n^{-(k_1+k_2)}\log(n)) \\
&=& O_p(n^{-\min\{k_1,k_2\}} \log^{1/2}(n)),
\EEqn
which implies that
\begin{equation*}
    \sqrt{n} \lrnorm{(\widehat\bp_n - \bp_n)(\phi_{f,g,h})}_{\Phi} 
  = O_p(n^{-\min\{k_1,k_2\}+\frac{1}{2}} \log^{1/2}(n)).
\end{equation*}
By Lemma \ref{Lemma:S1}, we have that 
\begin{equation*}
    \sup\limits_{x\in\br} 
    \lrabs{ \blrp{\sqrt{n}\lrnorm{(\bp_n-P)(\phi_{f,g,h})}_{\Phi} \le x} 
    - \blrp{\lrnorm{\bg(\phi_{f,g,h})}_{\Phi} \le x}}
  = o(1),
\end{equation*}
i.e. $\sqrt{n}\lrnorm{(\bp_n-P)(\phi_{f,g,h})}_{\Phi} \leadsto \|\bg(\phi_{f,g,h})\|_{\Phi}$, which implies that $\sqrt{n}\lrnorm{(\bp_n-P)(\phi_{f,g,h})}_{\Phi} = O_p(1)$.

When condition \eqref{Equ:PowerCondition} is satisfied, we have that
\begin{equation*}
    \lrnorm{P(\phi_{f,g,h})}_{\Phi}
  = \omega(n^{-\min\{k_1,k_2\}} \log^{1/2}(n)).
\end{equation*}
By plugging each term into Equation \eqref{Equ:Power_aux_1}, we have that
\BEqn
& & \sqrt{n} \widehat{U} = \sqrt{n} \|\widehat\bp_n(\phi_{f,g,h})\|_{\Phi} \\
&\geq& \omega(n^{-\min\{k_1,k_2\}+\frac{1}{2}} \log^{1/2}(n))
  - O_p(n^{-\min\{k_1,k_2\}+\frac{1}{2}} \log^{1/2}(n))
  - O_p(1) \\
&=& \omega_p(n^{-\min\{k_1,k_2\}+\frac{1}{2}} \log^{1/2}(n)).
\EEqn

In addition, it follows from Lemma \ref{Lemma:S1}, Lemma \ref{Lemma:S3} and Lemma \ref{Lemma:S4} that
\BEqn
& & \sup\limits_{x\in\br} 
    \lrabs{ \blrp{\sqrt{n}\lrnorm{(\bp_n-P)(\phi_{f,g,h})}_{\Phi} \le x} 
    - \blrp{\sqrt{n}\widehat{U}^{b} \le x \mid \cd, \cz}} \\
&\le& \sup\limits_{x\in\br} 
    \lrabs{ \blrp{\sqrt{n}\lrnorm{(\bp_n-P)(\phi_{f,g,h})}_{\Phi} \le x} 
    - \blrp{\lrnorm{\bg(\phi_{f,g,h})}_{\Phi} \le x}} \\
& & + \sup\limits_{x\in\br} 
    \lrabs{ \blrp{\lrnorm{\bg(\phi_{f,g,h})}_{\Phi} \le x}
    - \blrp{\sqrt{n}U^{b\ast} \le x \mid \cd, \cz^{c \ast}} } \\
& & + \sup\limits_{x\in\br} 
    \lrabs{ \blrp{\sqrt{n}U^{b\ast} \le x \mid \cd, \cz^{c \ast}}
    - \blrp{\sqrt{n}\widehat{U}^{b} \le x \mid \cd, \cz}} \\
&=& S_1 + S_3 + S_4 = o_p(1).
\EEqn
Recall that we have shown that $\sqrt{n}\lrnorm{(\bp_n-P)(\phi_{f,g,h})}_{\Phi} = O_p(1)$, then the conditional distribution of $\sqrt{n} \widehat{U}^{b}$ given $\cd,\cz$ is asymptotically tight in probability. Hence, its conditional $1-\alpha$ quantile satisfies that $\gamma_{1-\alpha} = O_p(1)$. Under Assumption \ref{Assumpt:double_robustness} with $0<k_1,k_2<\frac{1}{2}$, $\sqrt{n} \widehat{U} = \omega_p(n^{-\min\{k_1,k_2\}+\frac{1}{2}} \log^{1/2}(n))$ yields that $\sqrt{n}\widehat{U} \stra{p} \infty$, then we conclude that
\begin{equation*}
    \blrp{\sqrt{n}\widehat{U} \geq \gamma_{1-\alpha} \mid H_1} \rightarrow 1,
\end{equation*}
which completes the proof.
\end{Proof}

\section{Preliminary Results of Empirical Process Theory}\label{Appendix:EmpTheory}

To facilitate the following analysis, we first introduce some preliminary results of the empirical processes theory and establish some basic properties for the class of functions of interest in our setup.

\subsection{Uniform Entropy}

\begin{lemma}\label{Lemma:Envelope}
Let $u_{K,K'} > 0$ be the constant given by Assumption~\ref{Assumpt:kernel}\ref{Assumpt:kernel_bound} and let $u_{\ch,\ch'} = u_{K,K'}^{1/2}$. Under Assumption~\ref{Assumpt:kernel}, it holds that $\sup_{f\in\ch} \|f\|_{\infty} \le u_{\ch,\ch'}$ and $\sup_{h\in\ch'}\|h\|_\infty \le u_{\ch,\ch'}$.
\end{lemma}
\begin{Proof}
By the reproducing property, for any $f\in\ch$ and $z\in\br^{p+q}$,
\begin{equation*}
    |f(z)|
\le \|f\|_{\bh} \|K(z,\cdot)\|_{\bh}
  = \|f\|_{\bh} K^{1/2}(z,z).
\end{equation*}
Since $\ch$ is the unit ball of $\bh$, it follows that $\sup_{f\in\ch} \|f\|_{\infty} \le u_{K,K'}^{1/2}$. Similarly, $\sup_{h\in\ch'}\|h\|_\infty \le u_{K,K'}^{1/2}$. Therefore, we complete the proof for $f\in\ch$, and the analogous proof for $h\in\ch'$ follows similarly.
\end{Proof}

\begin{lemma}\label{Lemma:cover_num}
Let $\cp(\br^{2p+4q}, \cb(\br^{2p+4q}))$ denote the set of all probability measures on $(\br^{2p+4q}, \cb(\br^{2p+4q}))$. Under Assumption~\ref{Assumpt:kernel}, it holds that
\BEqn
& & \sup\limits_{P\in\cp(\br^{2p+4q}, \cb(\br^{2p+4q}))} N(\Phi, \|\cdot\|_{L_2(P)}, 4u_{\ch,\ch'}^3\varepsilon) \\
&\le& \lrp{\sup\limits_{P\in\cp(\br^{p+q},\cb(\br^{p+q}))} N\lrp{\ch, \|\cdot\|_{L_2(P)}, \frac{u_{\ch,\ch'}\varepsilon}{3}}}^4 
    \lrp{\sup\limits_{P\in\cp(\br^{2p},\cb(\br^{2p}))} N\lrp{\ch',\|\cdot\|_{L_2(P)}, \frac{u_{\ch,\ch'}\varepsilon}{3}}}.
\EEqn
\end{lemma}

\begin{Proof}
For each $f\in\ch$, $X_1\in\cx$ and $Y_1,Y_1'\in\cy$, define $\tilde{f}$ such that $\tilde{f}(X_1,Y_1,Y_1') = f(X_1,Y_1) - f(X_1,Y_1')$. Let $\tilde\cf = \{\tilde{f}: f\in\ch\}$, then for any $\tilde{f}_1,\tilde{f}_2\in\tilde{\cf}$, it is trivial that
\BEqn
& & \|\tilde{f}_1 - \tilde{f}_2\|_{L_2(P)}^2 \\
&=& \int_{\br^{2p+4q}} \lrcp{\bigp{f_1(X_1,Y_1) - f_1(X_1,Y_1')} - \bigp{f_2(X_1,Y_1) - f_2(X_1,Y_1')}}^2 dP \\
&\le& 2\int_{\br^{2p+4q}} \lrp{f_1(X_1,Y_1) - f_2(X_1,Y_1)}^2 dP 
    + 2\int_{\br^{2p+4q}} \lrp{f_1(X_1,Y_1') - f_2(X_1,Y_1')}^2 dP \\
&=& 2\int_{\br^{p+q}} \lrp{f_1(X_1,Y_1) - f_2(X_1,Y_1)}^2 dP_1 
    + 2\int_{\br^{p+q}} \lrp{f_1(X_1,Y_1') - f_2(X_1,Y_1')}^2 dP_1' \\
&=& 2\|f_1-f_2\|_{L_2(P_1)}^2 + 2\|f_1-f_2\|_{L_2(P_1')}^2, \label{Equ:aux_cover_num}
\EEqn
where $P=P_{X_1,Y_1,Y_1',X_2,Y_2,Y_2'}$ denote any joint measure probability on $(\br^{2p+4q},\cb(\br^{2p+4q}))$, and $P_1 = P_{X_1,Y_1}$ and $P_1' = P_{X_1, Y_1'}$ denote the two marginal probability associated with $P$.

We first show that 
\begin{equation*}
    \sup\limits_{P} N(\tilde\cf, \|\cdot\|_{L_2(P)}, \varepsilon)
  \le \lrp{\sup\limits_{P_1} N(\ch, \|\cdot\|_{L_2(P_1)}, \varepsilon/2)}^2,
  \qquad \forall~\varepsilon>0.
\end{equation*}
Let $v = \sup_{P_1} N(\ch, \|\cdot\|_{L_2(P_1)}, \varepsilon/2)$. For any joint measure probability $P$, there exist two sets of centers $f_1,\cdots,f_v$ and $f_1',\cdots,f_v'$, s.t. the union of $(\varepsilon/2)$-covers centered at each set of the centers covers $\ch$ with respect to the probability measures $P_1$ and $P_1'$ respectively. For each $1\le i,j\le v$, we define
\begin{equation*}
    \tilde{f}_{ij}(X_1,Y_1,Y_1') = f_i(X_1,Y_1) - f_j'(X_1,Y_1'),
\end{equation*}
then it suffices to show that the union of $(\varepsilon/2)$-covers centered at these $v^2$ centers form a covering of $\tilde{\cf}$. In fact, for any $f\in\ch$, there exits $1\le i\le v$ and $1\le j\le v$, s.t. 
\begin{equation*}
    \|f - f_i\|_{L_2(P_1)} \le \frac{\varepsilon}{2},
    \qquad
    \|f - f_j'\|_{L_2(P_1')} \le \frac{\varepsilon}{2}.
\end{equation*}
Then it follows from Equation (\ref{Equ:aux_cover_num}), that
\begin{equation*}
    \|\tilde{f} - \tilde{f}_{ij}\|_{L_2(P)}
\le \sqrt{2\|f - f_i\|_{L_2(P_1)}^2 + 2\|f - f_j'\|_{L_2(P_1')}^2} 
\le \varepsilon
\end{equation*}
Since the inequality is valid for an arbitrary probability measure $P$, we arrive at the desired result.

Similarly, we define $\tilde\cg = \{\tilde{g}: g\in\ch\}$ with $\tilde{g}(X_2,Y_2,Y_2') = g(X_2,Y_2) - g(X_2,Y_2')$, then $\tilde\cg$ has the same covering number as $\tilde\cf$.

By the definition of $\Phi$, it is trivial that $\Phi = \tilde\cf \cdot \tilde\cg \cdot \ch' = \{\tilde{f}\tilde{g}h: \tilde{f}\in\tilde\cf, \tilde{g}\in\tilde{\cg},h\in\ch'\}$. By Lemma~\ref{Lemma:Envelope}, both $\tilde\cf$ and $\tilde\cg$ have the envelope of $2u_{\ch,\ch'}$ whereas $\ch'$ have the envelope of $u_{\ch,\ch'}$ under Assumption~\ref{Assumpt:kernel}. For any given measure probability $P = P_{X_1,Y_1,Y_1',X_2,Y_2,Y_2'}$ on $\br^{2p+4q}$ We define the marginal probability measures 
\begin{equation*}
    P_f = P_{X_1,Y_1,Y_1'}, \quad
    P_g = P_{X_2,Y_2,Y_2'}, \quad
    P_h = P_{X_1,X_2}.
\end{equation*}
Let $v_1 = \sup\limits_{P_f}N(\tilde\cf, \|\cdot\|_{L_2(P_f)}, \frac{2u_{\ch,\ch'}\varepsilon}{3})$, $v_2 = \sup\limits_{P_g}N(\tilde\cg, \|\cdot\|_{L_2(P_g)}, \frac{2u_{\ch,\ch'}\varepsilon}{3})$ and $v_3 = \sup\limits_{P_h}N(\ch', \|\cdot\|_{L_2(P_h)}, \frac{u_{\ch,\ch'}\varepsilon}{3})$. It follows from the definition of covering number that there exists
\begin{equation*}
    \cv_1 := \{\tilde{f}_1,\cdots,\tilde{f}_{v_1}\} \in \tilde\cf, \quad
    \cv_2 := \{\tilde{g}_1,\cdots,\tilde{g}_{v_2}\} \in \tilde\cg, \quad
    \cv_3 := \{h_1,\cdots,h_{v_3}\} \in \ch',
\end{equation*}
s.t the $(2u_{\ch,\ch'}\varepsilon/3)$-covers centered at $\cv_1$ covers $\tilde\cf$, the $(2u_{\ch,\ch'}\varepsilon/3)$-covers centered at $\cv_2$ covers $\tilde\cg$ and the $(u_{\ch,\ch'}\varepsilon/3)$-covers center at $\cv_3$ covers $\ch'$. 

Denote $\cv = \{\tilde{f}_i \tilde{g}_j h_k: 1\le i\le v_1,~1\le j\le v_2,~1\le k\le v_3\}$, then we show that the $(4u_{\ch,\ch'}^3\varepsilon)$-covers centered at $\cv$ covers $\Phi$. In particular, for any $\tilde{f}\tilde{g}h\in\Phi$, there exists $\tilde{f}_i\in\cv_1$, $\tilde{g}_j\in\cv_2$ and $h_k\in\cv_3$, s.t. 
\begin{equation*}
    \|\tilde{f} - \tilde{f}_i\|_{L_2(P_f)} \le \frac{2u_{\ch,\ch'}\varepsilon}{3}, \quad
    \|\tilde{g} - \tilde{g}_j\|_{L_2(P_g)} \le \frac{2u_{\ch,\ch'}\varepsilon}{3}, \quad
    \|h - h_k\|_{L_2(P_h)} \le \frac{u_{\ch,\ch'}\varepsilon}{3}.
\end{equation*}
Note that 
\BEqn
    \lrabs{\tilde{f}\tilde{g}h - \tilde{f}_i\tilde{g}_j h_k}
&\le& \lrabs{(\tilde{f}-\tilde{f}_i) \tilde{g}h} 
  + \lrabs{\tilde{f}_i(\tilde{g}-\tilde{g}_j)h}
  + \lrabs{\tilde{f}_i\tilde{g}_j(h-h_k)} \\
&\le& 2u_{\ch,\ch'}^2|\tilde{f}-\tilde{f}_i|
  + 2u_{\ch,\ch'}^2|\tilde{g}-\tilde{g}_j| 
  + 4u_{\ch,\ch'}^2|h-h_k|.
\EEqn
It follows that 
\BEqn
& & \|\tilde{f}\tilde{g}h - \tilde{f}_i\tilde{g}_j h_k\|_{L_2(P)}^2 \\
&\le& \int \lrp{  2u_{\ch,\ch'}^2|\tilde{f}-\tilde{f}_i|
                + 2u_{\ch,\ch'}^2|\tilde{g}-\tilde{g}_j|
                + 4u_{\ch,\ch'}^2|h-h_k|}^2 dP \\
&\le& 3\int_{\br^{p+2q}} \lrp{2u_{\ch,\ch'}^2|\tilde{f}-\tilde{f}_i|}^2 dP_f 
    + 3\int_{\br^{p+2q}} \lrp{2u_{\ch,\ch'}^2|\tilde{g}-\tilde{g}_j|}^2 dP_g
    + 3\int_{\br^{2p}} \lrp{4u_{\ch,\ch'}^2|h-h_k|}^2 dP_h \\
&\le& 12u_{\ch,\ch'}^4 \|\tilde{f}-\tilde{f}_i\|_{L_2(P_f)}^2 
    + 12u_{\ch,\ch'}^4 \|\tilde{g}-\tilde{g}_j\|_{L_2(P_g)}^2 
    + 48u_{\ch,\ch'}^4 \|h-h_k\|_{L_2(P_h)}^2 \\
&\le& 12u_{\ch,\ch'}^4\lrp{\frac{2u_{\ch,\ch'}\varepsilon}{3}}^2 + 12u_{\ch,\ch'}^4\lrp{\frac{2u_{\ch,\ch'}\varepsilon}{3}}^2 + 48u_{\ch,\ch'}^4\lrp{\frac{u_{\ch,\ch'}\varepsilon}{3}}^2 \\
&=& 16u_{\ch,\ch'}^6\varepsilon^2,
\EEqn
which implies that $\tilde{f}\tilde{g}h$ is inside the $(4u_{\ch,\ch'}^3\varepsilon)$-cover centered at $\tilde{f}_i\tilde{g}_j h_k$. Due to the arbitrariness of $\tilde{f}\tilde{g}h\in\Phi$, we conclude that 
\BEqn
& & N(\Phi, \|\cdot\|_{L_2(P)}, 4u_{\ch,\ch'}^3\varepsilon) \\
&\le& \sup\limits_{P_f} N\lrp{\tilde\cf, \|\cdot\|_{L_2(P_f)}, \frac{2u_{\ch,\ch'}\varepsilon}{3}}
    \cdot \sup\limits_{P_g} N\lrp{\tilde\cg, \|\cdot\|_{L_2(P_g)}, \frac{2u_{\ch,\ch'}\varepsilon}{3}}
    \cdot \sup\limits_{P_h} N\lrp{\ch', \|\cdot\|_{L_2(P_h)}, \frac{u_{\ch,\ch'}\varepsilon}{3}}.
\EEqn
We further notice that the inequality holds for arbitrary $P$, then together with the earlier results, we conclude that for any $0<\varepsilon<1$, 
\BEqn
& & \sup\limits_{P} N(\Phi, \|\cdot\|_{L_2(P)}, 4u_{\ch,\ch'}^3\varepsilon) \\
&\le& \sup\limits_{P_f} N\lrp{\tilde\cf, \|\cdot\|_{L_2(P_f)}, \frac{2u_{\ch,\ch'}\varepsilon}{3}}
    \cdot \sup\limits_{P_g} N\lrp{\tilde\cg, \|\cdot\|_{L_2(P_g)}, \frac{2u_{\ch,\ch'}\varepsilon}{3}}
    \cdot \sup\limits_{P_h} N\lrp{\ch', \|\cdot\|_{L_2(P_h)}, \frac{u_{\ch,\ch'}\varepsilon}{3}} \\
&\le& \lrp{\sup\limits_{P\in\cp(\br^{p+q},\cb(\br^{p+q}))} N\lrp{\ch, \|\cdot\|_{L_2(P)}, \frac{u_{\ch,\ch'}\varepsilon}{3}}}^4 
    \lrp{\sup\limits_{P\in\cp(\br^{2p},\cb(\br^{2p}))} N\lrp{\ch',\|\cdot\|_{L_2(P)}, \frac{u_{\ch,\ch'}\varepsilon}{3}}}.
\EEqn
which completes the proof.
\end{Proof}

\begin{lemma}\label{Lemma:Unif_Entropy}
Under Assumption~\ref{Assumpt:kernel} and Assumption~\ref{Assumpt:entropy}, it holds that 
\begin{equation*}
    \int_{0}^{\infty} \sqrt{\sup\limits_{P \in \cp(\br^{2p+4q},\cb(\br^{2p+4q}))}
    \log N(\Phi, \|\cdot\|_{L_2(P)}, 4u_{\ch,\ch'}^3\varepsilon)}d\varepsilon
  < \infty.
\end{equation*}
\end{lemma}
\begin{Proof}
Under Assumption~\ref{Assumpt:kernel}, $\ch,\ch'$ both have a uniform envelope $u_{\ch,\ch'}$ by Lemma~\ref{Lemma:Envelope}. By the definition of $\phi_{f,g,h}\in\Phi$, the class $\Phi$ has an envelope function of $4u_{\ch,\ch'}^3$. Therefore, for any $\varepsilon>1$, $\Phi$ can be covered by one ball centered at zero with radius $(4u_{\ch,\ch'}^3\varepsilon)$, implying that
\begin{equation*}
    \sup\limits_{P} N(\Phi, \|\cdot\|_{L_2(P)}, 4u_{\ch,\ch'}^3\varepsilon) = 1, 
    \quad\forall~\varepsilon>1.
\end{equation*}

For any $\varepsilon\in(0,1)$, it follows from Assumption~\ref{Assumpt:entropy} and Lemma \ref{Lemma:cover_num} that there exist deterministic positive constants $C>0$ and $A>0$ and a constant $v\geq 1$ that solely depends on the dimensions $p,q$, s.t.
\BEqn
& & \sup\limits_{P\in\cp(\br^{2p+4q},\cb(\br^{2p+4q}))} \log N\lrp{\Phi, \|\cdot\|_{L_2(P)}, 4u_{\ch,\ch'}^3\varepsilon} \\
&\le& 4\sup\limits_{P\in\cp(\br^{p+q},\cb(\br^{p+q}))} \log N\lrp{\ch, \|\cdot\|_{L_2(P)}, \frac{u_{\ch,\ch'}\varepsilon}{3}} 
    + \sup\limits_{P\in\cp(\br^{2p},\cb(\br^{2p}))} \log N\lrp{\ch', \|\cdot\|_{L_2(P)}, \frac{u_{\ch,\ch'}\varepsilon}{3}} \\
&\le& C \log^{v}\lrp{\frac{A}{u_{\ch,\ch'}^{1/2}\varepsilon}},
\EEqn
where we use the equality $u_{\ch,\ch'} = u_{K,K'}^{1/2}$ from Lemma~\ref{Lemma:Envelope}.

Then it follows from direct calculations that
\BEqn
& & \int_{0}^{\infty} \sqrt{\sup\limits_{P\in\cp(\br^{2p+4q},\cb(\br^{2p+4q}))} \log N(\Phi, \|\cdot\|_{L_2(P)}, 4u_{\ch,\ch'}^3\varepsilon)}d\varepsilon \\
&=& \int_{0}^{1} \sqrt{\sup\limits_{P\in\cp(\br^{2p+4q},\cb(\br^{2p+4q}))} \log N(\Phi, \|\cdot\|_{L_2(P)}, 4u_{\ch,\ch'}^3\varepsilon)}d\varepsilon \\
&\le&  \int_{0}^{1} \sqrt{C \log^v\lrp{\frac{A}{u_{\ch,\ch'}\varepsilon}}} d\varepsilon 
 \le  \int_{A/u_{\ch,\ch'}}^{\infty} \frac{A\sqrt{C}\log^{v/2}(t)}{u_{\ch,\ch'}t^2} dt \\
&<& \infty,
\EEqn
where the last inequality follows from the fact that $\int_{A/u_{\ch,\ch'}}^{\infty} \frac{\log^{v/2}(t)}{t^2} dt < \infty$, and thus completes the proof.
\end{Proof}

\subsection{Donsker Property}

We follow \cite{van1996weak} and \cite{gine1984some} to introduce some basic notations and preliminary results.

\begin{defi}\label{Def:Gaussian}
A stochastic process $X$ is called Gaussian if each of its finite-dimensional marginals $(X(t_1), X(t_2),\cdots, X(t_k))$ has a multivariate normal distribution.
\end{defi}

\begin{defi}[Section 2.2.3 of \cite{kosorok2008introduction}]\label{Def:WeakConvg}
Let $\{X_n\}_{n=1}^{\infty}$ be a sequence of random processes in $\cd$. For some tight process $X$ in $\bd$, we use the notation $X_n\leadsto X$ to denote the weak convergence of $X_n$ to $X$ in the metric space $(\bd,d)$, and it holds if and only if
\begin{equation*}
    \sup\limits_{h\in BL_1} \lrabs{\be^{\ast}[h(X_n)] - \be[h(X)]} \rightarrow 0,
\end{equation*}
where $BL_1$ is the space of functions $f:\bd\mapsto\br$ with Lipschitz norm bounded by 1, i.e. $\|f\|_{\infty} \le 1$ and $\lrabs{f(x)-f(y)}\le d(x,y)$ for any $x,y\in\bd$ and where $\|\cdot\|_{\infty}$ is the uniform norm in $\bd$.
\end{defi}

\begin{defi}[\cite{van1996weak} page 81; \cite{gine1984some}]\label{Def:Donsker}
Given a measurable space $(S,\cs)$ and an $S$-valued random sample $\{X_i\}_{i=1}^{n}$ with the common law $P$ on $(S,\cs)$. Let $\cf$ denote a class of functions with empirical process $\bg_n := \{G_n(f): f\in\cf\}$ indexed by $\cf$ given by
\begin{equation*}
    G_n(f) = \sqrt{n} (\bp_n-P)(f) = \frac{1}{\sqrt{n}} \sum\limits_{i=1}^{n} \lrp{f(X_i)-\be[f(X)]},
\end{equation*}
Then $\cf$ is called a Donsker class or a $P$-Donsker class if its empirical process $\{\bg_nf: f\in\cf\}$ satisfies that
\begin{equation*}
    \bg_n = \sqrt{n} (\bp_n - P) \leadsto \bg \quad\mbox{in }\ell^{\infty}(\cf),
\end{equation*}
where $\bg = \{G(f):f\in\ch\}$ is a tight and centered Gaussian process indexed by $\cf$ with mean zero and covariance
\begin{equation*}
    \be[G(f)G(g)] = \int fgdP - \int fdP \int gdP
\end{equation*}
for any $f,g\in\cf$.
\end{defi}

\begin{defi}[Section 2.2.3 of \cite{kosorok2008introduction}]\label{Def:WeakConvg_cond}
Let $\widehat{X}_n$ be a sequence of bootstrapped processes in $\bd$ with random weights denoted by $M$. For some tight process $X$ in $\bd$, we use the notation $\widehat{X}_n \leadsto_{M}^{P} X$ to mean that 
\begin{enumerate}[label=(\roman*)]
    \item $\sup_{h\in BL_1} \lrabs{\be_M[h(\widehat{X}_n)] - \be[h(X)]} \stra{p} 0$;
    \item $\be_M[h(\widehat{X}_n)^{\ast}] - \be_M[h(\widehat{X}_n)_{\ast}] \stra{p} 0$ for all $h\in BL_1$,
\end{enumerate}
where $BL_1$ is defined as in Definition \ref{Def:WeakConvg} and $h(\widehat{X}_n)^{\ast}$ and $h(\widehat{X}_n)_{\ast}$ denote the measurable majorants and minorants with respect to the joint data including the weights $M$.
\end{defi}

\begin{lemma}[Theorem 10.4 of \cite{kosorok2008introduction}]\label{Lemma:Donsker_boot}
Let $\cf$ be a class of measurable functions and let $\{\varepsilon_i\}_{i=1}^{n}$ be iid random variables with mean zero, variance 1, and $\|\varepsilon_1\|_{2,1} = \int_0^{\infty}\sqrt{\bp(|\varepsilon|>t)}dt < \infty$ independent of the sample data $\{X_i\}_{i=1}^{n}$. Let
\begin{equation*}
    \bg_n' = \frac{1}{\sqrt{n}} \sum\limits_{i=1}^{n} \varepsilon_i (\delta_{X_i} - P), \quad
    \bg_n'' = \frac{1}{\sqrt{n}} \sum\limits_{i=1}^{n} (\varepsilon_i-\bar\varepsilon) \delta_{X_i},
\end{equation*}
where $\bar\varepsilon = \frac{1}{n}\sum\limits_{i=1}^{n}\varepsilon_i$. Then the following are equivalent:
\begin{enumerate}[label=(\roman*)]
    \item $\cf$ is $P$-Donsker;
    \item $\bg_n' \leadsto_\varepsilon^P \bg$ in $\ell^{\infty}(\cf)$ and $\bg_n'$ is asymptotically measurable;
    \item $\bg_n'' \leadsto_\varepsilon^P \bg$ in $\ell^{\infty}(\cf)$ and $\bg_n''$ is asymptotically measurable;
\end{enumerate}
\end{lemma}

\begin{defi}[Example 2.3.4 of \cite{van1996weak}]\label{Def:Pt_measurable}
The class $\cf$ of functions is said to be pointwise-measurable if $\cf$ contains a countable subset $\cg$ such that for every $f\in\ch$, there exists a sequence $g_m$ in $\cg$ with $g_m(x) \rightarrow f(x)$ for every $x$.
\end{defi}

\begin{lemma}[Theorem 11.6 of \cite{sen2018gentle}]\label{Lemma:Donsker_cover}
Let $\cf$ be a pointwise-measurable class of measurable functions with a measurable envelope $F$ such that $P(F^2) < \infty$. If 
\begin{equation}\label{Equ:Donsker_cover}
    \int_{0}^{\infty} \sup\limits_{Q} \sqrt{\log N(\cf, \|\cdot\|_{L_2(Q)}, \varepsilon\|F\|_{L_2(Q)})}
    d \varepsilon
    < \infty,
\end{equation}
then $\cf$ is $P$-Donsker. 
\end{lemma}

\begin{lemma}[Theorem 11.17 of \cite{ledoux1991probability}]\label{Lemma:DudleyContinuity}
Let $X = (X_t)_{t\in\ct}$ be a Gaussian process. Then 
\begin{equation*}
    \blre{\sup\limits_{t\in\ct} X_t}
\le 24 \int_{0}^{\infty} \sqrt{\log N(\ct, d, \varepsilon)} d\varepsilon.
\end{equation*}
Furthermore, if this entropy integral converges, $X$ has a version with almost all sample paths bounded and (uniformly) continuous on $(\ct, d)$.
\end{lemma}

\begin{lemma}\label{Lemma:Phi_Donsker}
Let $P$ denote the probability measure of $(X_1,Y_1,Y_1^{c\ast},X_2,Y_2,Y_2^{c\ast})$, then it holds under Assumption~\ref{Assumpt:kernel} and Assumption~\ref{Assumpt:entropy} that $\Phi$ is $P$-Donsker.
\end{lemma}
\begin{Proof}
We prove the statement using Lemma~\ref{Lemma:Donsker_cover}. Under Assumption~\ref{Assumpt:kernel}\ref{Assumpt:kernel_bound}, it is shown by Lemma~\ref{Lemma:Envelope} that $\ch,\ch'$ have a common envelope of $u_{\ch,\ch'}$. By the definition of $\Phi$, it holds that $\Phi$ has an envelope of $4u_{\ch,\ch'}^3$, which is square-integrable under $P$. Also note that condition (\ref{Equ:Donsker_cover}) has been verified in Lemma \ref{Lemma:Unif_Entropy}, then it remains to verify that $\Phi$ is a pointwise-measurable class of functions.

Under Assumption \ref{Assumpt:kernel}, the kernel $K:\br^{p+q}\times\br^{p+q} \rightarrow \br$ is continuous, then the associated RKHS $\bh$ is separable. Recall that $\ch$ is the unit ball in $\bh$, then there exists a countable subset $\cg\subseteq\ch$, s.t. for any $f\in\ch$, there exists a sequence $\{f_n\}_{n\in\bn}\subseteq\cg$, s.t. $\|f_n-f\|_{\bh} \rightarrow 0$. Note that for each $f_n$, it follows from the property of RKHS that
\begin{equation*}
    |f_n(x) - f(x)| 
  = |\lrag{f_n-f, K(x,\cdot)}_{\ch}| 
  \le \|f_n-f\|_{\bh} \|K(x,\cdot)\|_{\bh}.
\end{equation*}
Under Assumption \ref{Assumpt:kernel}\ref{Assumpt:kernel_bound}, the kernel $K$ is uniformly bounded, then $\sup\limits_{x\in\br^{p+q}} \|K(x,\cdot)\|_{\bh} < \infty$. Therefore, $\|f_n-f\|_\bh\rightarrow0$ implies $f_n(x) \rightarrow f(x)$ for any $x\in\br^{p+q}$, and the arbitrariness of $f$ leads to the pointwise-measurability of $\ch$. Similarly, we can show that $\ch'$ is also pointwise-measurable with the countable subset $\cg'$. 

Define 
\begin{equation*}
    \Phi' = \{\phi_{f,g,h}: f,g\in\cg, h\in\cg'\},
\end{equation*}
then $\Phi'$ is a countable subset of $\Phi$. Furthermore, for any $\phi_{f,g,h}\in\Phi$, there exist $\{f_n\}_{n\in\bn}\subseteq\cg, \{g_n\}_{n\in\bn}\subseteq\cg$ and $\{h_n\}_{n\in\bn}\subseteq\cg'$, s.t. $f_n\rightarrow f$, $g_n\rightarrow g$ and $h_n\rightarrow h$ pointwise. Since $\{\phi_{f_n,g_n,h_n}\}_n$ is a sequence in $\Phi'$, and it is trivial that for every $(x_1,y_1,y_1',x_2,y_2,y_2')\in\br^{2p+4q}$, it holds that 
\BEqn
& & \phi_{f_n,g_n,h_n}(x_1,y_1,y_1',x_2,y_2,y_2') \\
&=& \bigp{f_n(x_1,y_1) - f_n(x_1,y_1')}
    \bigp{g_n(x_2,y_2) - g_n(x_2,y_2')}
    h_n(x_1,x_2) \\
&\rightarrow&
  \bigp{f(x_1,y_1) - f(x_1,y_1')}
  \bigp{g(x_2,y_2) - g(x_2,y_2')}
  h(x_1,x_2) \\
&=& \phi_{f,g,h}(x_1,y_1,y_1',x_2,y_2,y_2'),    
\EEqn
Hence, by Lemma \ref{Lemma:Donsker_cover}, $\Phi$ is $P$-Donsker.
\end{Proof}

\section{Auxiliary Results for Asymptotic Boundedness}\label{Appendix:AuxDoubleRobustness}

\subsection{Auxiliary Results for Theorem \ref{Thm:Double-robustness}}

\begin{lemma}\label{Lemma:double_robustness_aux}
Let $\{\ci_\ell\}_{\ell=1}^{L}$, $\widehat{G}_1^{\pell}$ and $\widehat{G}_2^{\pell}$ be the notations defined in Section \ref{Sec:Method_TestStat}. 
For each $\ell=1,\ldots,L$, let $(X_1,Y_1^{c\ast},X_2,Y_2^{c\ast})$ denote generic held-out observations independent of $\cd^{\pmell}$, and
\begin{equation*}
    \widehat{Y}_1^{\pell} = \widehat{G}_2^{\pell}(X_1, Z_1),
    \qquad
    \widehat{Y}_2^{\pell} = \widehat{G}_1^{\pell}(X_2, Z_2),
\end{equation*}
where $Z_1,Z_2$ are the input noise independent of $\cd$.
Under Assumption \ref{Assumpt:double_robustness}, for each $n\ge1$, there exists a set $\ca_n$ of training sets satisfying $\bp(\ca_n) \rightarrow 1$, such that, for all $\cd \in \ca_n$,
\BEqn
& & \max\limits_{1\le\ell\le L}    
    \lrp{\bbe{ \sup\limits_{f\in\ch} 
               \bigp{f(X_1,Y_1^{c\ast}) - f(X_1,\widehat{Y}_1^{\pell})}^2 
         \mid \cd^{\pmell}}}^{1/2} 
    \le M_0 n^{-k_1}\log^{1/2}(n), \\
& & \max\limits_{1\le\ell\le L}    
    \lrp{\bbe{ \sup\limits_{g\in\ch}
               \bigp{g(X_2,Y_2^{c\ast}) - g(X_2,\widehat{Y}_2^{\pell})}^2
         \mid \cd^{\pmell}}}^{1/2} 
    \le M_0 n^{-k_2}\log^{1/2}(n), 
\EEqn
where $M_0 = L^{1/2} M$.
\end{lemma}

\begin{Proof}
For each $x\in\br^{p+q}$ and $f\in\ch$, we have $f(x) = \lrag{f,K(x,\cdot)}_\ch$ by the definition of RKHS. Then for each $\ell=1,\cdots,L$, it holds that
\BEqn
& & \blre{ \sup\limits_{f\in\ch} 
    \bigp{f(X_1,Y_1^{c\ast}) - f(X_1,\widehat{Y}_1^{\pell})}^2 
    \mid \cd^{\pmell}} \\ 
&=& \blre{ \sup\limits_{f\in\ch} 
    \lrag{f, K((X_1,Y_1^{c\ast}), \cdot) - K((X_1,\widehat{Y}_1^{\pell}), \cdot)}_{\ch}^2 
    \mid \cd^{\pmell}} \\ 
&\le& \blre{ \sup\limits_{f\in\ch} 
    \|f\|_\bh^2 \|K((X_1,Y_1^{c\ast}), \cdot) - K((X_1,\widehat{Y}_1^{\pell}), \cdot)\|_{\bh}^2
    \mid \cd^{\pmell}} \\ 
&\le& \blre{\|K((X_1,Y_1^{c\ast}), \cdot) - K((X_1,\widehat{Y}_1^{\pell}), \cdot)\|_{\bh}^2
    \mid \cd^{\pmell}},
\EEqn
where the second-to-last step follows from Cauchy-Schwarz inequality, and the last step follows from the fact that $\ch$ is a unit ball of $\bh$.

Furthermore, we have that
\BEqn
& & \blrp{\max\limits_{1\le\ell\le L}
    \bbe{ \sup\limits_{f\in\ch} \bigp{f(X_1,Y_1^{c\ast}) - f(X_1,\widehat{Y}_1^{\pell})}^2 
    \mid \cd^{\pmell}} \geq L M^2 n^{-2k_1}\log(n)} \\
&\le& \blrp{\max\limits_{1\le\ell\le L}    
    \bbe{\|K((X_1,Y_1^{c\ast}), \cdot) - K((X_1,\widehat{Y}_1^{\pell}), \cdot)\|_{\bh}^2
    \mid \cd^{\pmell}} \geq L M^2 n^{-2k_1}\log(n)} \\
&\le& \sum\limits_{\ell=1}^{L} 
    \blrp{\bbe{\|K((X_1,Y_1^{c\ast}), \cdot) - K((X_1,\widehat{Y}_1^{\pell}), \cdot)\|_{\bh}^2
    \mid \cd^{\pmell}} \geq L M^2 n^{-2k_1}\log(n)} \\
&\le& \sum\limits_{\ell=1}^{L} \frac{1}{L M^2 n^{-2k_1}\log(n)} 
    \blre{\blre{\|K((X_1,Y_1^{c\ast}), \cdot) - K((X_1,\widehat{Y}_1^{\pell}), \cdot)\|_{\bh}^2
    \mid \cd^{\pmell}}} \\
&=& \sum\limits_{\ell=1}^{L} \frac{1}{L M^2 n^{-2k_1}\log(n)} 
    \blre{\|K((X_1,Y_1^{c\ast}), \cdot) - K((X_1,\widehat{Y}_1^{\pell}), \cdot)\|_{\bh}^2} \\
&\le& \frac{1}{M^2 n^{-2k_1}\log(n)}
    \max\limits_{1\le\ell\le L} 
    \blre{ \|K\bigp{(X_1, Y_1^{c\ast}), \cdot} - K\bigp{(X_1, \widehat{Y}_1^{\pell}), \cdot}\|_{\bh}^2} \\
&\le& \frac{M^2 n^{-2k_1}}{M^2 n^{-2k_1}\log(n)} \\
&=& \log^{-1}(n),
\EEqn
where the third inequality follows from Markov's inequality, the equality follows from the law of total expectation, and the second-to-last inequality follows from \eqref{Equ:double_robustness_f} in Assumption \ref{Assumpt:double_robustness}. 

Equivalently, we have that
\begin{equation*}
    \blrp{\max\limits_{1\le\ell\le L}    
    \bigcp{\bbe{\sup\limits_{f\in\ch} \bigp{f(X_1,Y_1^{c\ast}) - f(X_1,\widehat{Y}_1^{\pell})}^2 
    \mid \cd^{\pmell}}}^{1/2} \geq L^{1/2} M n^{-k_1}\log^{1/2}(n)}
\le \log^{-1}(n).
\end{equation*}
Similarly, we have the counterpart statement over the supremum of $g\in\ch$, i.e.
\begin{equation*}
    \blrp{\max\limits_{1\le\ell\le L}    
    \bigcp{\bbe{\sup\limits_{g\in\ch} \bigp{g(X_2,Y_2^{c\ast}) - g(X_2,\widehat{Y}_2^{\pell})}^2 
    \mid \cd^{\pmell}}}^{1/2} \geq L^{1/2} M n^{-k_2}\log^{1/2}(n)}
\le \log^{-1}(n).
\end{equation*}

For each $n$, let $\ca_{n,f}$ and $\ca_{n,g}$ denote the events on which the two displayed inequalities hold, i.e.
{\small\begin{equation*}
    \ca_{n,f}
  = \lrcp{\cd: \max\limits_{1\le\ell\le L}    
    \bigp{\bbe{\sup\limits_{f\in\ch} \bigp{f(X_1,Y_1^{c\ast}) - f(X_1,\widehat{Y}_1^{\pell})}^2 
    \mid \cd^{\pmell}}}^{1/2} \le L^{1/2} M n^{-k_1}\log^{1/2}(n)},
\end{equation*}}
and
{\small\begin{equation*}
    \ca_{n,g}
  = \lrcp{\cd: \max\limits_{1\le\ell\le L}    
    \bigp{\bbe{\sup\limits_{g\in\ch} \bigp{g(X_2,Y_2^{c\ast}) - g(X_2,\widehat{Y}_2^{\pell})}^2 
    \mid \cd^{\pmell}}}^{1/2} \le L^{1/2} M n^{-k_2}\log^{1/2}(n)}.
\end{equation*}}
Let $\ca_n = \ca_{n,f} \cap \ca_{n,g}$, it follows that
\begin{equation*}
    \bp(\ca_n)
\geq 1 - 2\log^{-1}(n)
\rightarrow 1,
\end{equation*}
which completes the proof.
\end{Proof}

\begin{lemma}\label{Lemma:double_robustness}
Let $\ca_n$ denote the event defined in Lemma \ref{Lemma:double_robustness_aux}. Under Assumption~\ref{Assumpt:double_robustness}, the following bounds hold on the event $\ca_n$ for every $i=1,\cdots,n$,
\BEqn
& & \max\limits_{1\le\ell\le L}    
    \lrp{\bbe{ \sup\limits_{f\in\ch} 
               \bigp{f(X_{1i},Y_{1i}^{c\ast}) - f(X_{1i},\widehat{Y}_{1i})}^2 
               \bone\{i\in\ci_{\ell}\}
         \mid \cd^{\pmell}}}^{1/2} 
    \le M_0 n^{-k_1}\log^{1/2}(n), \\
& & \max\limits_{1\le\ell\le L}    
    \lrp{\bbe{ \sup\limits_{g\in\ch}
               \bigp{g(X_{2i},Y_{2i}^{c\ast}) - g(X_{2i},\widehat{Y}_{2i})}^2
               \bone\{i\in\ci_{\ell}\}
         \mid \cd^{\pmell}}}^{1/2} 
    \le M_0 n^{-k_2}\log^{1/2}(n), 
\EEqn
where $M_0 = L^{1/2} M$.
\end{lemma}
\begin{Proof}
For each fixed $i=1,\cdots,n$, let $\ell_i$ be the unique fold such that $i \in \ci_{\ell_i}$. For $\ell \neq \ell_i$, the indicator $\bone\{i\in\ci_\ell\}$ is zero, and hence the corresponding conditional expectation is zero.

For $\ell = \ell_i$, conditional on $\cd^{(-\ell_i)}$, the estimator $\widehat{G}_j^{(\ell_i)}$ is fixed, and the held-out observation
$(X_{ji}, Y_{ji}, Y_{ji}^{c\ast}, Z_{ji})$ is independent of $\cd^{(-\ell_i)}$ and has the same distribution as the generic held-out observation used in Lemma \ref{Lemma:double_robustness_aux}. Therefore, the conditional expectation in the display equals the corresponding fold-level conditional expectation in Lemma \ref{Lemma:double_robustness_aux}. The desired bounds then follow on $\ca_n$.
\end{Proof}

\begin{lemma}[Theorem 4.5 of \cite{sen2018gentle} with minor modifications]\label{Lemma:Dudley}
Let $(T,d)$ be a separable metric space and let $\{X_t:t\in T\}$ be a separable stochastic process. Suppose that for every $s,t\in T$ and every $u\geq0$, we have
\begin{equation*}
    \blrp{|X_s-X_t|\geq u} \le 2\exp\lrp{-\frac{u^2}{2d^2(s,t)}}. 
\end{equation*}
Then for every $t_0\in T$, we have
\begin{equation*}
    \blre{\sup\limits_{t\in T}|X_t-X_{t_0}|}
\le C\int_{0}^{\infty} \sqrt{\log N(T, d, \varepsilon)} d\varepsilon.
\end{equation*}
\end{lemma}

\begin{lemma}[Corollary 3.4 of \cite{talagrand1994sharper}]\label{Lemma:Talagrand}
Consider a class $\cf$ of functions on a probability space and assume that $-1\le f\le 1$ for all $f\in\cf$. Set $\sigma^2 = \sup\limits_{f\in\cf} \be[f^2]$, then 
\begin{equation*}
    \blre{\lrnorm{\sum\limits_{i=1}^{n} f^2(X_i)}_{\cf}}
\le n\sigma^2 + 8\blre{\lrnorm{\sum\limits_{i=1}^{n} \xi_i f(X_i)}_{\cf}},
\end{equation*}
where $\{\xi_i\}_{i=1}^{n}$ is a random sample of Rademacher random variables independent of $\{X_i\}_{i=1}^{n}$, and $\|\cdot\|_{\cf}$ denotes the supremum over $\cf$.
\end{lemma}


\begin{lemma}[Theorem 3.17 of \cite{sen2018gentle}]\label{Lemma:Symmetrization}
Let $\{X_i\}_{i=1}^{n}$ be a random sample with a common law. For any class of measurable functions, it holds that
\begin{equation*}
    \blre{\lrnorm{\frac{1}{n}\sum\limits_{i=1}^{n} \bigp{f(X_i)-\be[f(X_1)]}}_{\cf}}
\le 2 \blre{\lrnorm{\frac{1}{n}\sum\limits_{i=1}^{n} \xi_i f(X_i)}_{\cf}},
\end{equation*}
where $\{\xi_i\}_{i=1}^{n}$ is a random sample of Rademacher variables independent of $\{X_i\}_{i=1}^{n}$.
\end{lemma}

\begin{lemma}\label{Lemma:Rad_VC}
Let $\cf$ be a measurable uniformly bounded class of functions and $\{X_i\}_{i=1}^{n}\stsim{iid}X\in\br^{p}$ be a random sample following the common law $P$. Let $\sigma^2, u$ be any numbers satisfying that
\begin{equation*}
    \sup\limits_{f\in\cf} \be[f^2(X)] \le \sigma^2,\qquad
    \sup\limits_{f\in\cf} \|f\|_{\infty} \le u, \qquad
    0<\sigma\le u.
\end{equation*}
Assume that there exists an integer $v\geq1$ that solely depends on the dimension $p$ and some deterministic constants $a>3u\exp(v/2), K>0$, s.t.
\begin{equation}\label{Equ:Rad_VC_condition}
    \sup\limits_{P} 
    \log N(\cf\cup\{0\}, \|\cdot\|_{L_2(P)}, u\varepsilon)
\le K \log^{v}(\frac{a}{u\varepsilon}),
\qquad \forall~ 0<\varepsilon<1.
\end{equation}
Then there exists some constant $C>0$ that solely depends on the dimension $p$, s.t.
\begin{equation*}
    \blre{\lrnorm{\sum\limits_{i=1}^{n} \xi_i f(X_i)}_{\cf}} 
\le C\lrp{u\log^v\lrp{\frac{a}{\sigma}} + \sqrt{n\sigma^2\log^v\lrp{\frac{a}{\sigma}}}},
\end{equation*}
where $\{\xi_i\}_{i=1}^{n}$ is a random sample of Rademacher variables independent of $\{X_i\}_{i=1}^{n}$.
\end{lemma}
\begin{Proof}
Conditioning on $\{X_i\}_{i=1}^{n}$, the $L_2(P_n)$-norm of any $f\in\cf$ is given by
\begin{equation*}
    \|f\|_{L_2(P_n)}^2 = \frac{1}{n}\sum\limits_{i=1}^{n}f^2(X_i).
\end{equation*}
Then it follows from Hoeffding's inequality for Rademacher variables that for any $x\geq0$, we have that
\BEqn
    \blrp{\frac{1}{\sqrt{n}}\lrabs{\sum\limits_{i=1}^{n} \xi_i f(X_i) - \sum\limits_{i=1}^{n} \xi_i g(X_i)} \geq x}
&=& \blrp{\frac{1}{\sqrt{n}}\lrabs{\sum\limits_{i=1}^{n} \xi_i\lrp{f(X_i)-g(X_i)}} \geq x} \\
&\le& 2\exp\lrp{-\frac{x^2}{\frac{2}{n} \sum\limits_{i=1}^{n} \lrp{f(X_i) - g(X_i)}^2}} \\
&=& 2\exp\lrp{-\frac{x^2}{2\|f-g\|_{L_2(P_n)}^2}}.
\EEqn
This implies that $\{\frac{1}{\sqrt{n}}\sum\limits_{i=1}^{n} \xi_i f(X_i): f\in\cf\}$ is sub-Gaussian with respect to the metric $\|\cdot\|_{L_2(P_n)}$ when conditioning on $\{X_i\}_{i=1}^{n}$. 

By Dudley's entropy bound for Rademacher processes stated in Lemma \ref{Lemma:Dudley}, we can find the following upper bound for the conditional expectation given $\{X_i\}_{i=1}^{n}$, that is,
\BEqn
    \blreft{\xi}{\lrnorm{ \frac{1}{\sqrt{n}} \sum\limits_{i=1}^{n} \xi_i f(X_i) }_{\cf}}
&\le& C\int_{0}^{\infty} \sqrt{\log N(\cf\cup\{0\}, \|\cdot\|_{L_2(P_n)}, \varepsilon)} d\varepsilon \\
&\le& C\sqrt{K} \int_{0}^{\sup\limits_{f\in\cf} \|f\|_{L_2(P_n)}} \log^{v/2}\lrp{\frac{a}{\varepsilon}} d\varepsilon \\
&=& aC\sqrt{K} \int_{a/\sup\limits_{f\in\cf} \|f\|_{L_2(P_n)}}^{\infty} \log^{v/2}(t)t^{-2}dt,
\EEqn
where the second inequality follows from the rescaling of Equation \eqref{Equ:Rad_VC_condition}, and the last step follows from the change of variable.

Next, we find the upper bound of the integral. Since $\sup\limits_{f\in\cf} \|f\|_{\infty} \le u$, it follows that $\sup\limits_{f\in\cf} \|f\|_{L_2(P_n)} \le u$. Define
\begin{equation*}
    G(t) = - \frac{\log^{v/2}(t)}{t}, \qquad \forall~ t \geq a/u.
\end{equation*}
By simple calculations, we have that 
\begin{equation*}
    G'(t)
  = \lrp{1 - \frac{v}{2\log(t)}} \frac{\log^{v/2}(t)}{t^2}
  \geq \lrp{1 - \frac{v}{2\log(a/u)}} \frac{\log^{v/2}(t)}{t^2}, 
  \qquad \forall~ t \geq a/u.
\end{equation*}
Since $a > 3u\exp(v/2)$, it holds that $\log(a/u) > \log(3\exp(v/2)) > v/2$, and $0 < 1 - \frac{v}{2\log(a/u)} < 1$. Furthermore, it follows that
\BEqn
& & \int_{a/\sup\limits_{f\in\cf} \|f\|_{L_2(P_n)}}^{\infty}
    \log^{v/2}(t) t^{-2} dt \\
&\le& \lrp{1 - \frac{v}{2\log(a/u)}}^{-1} 
    \int_{a/\sup\limits_{f\in\cf} \|f\|_{L_2(P_n)}}^{\infty} G'(t) dt \\
&=& \left.\lrp{1 - \frac{v}{2\log(a/u)}}^{-1} G(t)\right|_{a/\sup\limits_{f\in\cf} \|f\|_{L_2(P_n)}}^{\infty} \\
&=& \lrp{1 - \frac{v}{2\log(a/u)}}^{-1} \lrp{\frac{a}{\sup\limits_{f\in\cf}\|f\|_{L_2(P_n)}}}^{-1} \log^{v/2}\lrp{\frac{a}{\sup\limits_{f\in\cf}\|f\|_{L_2(P_n)}}},
\EEqn
which implies that
\begin{equation*}
    \blreft{\xi}{\lrnorm{ \frac{1}{\sqrt{n}} \sum\limits_{i=1}^{n} \xi_i f(X_i) }_{\cf}}
\le C' \sqrt{\sup\limits_{f\in\cf} \|f\|_{L_2(P_n)}^2 \log^v\lrp{\frac{a^2}{\sup\limits_{f\in\cf} \|f\|_{L_2(P_n)}^2}}},
\end{equation*}
where $C'>0$ is a constant.

Note that $y=\sqrt{x}$ is a concave function. In addition, $y=x\log^{v}(a^2/x)$ is concave and increasing over $(0,9u^2]$ since $a>3u\exp(v/2)$. By taking expectations on both sides and applying Jensen's inequality twice, we obtain that
\BEqn
& & \blre{\lrnorm{ \frac{1}{\sqrt{n}} \sum\limits_{i=1}^{n} \xi_i f(X_i)}_{\cf} } \\
&\le& C' \sqrt{\blre{\sup\limits_{f\in\cf} \|f\|_{L_2(P_n)}^2 
    \log^v\lrp{\frac{a^2}{\sup\limits_{f\in\cf} \|f\|_{L_2(P_n)}^2}}}} \\
&\le& C' \sqrt{\bbe{\sup\limits_{f\in\cf} \|f\|_{L_2(P_n)}^2} 
    \log^v\lrp{\frac{a^2}{\bbe{\sup\limits_{f\in\cf} \|f\|_{L_2(P_n)}^2}}}}.
\EEqn
Applying Lemma \ref{Lemma:Talagrand} to the rescaled class $\cf/u$ gives    
\begin{equation*}
    \blre{\sup\limits_{f\in\cf} \|f\|_{L_2(P_n)}^2}
  \le \sigma^2 
  + 8u\blre{\lrnorm{ \frac{1}{n}\sum\limits_{i=1}^{n}\xi_i f(X_i) }_{\cf}}.
\end{equation*}
By noting that $\sigma^2 \le u^2$ and $\blre{\lrnorm{ \frac{1}{n}\sum\limits_{i=1}^{n}\xi_i f(X_i) }_{\cf}} \le u$, we have that $\blre{\sup\limits_{f\in\cf} \|f\|_{L_2(P_n)}^2} \le 9u^2$. Recall that $y=x\log^{v}(a^2/x)$ is increasing over $(0,9u^2]$, then it follows that 
\BEqn
& & \blre{\lrnorm{\frac{1}{\sqrt{n}} \sum\limits_{i=1}^{n} \xi_i f(X_i)}_{\cf}} \\
&\le& C' \sqrt{\lrp{\sigma^2 + \frac{8u}{\sqrt{n}} \blre{\lrnorm{\frac{1}{\sqrt{n}}\sum\limits_{i=1}^{n}\xi_i f(X_i)}_{\cf}}} \log^v\lrp{\frac{a^2}{\sigma^2 + 8u\blre{\lrnorm{ \frac{1}{n}\sum\limits_{i=1}^{n}\xi_i f(X_i) }_{\cf}}}}} \\
&\le& C' \sqrt{\lrp{\sigma^2 + \frac{8u}{\sqrt{n}} \blre{\lrnorm{\frac{1}{\sqrt{n}}\sum\limits_{i=1}^{n}\xi_i f(X_i)}_{\cf}}} \log^v\lrp{\frac{a^2}{\sigma^2}}},
\EEqn
which is equivalent to a quadratic inequality of $Z := \blre{\lrnorm{\sum\limits_{i=1}^{n} \xi_i f(X_i)}_{\cf}}$, i.e.
\begin{equation*}
    Z^2 
\le C_1 n\sigma^2\log^v\lrp{\frac{a}{\sigma}} 
  + C_2 u \log^v\lrp{\frac{a}{\sigma}} Z,
\end{equation*}
where $C_1,C_2$ are some constants. Then we can conclude that the larger root of the corresponding equation is an upper bound of $Z$, that is, 
\begin{equation*}
    \frac{C_2 u \log^v\lrp{\frac{a}{\sigma}} + \sqrt{C_2^2 u^2 \log^{2v}\lrp{\frac{a}{\sigma}} + 4 C_1 n\sigma^2\log^v\lrp{\frac{a}{\sigma}} }}{2}.
\end{equation*}
By using the basic inequality $\sqrt{a+b}\le \sqrt{a} + \sqrt{b}$ for any $a,b\geq0$, we conclude that there exists some constant $C$ that depends only on the dimension of the domain of $\cf$, s.t.
\begin{equation*}
    \blre{\lrnorm{\sum\limits_{i=1}^{n} \xi_i f(X_i)}_{\cf}}
\le C\lrp{u\log^v\lrp{\frac{a}{\sigma}} + \sqrt{n\sigma^2\log^v\lrp{\frac{a}{\sigma}}}},
\end{equation*}
which completes the proof.
\end{Proof}

\begin{rmk}
Lemma \ref{Lemma:Rad_VC} provides a Rademacher complexity bound beyond the classical finite-VC setting. In particular, both Proposition 2.1 of \cite{gine2001consistency} and Theorem 7.13 of \cite{sen2018gentle} establish related bounds for measurable uniformly bounded VC classes of functions, for which the covering number typically satisfies a polynomial bound, equivalently,
\begin{equation*}
    \log N(\cf, \|\cdot\|_{L_2(P)}, \varepsilon)
    \lesssim
    v \log(1/\varepsilon).
\end{equation*}
In the present work, the function classes arising from RKHS unit balls need not have finite VC dimension. We therefore replace the VC entropy condition by the logarithmic-power entropy condition \eqref{Equ:Rad_VC_condition}, allowing $\cf$ to have infinite VC dimension. This extension is useful for the RKHS-based function classes considered in this article, where the complexity of the class is controlled through entropy bounds rather than finite-dimensional combinatorial structures.

From a technical perspective, the proof follows the standard empirical-process strategy used in \cite{gine2001consistency} and \cite{sen2018gentle}, based on symmetrization, Dudley's entropy bound, Talagrand's inequality, and a quadratic self-bounding argument. The main additional step is the explicit control of the entropy integral under the faster logarithmic-power entropy growth allowed in \eqref{Equ:Rad_VC_condition}.
\end{rmk}

\begin{lemma}\label{Lemma:decomp}
It holds that 
\begin{equation*}
    |\widehat{U}-U^{\ast}| 
\le \max\limits_{1\le\ell\le L} U_1^{\pell} 
  + \max\limits_{1\le\ell\le L} U_2^{\pell} 
  + \max\limits_{1\le\ell\le L} U_3^{\pell},
\end{equation*}
where
\begin{eqnarray*}
    U_1^{\pell}
&=& \lrnorm{\frac{1}{n_0} \sum\limits_{i\in\ci_\ell} 
    \phi_{f,g,h}(X_{1i},Y_{1i}^{c\ast},\widehat{Y}_{1i},X_{2i},Y_{2i},Y_{2i}^{c\ast})}_{\Phi}, \\
    U_2^{\pell}
&=& \lrnorm{\frac{1}{n_0} \sum\limits_{i\in\ci_\ell} 
    \phi_{f,g,h}(X_{1i},Y_{1i},Y_{1i}^{c\ast},X_{2i},Y_{2i}^{c\ast},\widehat{Y}_{2i})}_{\Phi}, \\
    U_3^{\pell}
&=& \lrnorm{\frac{1}{n_0} \sum\limits_{i\in\ci_\ell} 
    \phi_{f,g,h}(X_{1i},Y_{1i}^{c\ast},\widehat{Y}_{1i},X_{2i},Y_{2i}^{c\ast},\widehat{Y}_{2i})}_{\Phi}.
\end{eqnarray*}
\end{lemma}
\begin{Proof}
For any $\phi_{f,g,h}\in\Phi$, recall Equation \eqref{Equ:phi},
\begin{equation*}
    \phi_{f,g,h}(x_1, y_1, y_1', x_2, y_2, y_2') 
  = \bigp{f(x_1,y_1) - f(x_1,y_1')} \bigp{g(x_2,y_2) - g(x_2,y_2')} h(x_1, x_2).
\end{equation*}
It follows that
\BEqn
& & \phi_{f,g,h}(x_1,y_1,\widehat{y}_1,x_2,y_2,\widehat{y}_2) 
  - \phi_{f,g,h}(x_1,y_1,y_1^{c\ast},x_2,y_2,y_2^{c\ast}) \\
&=& \lrbk{\bigcp{f(x_1,y_1) - f(x_1,\widehat{y}_1)}
    \bigcp{g(x_2,y_2) - g(x_2,\widehat{y}_2)}
    - \bigcp{f(x_1,y_1) - f(x_1,y_1^{c\ast})}
    \bigcp{g(x_2,y_2) - g(x_2,y_2^{c\ast})}}
    h(x_1,x_2) \\
&=& \bigcp{f(x_1,y_1^{c\ast}) - f(x_1,\widehat{y}_1)}
    \bigcp{g(x_2,y_2) - g(x_2,y_2^{c\ast})}
    h(x_1,x_2) \\
& & + \bigcp{f(x_1,y_1) - f(x_1,y_1^{c\ast})}
    \bigcp{g(x_2,y_2^{c\ast}) - g(x_2,\widehat{y}_2)}    
    h(x_1,x_2) \\
& & + \bigcp{f(x_1,y_1^{c\ast}) - f(x_1,\widehat{y}_1)}
    \bigcp{g(x_2,y_2^{c\ast}) - g(x_2,\widehat{y}_2)} 
    h(x_1,x_2) \\
&=& \phi_{f,g,h}(x_1,y_1^{c\ast},\widehat{y}_1,x_2,y_2,y_2^{c\ast})
  + \phi_{f,g,h}(x_1,y_1,y_1^{c\ast},x_2,y_2^{c\ast},\widehat{y}_2)
  + \phi_{f,g,h}(x_1,y_1^{c\ast},\widehat{y}_1,x_2,y_2^{c\ast},\widehat{y}_2).
\EEqn

Note the triangle inequality that $\lrabs{\sup_{\phi\in\Phi}|A_\phi| - \sup_{\phi\in\Phi}|B_\phi|} \le \sup_{\phi\in\Phi}|A_\phi - B_\phi|$, then by the definition of $\widehat{U}$ and $U^{\ast}$, we further have that 
\BEqn
& & |\widehat{U} - U^{\ast}| \\
&\le& \sup\limits_{\phi_{f,g,h}\in\Phi} 
    \lrabs{\frac{1}{n} \sum\limits_{i=1}^{n} 
    \lrp{\phi_{f,g,h}(X_{1i},Y_{1i},\widehat{Y}_{1i},X_{2i},Y_{2i},\widehat{Y}_{2i})
    - \phi_{f,g,h}(X_{1i},Y_{1i},Y_{1i}^{c\ast},X_{2i},Y_{2i},Y_{2i}^{c\ast})}} \\
&\le& \sup\limits_{\phi_{f,g,h}\in\Phi} 
    \lrabs{\frac{1}{n} \sum\limits_{i=1}^{n} 
    \phi_{f,g,h}(X_{1i},Y_{1i}^{c\ast},\widehat{Y}_{1i},X_{2i},Y_{2i},Y_{2i}^{c\ast})}
  + \sup\limits_{\phi_{f,g,h}\in\Phi} 
    \lrabs{\frac{1}{n} \sum\limits_{i=1}^{n} 
    \phi_{f,g,h}(X_{1i},Y_{1i},Y_{1i}^{c\ast},X_{2i},Y_{2i}^{c\ast},\widehat{Y}_{2i})} \\
& & + \sup\limits_{\phi_{f,g,h}\in\Phi} 
    \lrabs{\frac{1}{n} \sum\limits_{i=1}^{n} 
    \phi_{f,g,h}(X_{1i},Y_{1i}^{c\ast},\widehat{Y}_{1i},X_{2i},Y_{2i}^{c\ast},\widehat{Y}_{2i})} \\
&=& \sup\limits_{\phi_{f,g,h}\in\Phi} 
    \lrabs{\frac{1}{L} \sum\limits_{\ell=1}^{L}
    \frac{1}{n_0} \sum\limits_{i\in\ci_\ell} 
    \phi_{f,g,h}(X_{1i},Y_{1i}^{c\ast},\widehat{Y}_{1i},X_{2i},Y_{2i},Y_{2i}^{c\ast})} \\
& & + \sup\limits_{\phi_{f,g,h}\in\Phi} 
    \lrabs{\frac{1}{L} \sum\limits_{\ell=1}^{L}
    \frac{1}{n_0} \sum\limits_{i\in\ci_\ell} 
    \phi_{f,g,h}(X_{1i},Y_{1i},Y_{1i}^{c\ast},X_{2i},Y_{2i}^{c\ast},\widehat{Y}_{2i})} \\
& & + \sup\limits_{\phi_{f,g,h}\in\Phi} 
    \lrabs{\frac{1}{L} \sum\limits_{\ell=1}^{L}
    \frac{1}{n_0} \sum\limits_{i\in\ci_\ell} 
    \phi_{f,g,h}(X_{1i},Y_{1i}^{c\ast},\widehat{Y}_{1i},X_{2i},Y_{2i}^{c\ast},\widehat{Y}_{2i})} \\
&\le& \frac{1}{L} \sum\limits_{\ell=1}^{L}
    \sup\limits_{\phi_{f,g,h}\in\Phi} 
    \lrabs{\frac{1}{n_0} \sum\limits_{i\in\ci_\ell} 
    \phi_{f,g,h}(X_{1i},Y_{1i}^{c\ast},\widehat{Y}_{1i},X_{2i},Y_{2i},Y_{2i}^{c\ast})} \\
& & + \frac{1}{L} \sum\limits_{\ell=1}^{L}
    \sup\limits_{\phi_{f,g,h}\in\Phi} 
    \lrabs{\frac{1}{n_0} \sum\limits_{i\in\ci_\ell} 
    \phi_{f,g,h}(X_{1i},Y_{1i},Y_{1i}^{c\ast},X_{2i},Y_{2i}^{c\ast},\widehat{Y}_{2i})} \\
& & + \frac{1}{L} \sum\limits_{\ell=1}^{L}
    \sup\limits_{\phi_{f,g,h}\in\Phi} 
    \lrabs{\frac{1}{n_0} \sum\limits_{i\in\ci_\ell} 
    \phi_{f,g,h}(X_{1i},Y_{1i}^{c\ast},\widehat{Y}_{1i},X_{2i},Y_{2i}^{c\ast},\widehat{Y}_{2i})} \\
&\le& \max\limits_{1\le\ell\le L}  
    \lrnorm{\frac{1}{n_0} \sum\limits_{i\in\ci_\ell} 
    \phi_{f,g,h}(X_{1i},Y_{1i}^{c\ast},\widehat{Y}_{1i},X_{2i},Y_{2i},Y_{2i}^{c\ast})}_{\Phi}
   + \max\limits_{1\le\ell\le L}  
    \lrnorm{\frac{1}{n_0} \sum\limits_{i\in\ci_\ell} 
    \phi_{f,g,h}(X_{1i},Y_{1i},Y_{1i}^{c\ast},X_{2i},Y_{2i}^{c\ast},\widehat{Y}_{2i})}_{\Phi} \\
& & + \max\limits_{1\le\ell\le L}  
    \lrnorm{\frac{1}{n_0} \sum\limits_{i\in\ci_\ell} 
    \phi_{f,g,h}(X_{1i},Y_{1i}^{c\ast},\widehat{Y}_{1i},X_{2i},Y_{2i}^{c\ast},\widehat{Y}_{2i})}_{\Phi} \\
&=& \max\limits_{1\le\ell\le L} U_1^{\pell} 
  + \max\limits_{1\le\ell\le L} U_2^{\pell} 
  + \max\limits_{1\le\ell\le L} U_3^{\pell},
\EEqn
which completes the proof.
\end{Proof}

\begin{lemma}\label{Lemma:Pf_mu12}
For $\ell=1,\cdots,L$ and $i=1,\cdots,n$, define
\BEqn
& & \mu_1^{\pell}(\phi_{f,g,h},\cd)
 := \blre{
    \phi_{f,g,h}(X_{1i},Y_{1i}^{c\ast},\widehat{Y}_{1i},X_{2i},Y_{2i},Y_{2i}^{c\ast})
    \bone\{i\in\ci_\ell\} \mid \cd^{\pmell}}, \\
& & \mu_2^{\pell}(\phi_{f,g,h},\cd)
 := \blre{
    \phi_{f,g,h}(X_{1i},Y_{1i},Y_{1i}^{c\ast},X_{2i},Y_{2i}^{c\ast},\widehat{Y}_{2i})
    \bone\{i\in\ci_\ell\} \mid \cd^{\pmell}},
\EEqn
where $Y_{1i}^{c\ast}, Y_{2i}^{c\ast}$ are given by Equation \eqref{Equ:Y_ast} and $\widehat{Y}_{1i}, \widehat{Y}_{2i}$ are given by Equation \eqref{Equ:Y_hat}. 

Under Assumption \ref{Assumpt:kernel}--\ref{Assumpt:X1X2} and Assumption \ref{Assumpt:double_robustness}, for any $n\geq 1$ and $\cd=\{(X_{1i},Y_{1i},X_{2i},Y_{2i})\}_{i=1}^{n}\in\ca_n$, it holds that
\BEqn
& & \max\limits_{1\le\ell\le L}
    \sup\limits_{\phi_{f,g,h}\in\Phi}
    \lrabs{\mu_1^{\pell}(\phi_{f,g,h},\cd)}
\le \mu_1 := 2u_{\ch,\ch'}^2 M_0 n^{-k_1} \log^{1/2}(n), \\
& & \max\limits_{1\le\ell\le L}
    \sup\limits_{\phi_{f,g,h}\in\Phi}
    \lrabs{\mu_2^{\pell}(\phi_{f,g,h},\cd)} 
\le \mu_2 := 2u_{\ch,\ch'}^2 M_0 n^{-k_2} \log^{1/2}(n), 
\EEqn
where $k_1,k_2$ are defined as in Assumption \ref{Assumpt:double_robustness}, $u_{\ch,\ch'}$ is defined as in Lemma~\ref{Lemma:Envelope}, and $M_0,\ca_n$ are defined as in Lemma \ref{Lemma:double_robustness_aux}.

In addition, it holds under the null for any $n\geq1$ that 
\begin{equation*}
    \max\limits_{1\le\ell\le L}
    \sup\limits_{\phi_{f,g,h}\in\Phi}
    \lrabs{\mu_1^{\pell}(\phi_{f,g,h},\cd)}
  = \mu_{1,H_0} := 0, \qquad
    \max\limits_{1\le\ell\le L}
    \sup\limits_{\phi_{f,g,h}\in\Phi}
    \lrabs{\mu_2^{\pell}(\phi_{f,g,h},\cd)} 
  = \mu_{2,H_0} := 0. 
\end{equation*}
\end{lemma}
\begin{Proof}
We only prove the statements for $\mu_1^{\pell}(\phi_{f,g,h},\cd)$. Under Assumption \ref{Assumpt:kernel}\ref{Assumpt:kernel_bound}, it follows from Lemma~\ref{Lemma:Envelope} that $\lrabs{g(X_{2i},Y_{2i}) - g(X_{2i},Y_{2i}^{c\ast})} \le 2u_{\ch,\ch'}$ for any $g\in\ch$ and $(X_{2i},Y_{2i},Y_{2i}^{c\ast})$. In addition, we have that $|h(X_{1i},X_{2i})|\le u_{\ch,\ch'}$ for any $h\in\ch'$ and any $(X_{1i},X_{2i})$. Then it follows from Jensen's inequality that
\BEqn
& & \lrabs{\mu_1^{\pell}(\phi_{f,g,h},\cd)} \\
&=& \lrabs{\bbe{
    \bigp{f(X_{1i},Y_{1i}^{c\ast}) - f(X_{1i},\widehat{Y}_{1i})} 
    \bigp{g(X_{2i},Y_{2i}) - g(X_{2i}, Y_{2i}^{c\ast})} 
    h(X_{1i},X_{2i}) \bone\{i\in\ci_\ell\} \mid \cd^{\pmell}}} \\
&\le& 2u_{\ch,\ch'}^2 \cdot
    \bbe{ |f(X_{1i},Y_{1i}^{c\ast}) - f(X_{1i},\widehat{Y}_{1i})| 
    \bone\{i\in\ci_\ell\} \mid \cd^{\pmell}} \\
&\le& 2u_{\ch,\ch'}^2 \cdot
    \lrp{\bbe{ \bigp{f(X_{1i},Y_{1i}^{c\ast}) - f(X_{1i},\widehat{Y}_{1i})}^2 
    \bone\{i\in\ci_\ell\} \mid \cd^{\pmell}}}^{1/2}.
\EEqn
Under Assumption \ref{Assumpt:double_robustness}, it follows from Lemma \ref{Lemma:double_robustness} that for any $n\geq 1$ and $\cd\in\ca_n$, we have that
\BEqn
& & \max\limits_{1\le\ell\le L}
    \sup\limits_{\phi_{f,g,h} \in \Phi}
    \lrabs{\mu_1^{\pell}(\phi_{f,g,h},\cd)} \\
&\le& 2u_{\ch,\ch'}^2 \max\limits_{1\le\ell\le L}    
    \lrp{\bbe{ \sup\limits_{f\in\ch} \bigp{f(X_{1i},Y_{1i}^{c\ast}) - f(X_{1i},\widehat{Y}_{1i})}^2 \bone\{i\in\ci_\ell\}
    \mid \cd^{\pmell}}}^{1/2} \\
&\le& 2u_{\ch,\ch'}^2 M_0 n^{-k_1}\log^{1/2}(n),
\EEqn
which implies that $\max\limits_{1\le\ell\le L} \sup\limits_{\phi_{f,g,h}\in\Phi} \lrabs{\mu_1^{\pell}(\phi_{f,g,h},\cd)} \le \mu_1$ with $\mu_1 = 2u_{\ch,\ch'}^2 M_0 n^{-k_1}\log^{1/2}(n)$.

Under Assumption \ref{Assumpt:kernel} and Assumption~\ref{Assumpt:X1X2}, it is shown by Theorem~\ref{Thm:Validity} that $P_{Y_1|X_1}(\cdot|x) = P_{Y_2|X_2}(\cdot|x)$ for both $P_{X_1}$- and $P_{X_2}$-almost every $x\in\cx$ under the null. Recall that $Y_{2i} = G_2^{\ast}(X_{2i},Z_{2i}^{\ast})$ and $Y_{2i}^{c\ast} = G_1^{\ast}(X_{2i}, Z_{2i}^{c \ast})$, then $(X_{2i},Y_{2i}) =^d (X_{2i},Y_{2i}^{c \ast})$ under the null. It follows from the tower property of the conditional expectation that
{\small\BEqn
& & \mu_1^{\pell}(\phi_{f,g,h},\cd) \\
&=& \blre{
    \bbe{\bigp{f(X_{1i},Y_{1i}^{c\ast}) - f(X_{1i},\widehat{Y}_{1i})} 
    \bigp{g(X_{2i},Y_{2i}) - g(X_{2i}, Y_{2i}^{c\ast})}
    h(X_{1i},X_{2i}) \bone\{i\in\ci_\ell\} \mid X_{1i},X_{2i},\cd^{\pmell}}
    \mid \cd^{\pmell}} \\
&=& \blre{
    \bbe{ f(X_{1i},Y_{1i}^{c\ast}) - f(X_{1i},\widehat{Y}_{1i}) \mid X_{1i}, X_{2i}, \cd^{\pmell}}
    \cdot
    \bbe{ g(X_{2i},Y_{2i}) - g(X_{2i}, Y_{2i}^{c\ast}) \mid X_{1i}, X_{2i} }
    \cdot
    h(X_{1i},X_{2i}) 
    \mid \cd^{\pmell}} \\
&=& 0,
\EEqn}
where the second equality follows from the conditional independence of the two factors given $X_{1i}, X_{2i}$ and $\cd^{\pmell}$, and the last equality follows from the fact that 
\begin{equation*}
    \bbe{g(X_{2i},Y_{2i}) - g(X_{2i}, Y_{2i}^{c\ast}) \mid X_{1i},X_{2i}} = 0, \qquad \forall~g\in\ch.
\end{equation*}
This implies that $\max\limits_{1\le\ell\le L} \sup\limits_{\phi_{f,g,h} \in \Phi} \lrabs{\mu_1^{\pell}(\phi_{f,g,h},\cd)} = \mu_{1,H_0} = 0$ under the null. The statements for $\mu_2$ and $\mu_{2,H_0}$ follow by the same argument, with the roles of the two samples interchanged, and we spare the details.
\end{Proof}

\begin{lemma}\label{Lemma:Pf_mu3}
For $\ell=1,\cdots,L$ and $i=1,\cdots,n$, define
\begin{equation*}
    \mu_3^{\pell}(\phi_{f,g,h},\cd)
 := \blre{
    \phi_{f,g,h}(X_{1i},Y_{1i}^{c\ast},\widehat{Y}_{1i},X_{2i},Y_{2i}^{c\ast},\widehat{Y}_{2i})
    \bone\{i\in\ci_\ell\} \mid \cd^{\pmell}},
\end{equation*}
where $Y_{1i}^{c\ast}, Y_{2i}^{c\ast}$ are given by Equation \eqref{Equ:Y_ast} and $\widehat{Y}_{1i}, \widehat{Y}_{2i}$ are given by Equation \eqref{Equ:Y_hat}. 

Under Assumption \ref{Assumpt:kernel} and Assumption \ref{Assumpt:double_robustness}, for any $n\geq 1$ and $\cd=\{(X_{1i},Y_{1i},X_{2i},Y_{2i})\}_{i=1}^{n}\in\ca_n$, it holds that
\begin{equation*}
    \max\limits_{1\le\ell\le L}
    \sup\limits_{\phi_{f,g,h} \in \Phi}
    \lrabs{\mu_3^{\pell}(\phi_{f,g,h},\cd)} 
\le \mu_3 = u_{\ch,\ch'} M_0^2n^{-(k_1+k_2)}\log(n),
\end{equation*}
where $k_1,k_2$ are defined as in Assumption \ref{Assumpt:double_robustness}, $u_{\ch,\ch'}$ is defined as in Lemma~\ref{Lemma:Envelope}, and $M_0,\ca_n$ are defined as in Lemma \ref{Lemma:double_robustness_aux}.
\end{lemma}
\begin{Proof}
It follows from the similar techniques used for Lemma \ref{Lemma:Pf_mu12} that
\BEqn
& & \mu_3^{\pell}(\phi_{f,g,h},\cd) \\
&=& \blre{\bbe{
    \bigp{f(X_{1i},Y_{1i}^{c\ast}) - f(X_{1i},\widehat{Y}_{1i})} 
    \bigp{g(X_{2i},Y_{2i}^{c\ast}) - g(X_{2i}, \widehat{Y}_{2i})}
    h(X_{1i}, X_{2i}) \bone\{i\in\ci_\ell\} \mid X_{1i}, X_{2i}, \cd^{\pmell}} 
    \mid \cd^{\pmell}} \\
&=& \be\Big[
    \bbe{f(X_{1i},Y_{1i}^{c\ast}) - f(X_{1i},\widehat{Y}_{1i}) \bone\{i\in\ci_\ell\} \mid X_{1i}, X_{2i}, \cd^{\pmell}} \\
& & \hspace{1.5em}    
    \times
    \bbe{g(X_{2i},Y_{2i}^{c\ast}) - g(X_{2i},\widehat{Y}_{2i}) \bone\{i\in\ci_\ell\} \mid X_{1i}, X_{2i}, \cd^{\pmell}}
    \cdot
    h(X_{1i}, X_{2i})
    \mid \cd^{\pmell} \Big].
\EEqn

By Lemma~\ref{Lemma:Envelope}, we have $|h(X_{1i},X_{2i})|\le u_{\ch,\ch'}$ for any $h\in\ch'$ and any $(X_{1i},X_{2i})$ under Assumption \ref{Assumpt:kernel}\ref{Assumpt:kernel_bound}. Together with the conditional independence and Jensen's inequality, we have that
\BEqn
& & \lrabs{\mu_3^{\pell}(\phi_{f,g,h},\cd)} \\
&\le& u_{\ch,\ch'}
    \lrp{\blre{ \bbe{\bigp{f(X_{1i},Y_{1i}^{c\ast}) - f(X_{1i},\widehat{Y}_{1i})}^2 
    \bone\{i\in\ci_\ell\} \mid X_{1i}, X_{2i}, \cd^{\pmell}} \mid \cd^{\pmell}}}^{1/2} \\
& & \times 
    \lrp{\blre{ \bbe{\bigp{g(X_{2i},Y_{2i}^{c\ast}) - g(X_{2i},\widehat{Y}_{2i})}^2 
    \bone\{i\in\ci_\ell\} \mid X_{1i}, X_{2i}, \cd^{\pmell}} \mid \cd^{\pmell}}}^{1/2} \\    
&=& u_{\ch,\ch'}
    \lrp{\blre{ \bigp{f(X_{1i},Y_{1i}^{c\ast}) - f(X_{1i},\widehat{Y}_{1i})}^2
    \bone\{i\in\ci_\ell\} \mid \cd^{\pmell}}}^{1/2} \\
& & \hspace{1em} \times
    \lrp{\blre{ \bigp{g(X_{2i},Y_{2i}^{c\ast}) - g(X_{2i},\widehat{Y}_{2i})}^2 
    \bone\{i\in\ci_\ell\} \mid \cd^{\pmell}}}^{1/2}.
\EEqn
Furthermore, under Assumption \ref{Assumpt:double_robustness}, it follows from Lemma \ref{Lemma:double_robustness} that for any $n\geq 1$ and $\cd\in\ca_n$, we have that
\BEqn
& & \max\limits_{1\le\ell\le L}
    \sup\limits_{\phi_{f,g,h} \in \Phi}
    \lrabs{\mu_3^{\pell}(\phi_{f,g,h},\cd)} \\
&\le& u_{\ch,\ch'} \max\limits_{1\le\ell\le L}    
    \lrp{\bbe{ \sup\limits_{f\in\ch} \bigp{f(X_{1i},Y_{1i}^{c\ast}) - f(X_{1i},\widehat{Y}_{1i})}^2 \bone\{i\in\ci_\ell\}
    \mid \cd^{\pmell}}}^{1/2} \\
& & \hspace{1em} \times
    \max\limits_{1\le\ell\le L}    
    \lrp{\bbe{ \sup\limits_{g\in\ch} \bigp{g(X_{2i},Y_{2i}^{c\ast}) - g(X_{2i},\widehat{Y}_{2i})}^2 \bone\{i\in\ci_\ell\}
    \mid \cd^{\pmell}}}^{1/2} \\
&\le& u_{\ch,\ch'} M_0^2 n^{-(k_1+k_2)}\log(n),
\EEqn
which completes the proof.
\end{Proof}

\begin{lemma}\label{Lemma:Pf_sigma12}
For $\ell=1,\cdots,L$ and $i=1,\cdots,n$, define 
\BEqn
& & \lrp{\sigma_1^{\pell}(\phi_{f,g,h},\cd)}^2
 := \blre{
    \phi_{f,g,h}^2(X_{1i},Y_{1i}^{c\ast},\widehat{Y}_{1i},X_{2i},Y_{2i},Y_{2i}^{c\ast})
    \bone\{i\in\ci_\ell\} \mid \cd^{\pmell}}, \\
& & \lrp{\sigma_2^{\pell}(\phi_{f,g,h},\cd)}^2
 := \blre{
    \phi_{f,g,h}^2(X_{1i},Y_{1i},Y_{1i}^{c\ast},X_{2i},Y_{2i}^{c\ast},\widehat{Y}_{2i})
    \bone\{i\in\ci_\ell\} \mid \cd^{\pmell}},
\EEqn
where $Y_{1i}^{c\ast}, Y_{2i}^{c\ast}$ are given by Equation \eqref{Equ:Y_ast} and $\widehat{Y}_{1i}, \widehat{Y}_{2i}$ are given by Equation \eqref{Equ:Y_hat}. 

Under Assumption \ref{Assumpt:kernel} and Assumption \ref{Assumpt:double_robustness}, for any $n\geq 1$ and $\cd=\{(X_{1i},Y_{1i},X_{2i},Y_{2i})\}_{i=1}^{n}\in\ca_n$, it holds that
\BEqn
& & \max\limits_{1\le\ell\le L}
    \sup\limits_{\phi_{f,g,h} \in \Phi}
    \lrp{\sigma_1^{\pell}(\phi_{f,g,h},\cd)}^2
\le \sigma_1^2 := 4u_{\ch,\ch'}^4 M_0^2 n^{-2k_1} \log(n), \\
& & \max\limits_{1\le\ell\le L}
    \sup\limits_{\phi_{f,g,h} \in \Phi}
    \lrp{\sigma_2^{\pell}(\phi_{f,g,h},\cd)}^2
\le \sigma_2^2 := 4u_{\ch,\ch'}^4 M_0^2 n^{-2k_2} \log(n),
\EEqn
where $k_1,k_2$ are defined as in Assumption \ref{Assumpt:double_robustness}, $u_{\ch,\ch'}$ is defined as in Lemma~\ref{Lemma:Envelope}, and $M_0,\ca_n$ are defined as in Lemma \ref{Lemma:double_robustness_aux}.
\end{lemma}
\begin{Proof}
Under Assumption \ref{Assumpt:kernel}\ref{Assumpt:kernel_bound}, by using the uniform boundedness property established in Lemma~\ref{Lemma:Envelope}, it holds for any $\phi_{f,g,h}\in\Phi$ that
\begin{equation*}
    \lrp{\sigma_1^{\pell}(\phi_{f,g,h},\cd)}^2
\le 4u_{\ch,\ch'}^4 
    \blre{ \bigp{f(X_{1i},Y_{1i}^{c\ast}) - f(X_{1i},\widehat{Y}_{1i})}^2
    \bone\{i\in\ci_\ell\} \mid \cd^{\pmell}}.
\end{equation*}
Under Assumption \ref{Assumpt:double_robustness}, for any $n\geq 1$ and $\cd=\{(X_{1i},Y_{1i},X_{2i},Y_{2i})\}_{i=1}^{n}\in\ca_n$, it follows from Lemma \ref{Lemma:double_robustness} that 
\BEqn
& & \max\limits_{1\le\ell\le L} \sup\limits_{\phi_{f,g,h} \in \Phi}
    \lrp{\sigma_1^{\pell}(\phi_{f,g,h},\cd)}^2 \\
&\le& 4u_{\ch,\ch'}^4 \max\limits_{1\le\ell\le L} \sup\limits_{f\in\ch} 
    \blre{ \bigp{f(X_{1i},Y_{1i}^{c\ast}) - f(X_{1i},\widehat{Y}_{1i})}^2
    \bone\{i\in\ci_\ell\} \mid \cd^{\pmell}} \\
&\le& 4u_{\ch,\ch'}^4 M_0^2 n^{-2k_1} \log(n) = \sigma_1^2.
\EEqn
Similarly, we can verify the counterpart for $\sigma_2^2$.
\end{Proof}

\begin{lemma}\label{Lemma:Pf_sigma3}
For $\ell=1,\cdots,L$ and $i=1,\cdots,n$, define 
\begin{equation*}
    \lrp{\sigma_3^{\pell}(\phi_{f,g,h},\cd)}^2
 := \blre{
    \phi_{f,g,h}^2(X_{1i}, Y_{1i}^{c\ast}, \widehat{Y}_{1i}, X_{2i}, Y_{2i}^{c\ast}, \widehat{Y}_{2i})
    \bone\{i\in\ci_\ell\} \mid \cd^{\pmell}},
\end{equation*}
where $Y_{1i}^{c\ast}, Y_{2i}^{c\ast}$ are given by Equation \eqref{Equ:Y_ast} and $\widehat{Y}_{1i}, \widehat{Y}_{2i}$ are given by Equation \eqref{Equ:Y_hat}. 

Under Assumption \ref{Assumpt:kernel} and Assumption \ref{Assumpt:double_robustness}, for any $n\geq 1$ and $\cd=\{(X_{1i},Y_{1i},X_{2i},Y_{2i})\}_{i=1}^{n}\in\ca_n$, it holds that
\BEqn
& & \max\limits_{1\le\ell\le L}
    \sup\limits_{\phi_{f,g,h} \in \Phi}
    \lrp{\sigma_3^{\pell}(\phi_{f,g,h},\cd)}^2
\le \sigma_3^2 := 4u_{\ch,\ch'}^4 M_0^2 n^{-(k_1+k_2)} \log(n),
\EEqn
where $k_1,k_2$ are defined as in Assumption \ref{Assumpt:double_robustness}, $u_{\ch,\ch'}$ is defined as in Lemma~\ref{Lemma:Envelope}, and $M_0,\ca_n$ are defined as in Lemma \ref{Lemma:double_robustness_aux}.
\end{lemma}
\begin{Proof}
Under Assumption \ref{Assumpt:kernel}\ref{Assumpt:kernel_bound}, by the envelope boundedness established in Lemma~\ref{Lemma:Envelope} and Cauchy-Schwarz's inequality, we have that 
\BEqn
& & \lrp{\sigma_3^{\pell}(\phi_{f,g,h},\cd)}^2 \\
&\le& 4u_{\ch,\ch'}^4
    \blre{|f(X_{1i},Y_{1i}^{c\ast}) - f(X_{1i}, \widehat{Y}_{1i})| \cdot
          |g(X_{2i},Y_{2i}^{c\ast}) - g(X_{2i}, \widehat{Y}_{2i})| \cdot
          \bone\{i\in\ci_\ell\} \mid \cd^{\pmell}} \\
&\le& 4u_{\ch,\ch'}^4
    \lrp{\blre{\bigp{f(X_{1i},Y_{1i}^{c\ast}) - f(X_{1i}, \widehat{Y}_{1i})}^2 
    \bone\{i\in\ci_\ell\} \mid \cd^{\pmell}}}^{1/2} \\
& & \hspace{1em} \times
    \lrp{\blre{\bigp{g(X_{2i},Y_{2i}^{c\ast}) - g(X_{2i}, \widehat{Y}_{2i})}^2 
    \bone\{i\in\ci_\ell\} \mid \cd^{\pmell}}}^{1/2}.    
\EEqn
Under Assumption \ref{Assumpt:double_robustness}, by using the established results in Lemma \ref{Lemma:double_robustness}, it holds for any $n\geq 1$ and $\cd = \{(X_{1i},Y_{1i},X_{2i},Y_{2i})\}_{i=1}^{n}\in\ca_n$ that
\BEqn
& & \max\limits_{1\le\ell\le L}
    \sup\limits_{\phi_{f,g,h} \in \Phi}
    \lrp{\sigma_3^{\pell}(\phi_{f,g,h},\cd)}^2 \\
&\le& 4u_{\ch,\ch'}^4
    \max\limits_{1\le\ell\le L}    
    \lrp{\bbe{ \sup\limits_{f\in\ch} \bigp{f(X_{1i},Y_{1i}^{c\ast}) - f(X_{1i},\widehat{Y}_{1i})}^2 \bone\{i\in\ci_\ell\}
    \mid \cd^{\pmell}}}^{1/2} \\
& & \hspace{1em} \times
    \max\limits_{1\le\ell\le L}    
    \lrp{\bbe{ \sup\limits_{g\in\ch} \bigp{g(X_{2i},Y_{2i}^{c\ast}) - g(X_{2i},\widehat{Y}_{2i})}^2 \bone\{i\in\ci_\ell\}
    \mid \cd^{\pmell}}}^{1/2} \\
&\le& 4u_{\ch,\ch'}^4 M_0^2 n^{-(k_1+k_2)} \log(n),
\EEqn
which arrives at the desired result.
\end{Proof}

\begin{lemma}\label{Lemma:Expectation_U}
Under Assumption \ref{Assumpt:kernel}--\ref{Assumpt:X1X2} and Assumption~\ref{Assumpt:entropy}--\ref{Assumpt:double_robustness}, it holds that
\BEqn
& & \max\limits_{1\le\ell\le L} \be[U_1^{\pell} \mid \cd^{\pmell}] \cdot \bone\{\cd\in\ca_n\}
\le C \lrp{4u_{\ch,\ch'}^3 n^{-1} \log^v\lrp{\frac{12au_{\ch,\ch'}^2}{\sigma_1}} + n^{-1/2}\sigma_1 \log^{v/2}\lrp{\frac{12au_{\ch,\ch'}^2}{\sigma_1}}} + \mu_1, \\
& & \max\limits_{1\le\ell\le L} \be[U_2^{\pell} \mid \cd^{\pmell}] \cdot \bone\{\cd\in\ca_n\}
\le C \lrp{4u_{\ch,\ch'}^3 n^{-1} \log^v\lrp{\frac{12au_{\ch,\ch'}^2}{\sigma_2}} + n^{-1/2}\sigma_2 \log^{v/2}\lrp{\frac{12au_{\ch,\ch'}^2}{\sigma_2}}} + \mu_2, \\
& & \max\limits_{1\le\ell\le L} \be[U_3^{\pell} \mid \cd^{\pmell}] \cdot \bone\{\cd\in\ca_n\}
\le C \lrp{4u_{\ch,\ch'}^3 n^{-1} \log^v\lrp{\frac{12au_{\ch,\ch'}^2}{\sigma_3}} + n^{-1/2}\sigma_3 \log^{v/2}\lrp{\frac{12au_{\ch,\ch'}^2}{\sigma_3}}} + \mu_3,
\EEqn
where $\ca_n$ is defined as in Lemma \ref{Lemma:double_robustness_aux}, $u_{\ch,\ch'}$ is defined as in Lemma~\ref{Lemma:Envelope}, $U_1^{\pell},U_2^{\pell},U_3^{\pell}$ are defined as in Lemma \ref{Lemma:decomp}, $\mu_1,\mu_2$ are defined as in Lemma \ref{Lemma:Pf_mu12}, $\mu_3$ is defined as in Lemma \ref{Lemma:Pf_mu3}, $\sigma_1,\sigma_2$ are defined as in Lemma \ref{Lemma:Pf_sigma12}, and $\sigma_3$ is defined as in Lemma \ref{Lemma:Pf_sigma3}.

In addition, the above bounds hold under the null with $\mu_1$ and $\mu_2$ replaced with $\mu_{1,H_0}$ and $\mu_{2,H_0}$.
\end{lemma}
\begin{Proof}
By the definition of $U_1^{\pell}$ in Lemma \ref{Lemma:decomp} and the triangle inequality, we have that
\BEqn
    U_1^{\pell}
&=& \lrnorm{\frac{1}{n_0} \sum\limits_{i\in\ci_\ell} 
    \phi_{f,g,h}(X_{1i},Y_{1i}^{c\ast},\widehat{Y}_{1i},X_{2i},Y_{2i},Y_{2i}^{c\ast})}_{\Phi} \\
&\le& \lrnorm{\frac{1}{n_0} \sum\limits_{i\in\ci_\ell} 
    \phi_{f,g,h}(X_{1i},Y_{1i}^{c\ast},\widehat{Y}_{1i},X_{2i},Y_{2i},Y_{2i}^{c\ast})
    - \mu_1^{\pell}(\phi_{f,g,h},\cd)}_{\Phi} 
 + \lrnorm{\mu_1^{\pell}(\phi_{f,g,h},\cd)}_{\Phi}.
\EEqn
Recall Lemma \ref{Lemma:Pf_mu12}, $\mu_1^{\pell}(\phi_{f,g,h},\cd)$ is measurable with respect to $\cd^{\pmell}$. By the cross-fitting construction, $\{(X_{1i}, Y_{1i}^{c\ast}, \widehat{Y}_{1i}, X_{2i}, Y_{2i}, Y_{2i}^{c\ast})\}_{i\in\ci_\ell}$ are conditionally iid given $\cd^{\pmell}$. By Lemma \ref{Lemma:Symmetrization}, we have that 
\BEqn
& & \max\limits_{1\le\ell\le L} 
    \blre{U_1^{\pell} \mid \cd^{\pmell}} \cdot \bone\{\cd\in\ca_n\} \\
&\le& \max\limits_{1\le\ell\le L}  \blre{
    \lrnorm{\frac{1}{n_0} \sum\limits_{i\in\ci_\ell} 
    \phi_{f,g,h}(X_{1i},Y_{1i}^{c\ast},\widehat{Y}_{1i}, X_{2i}, Y_{2i}, Y_{2i}^{c\ast})
    - \mu_1^{\pell}(\phi_{f,g,h},\cd)}_{\Phi} \mid \cd^{\pmell}} \cdot \bone\{\cd\in\ca_n\} \\
& & + \max\limits_{1\le\ell\le L} \sup\limits_{\phi_{f,g,h}\in\Phi} \lrabs{\mu_1^{\pell}(\phi_{f,g,h},\cd)} \cdot \bone\{\cd\in\ca_n\} \\
&\le& 2 \max\limits_{1\le\ell\le L} \blre{
    \lrnorm{\frac{1}{n_0} \sum\limits_{i\in\ci_\ell} 
    \xi_i \phi_{f,g,h}(X_{1i},Y_{1i}^{c\ast},\widehat{Y}_{1i}, X_{2i}, Y_{2i}, Y_{2i}^{c\ast})}_{\Phi}
    \mid \cd^{\pmell}} \cdot \bone\{\cd\in\ca_n\} \\
& & + \max\limits_{1\le\ell\le L} \sup\limits_{\phi_{f,g,h}\in\Phi} \lrabs{\mu_1^{\pell}(\phi_{f,g,h},\cd)} \cdot \bone\{\cd\in\ca_n\}.
\EEqn

Under Assumption \ref{Assumpt:kernel}\ref{Assumpt:kernel_bound}, it follows from Lemma~\ref{Lemma:Envelope} that $\sup\limits_{\phi\in\Phi} \|\phi\|_{\infty} \le 4u_{\ch,\ch'}^3$. By Lemma \ref{Lemma:cover_num}, we have that
\BEqn
& & \sup\limits_{P\in\cp(\br^{2p+4q}, \cb(\br^{2p+4q}))} N(\Phi, \|\cdot\|_{L_2(P)}, 4u_{\ch,\ch'}^3\varepsilon) \\
&\le& \lrp{\sup\limits_{P\in\cp(\br^{p+q},\cb(\br^{p+q}))} N\lrp{\ch, \|\cdot\|_{L_2(P)}, \frac{u_{\ch,\ch'}\varepsilon}{3}}}^4 
    \lrp{\sup\limits_{P\in\cp(\br^{2p},\cb(\br^{2p}))} N\lrp{\ch',\|\cdot\|_{L_2(P)}, \frac{u_{\ch,\ch'}\varepsilon}{3}}}.
\EEqn
Together with Assumption \ref{Assumpt:entropy}, we further have that
\begin{equation*}
    \sup\limits_{P\in\cp(\br^{2p+4q},\cb(\br^{2p+4q}))}
    \log N(\Phi, \|\cdot\|_{L_2(P)}, 4u_{\ch,\ch'}^3\varepsilon)
\le 5\tau \log^{v} \lrp{\frac{3a}{u_{\ch,\ch'}\varepsilon}}
  = \tau' \log^{v} \lrp{\frac{a'}{4u_{\ch,\ch'}^3 \varepsilon}},
\end{equation*}
where $\tau' = 5\tau, a' = 12au_{\ch,\ch'}^2$. Under Assumption \ref{Assumpt:entropy}, we have $a > u_{K,K'}^{1/2} \exp(v/2) = u_{\ch,\ch'} \exp(v/2)$, then $a' > 3(4u_{\ch,\ch'}^3 \exp(v/2))$. By Lemma \ref{Lemma:Rad_VC} and Lemma \ref{Lemma:Pf_mu12}, we obtain that 
\BEqn
& & \max\limits_{1\le\ell\le L} 
    \blre{U_1^{\pell} \mid \cd^{\pmell}} \cdot \bone\{\cd\in\ca_n\} \\
&\le& C \lrp{4u_{\ch,\ch'}^3 n^{-1} \log^v\lrp{\frac{12au_{\ch,\ch'}^2}{\sigma_1}} + n^{-1/2}\sigma_1 \log^{v/2}\lrp{\frac{12au_{\ch,\ch'}^2}{\sigma_1}}} + \mu_1,
\EEqn
where $\sigma_1$ is given in Lemma \ref{Lemma:Pf_sigma12}, and $C$ is a positive constant that only depends on the dimensions $p,q$. Here we use the fact that $n_0 = n/L$ with the fixed split number $L$. By using similar arguments, we can verify the remaining statements.

Under the null, it follows from Lemma \ref{Lemma:Pf_mu12} that,
\begin{equation*}
    \max\limits_{1\le\ell\le L}
    \sup\limits_{\phi_{f,g,h}\in\Phi}
    \lrabs{\mu_1^{\pell}(\phi_{f,g,h},\cd)}
  = \mu_{1,H_0} := 0, \qquad
    \max\limits_{1\le\ell\le L}
    \sup\limits_{\phi_{f,g,h}\in\Phi}
    \lrabs{\mu_2^{\pell}(\phi_{f,g,h},\cd)} 
  = \mu_{2,H_0} := 0. 
\end{equation*}
Then it is trivial that the desired bounds hold with $\mu_1$ and $\mu_2$ replaced with $\mu_{1,H_0}$ and $\mu_{2,H_0}$.
\end{Proof}

\begin{lemma}\label{Lemma:Rate_U12}
Under Assumption~\ref{Assumpt:kernel}--\ref{Assumpt:X1X2} and Assumption~\ref{Assumpt:entropy}--\ref{Assumpt:double_robustness}, it holds that 
\BEqn
& & \max\limits_{1\le\ell\le L} U_1^{\pell}
 =  O_p\lrp{n^{-k_1}\log^{1/2}(n)}, \\
& & \max\limits_{1\le\ell\le L} U_2^{\pell}
 =  O_p\lrp{n^{-k_2}\log^{1/2}(n)}.
\EEqn
In addition, it holds under the null that
\BEqn
& & \max\limits_{1\le\ell\le L} U_1^{\pell}
 =  O_p\lrp{n^{-(k_1+\frac{1}{2})}\log^{(v+1)/2}(n)}, \\
& & \max\limits_{1\le\ell\le L} U_2^{\pell}
 =  O_p\lrp{n^{-(k_2+\frac{1}{2})}\log^{(v+1)/2}(n)}.
\EEqn
\end{lemma}
\begin{Proof}
We only provide the detailed proof for $U_1^{\pell}$, and that for $U_2^{\pell}$ is highly symmetric. Under Assumption~\ref{Assumpt:kernel}--\ref{Assumpt:X1X2} and Assumption \ref{Assumpt:double_robustness}, we have shown in Lemma \ref{Lemma:Pf_mu12} and Lemma \ref{Lemma:Pf_sigma12} that
\begin{equation*}
    \mu_1 = 2u_{\ch,\ch'}^2 M_0 n^{-k_1} \log^{1/2}(n), \qquad
    \sigma_1 = 2u_{\ch,\ch'}^2 M_0 n^{-k_1} \log^{1/2}(n),
\end{equation*}
where $u_{\ch,\ch'}, M_0, a$ are all positive constants. Then there exists some deterministic constant $C>0$, such that
\begin{equation*}
    \log\lrp{\frac{12au_{\ch,\ch'}^2}{\sigma_1}} 
  = \log\lrp{\frac{6a n^{k_1}}{M_0 \log^{1/2}(n)}}
  \le C \log(n).
\end{equation*}
By Lemma \ref{Lemma:Expectation_U}, it holds that
\BEqn
& & \max\limits_{1\le\ell\le L} \be[U_1^{\pell} \mid \cd^{\pmell}] \cdot \bone\{\cd\in\ca_n\} \\
&\le& C \lrp{4u_{\ch,\ch'}^3 n^{-1} \log^v\lrp{\frac{12au_{\ch,\ch'}^2}{\sigma_1}} + n^{-1/2}\sigma_1 \log^{v/2}\lrp{\frac{12au_{\ch,\ch'}^2}{\sigma_1}}} + \mu_1 \\
&\le& C \lrp{n^{-1} \log^v(n) + n^{-(k_1+\frac{1}{2})} \log^{(v+1)/2}(n) + n^{-k_1} \log^{1/2}(n)},
\EEqn
where $C>0$ is some positive constant that depends only on the dimension $p,q$ but may vary from line to line. 

Recall that $v\geq1$ is a constant that depends only on $p,q$, and $0<k_1<\frac{1}{2}$ under Assumption \ref{Assumpt:double_robustness}. Note that
\begin{equation*}
    \frac{n^{-k_1} \log^{1/2}(n)}{n^{-1} \log^v(n)}
  = \frac{n^{1-k_1}}{\log^{v-1/2}(n)}
  \rightarrow \infty,
  \qquad
    \frac{n^{-k_1} \log^{1/2}(n)}{n^{-(k_1+\frac{1}{2})} \log^{(v+1)/2}(n)}
  = \frac{n^{1/2}}{\log^{v/2}(n)}
  \rightarrow \infty,
\end{equation*}
then $n^{-k_1} \log^{1/2}(n)$ is the asymptotically dominating term. Consequently, there exists some constant $C>0$ that depends on $p,q$, s.t.
\begin{equation*}
    \frac{\max\limits_{1\le\ell\le L} \be[U_1^{\pell} \mid \cd^{\pmell}] \cdot \bone\{\cd\in\ca_n\}}{{n^{-k_1} \log^{1/2}(n)}} 
\le C
\end{equation*}
for sufficiently large $n$.

Consequently, for any fixed $M>0$ and sufficiently large $n$, we have that
\BEqn
& & \blrp{\max\limits_{1\le\ell\le L} \frac{U_1^{\pell}}{n^{-k_1} \log^{1/2}(n)} \geq M} \\
&\le& \sum\limits_{\ell=1}^{L} \blrp{\frac{U_1^{\pell}}{n^{-k_1} \log^{1/2}(n)} \geq M} \\
&=& \sum\limits_{\ell=1}^{L} \blre{ \blrp{\frac{U_1^{\pell}}{n^{-k_1} \log^{1/2}(n)} \geq M \mid \cd^{\pmell}} \bone\{\cd\in\ca_n\}} \\
& & + \sum\limits_{\ell=1}^{L} \blre{ \blrp{\frac{U_1^{\pell}}{n^{-k_1} \log^{1/2}(n)} \geq M \mid \cd^{\pmell}} \bone\{\cd\notin\ca_n\}} \\
&\le& \sum\limits_{\ell=1}^{L} \blre{ \frac{1}{{Mn^{-k_1} \log^{1/2}(n)}} \blre{ U_1^{\pell} \mid \cd^{\pmell} } \bone\{\cd\in\ca_n\} } 
+ \sum\limits_{\ell=1}^{L} \bp(\ca_n^c) \\
&\le& L \blre{ \frac{\max\limits_{1\le\ell\le L} \be[U_1^{\pell} \mid \cd^{\pmell}] \cdot \bone\{\cd\in\ca_n\}}{{M n^{-k_1} \log^{1/2}(n)}} } 
+ L(1-\bp(\ca_n)) \\
&\le& \frac{CL}{M} + L(1-\bp(\ca_n)),
\EEqn
where the first equality follows from the tower property of conditional expectation together with the decomposition of $\bone\{\cd\in\ca_n\} + \bone\{\cd\notin\ca_n\} \equiv 1$. Note that we have shown in Lemma \ref{Lemma:double_robustness_aux} that $\bp(\ca_n)\rightarrow1$, then we can conclude that 
\begin{equation*}
    \max\limits_{1\le\ell\le L} U_1^{\pell} = O_p\lrp{n^{-k_1} \log^{1/2}(n)}.
\end{equation*}

Under the null, we have shown in Lemma \ref{Lemma:Pf_mu12} that $\mu_{1,H_0} = 0$, then following the same arguments, we have that 
\begin{equation*}
    \max\limits_{1\le\ell\le L} \be[U_1^{\pell} \mid \cd^{\pmell}] \cdot \bone\{\cd\in\ca_n\}
\le C \lrp{n^{-1} \log^v(n) + n^{-(k_1+\frac{1}{2})} \log^{(v+1)/2}(n)}.
\end{equation*}
With $0<k_1<\frac{1}{2}$, we observe that
\begin{equation*}
    \frac{n^{-(k_1+\frac{1}{2})} \log^{(v+1)/2}(n)}{n^{-1} \log^v(n)}
  = \frac{n^{\frac{1}{2}-k_1}}{\log^{(v-1)/2}(n)}
  \rightarrow \infty,
\end{equation*}
hence $n^{-(k_1+\frac{1}{2})} \log^{(v+1)/2}(n)$ dominates $n^{-1} \log^v(n)$ asymptotically, implying that
\begin{equation*}
    \frac{\max\limits_{1\le\ell\le L} \be[U_1^{\pell} \mid \cd^{\pmell}] \cdot \bone\{\cd\in\ca_n\}}{{n^{-(k_1+\frac{1}{2})} \log^{(v+1)/2}(n)}} 
\le C
\end{equation*}
for some constant $C>0$ and sufficiently large $n$. By repeating the previous analysis, we can show that it holds under the null that
\begin{equation*}
    \max\limits_{1\le\ell\le L} U_1^{\pell} = O_p\lrp{n^{-(k_1+\frac{1}{2})} \log^{(v+1)/2}(n)},
\end{equation*}
which completes the proof.
\end{Proof}

\begin{lemma}\label{Lemma:Rate_U3}
Under Assumption~\ref{Assumpt:kernel}--\ref{Assumpt:X1X2} and Assumption~\ref{Assumpt:entropy}--\ref{Assumpt:double_robustness}, it holds that 
\begin{equation*}
    \max\limits_{1\le\ell\le L} U_3^{\pell}
  = O_p(n^{-(k_1+k_2)}\log(n)).
\end{equation*}
\end{lemma}
\begin{Proof}
By Lemma \ref{Lemma:Pf_mu3} and Lemma \ref{Lemma:Pf_sigma3}, we have that 
\begin{equation*}
    \mu_3 = u_{\ch,\ch'}M_0^2n^{-(k_1+k_2)}\log(n), \quad
    \sigma_3 = 2u_{\ch,\ch'}^2 M_0 n^{-(k_1+k_2)/2} \log^{1/2}(n).
\end{equation*}
Then by Lemma \ref{Lemma:Expectation_U}, there exists a constant $C>0$ that depends only on $p,q$, s.t. it holds that
\BEqn
& & \max\limits_{1\le\ell\le L} \be[U_3^{\pell} \mid \cd^{\pmell}] \cdot \bone\{\cd\in\ca_n\} \\
&\le& C \lrp{4u_{\ch,\ch'}^3 n^{-1} \log^v\lrp{\frac{12au_{\ch,\ch'}^2}{\sigma_3}} + n^{-1/2}\sigma_3 \log^{v/2}\lrp{\frac{12au_{\ch,\ch'}^2}{\sigma_3}}} + \mu_3 \\
&\le& C \lrp{n^{-1} \log^v(n) + n^{-(k_1+k_2+1)/2} \log^{(v+1)/2}(n) + n^{-(k_1+k_2)} \log(n)}.
\EEqn
Under Assumption \ref{Assumpt:double_robustness}, we have $0<k_1,k_2<\frac{1}{2}$ and thus $k_1+k_2<1$. Together with $v\geq1$, it holds that
\begin{equation*}
    \frac{n^{-(k_1+k_2)} \log(n)}{n^{-1} \log^v(n)}
  = \frac{n^{1-k_1-k_2}}{\log^{v-1}(n)}
  \rightarrow \infty,
\end{equation*}
and
\begin{equation*}
    \frac{n^{-(k_1+k_2)} \log(n)}{n^{-(k_1+k_2+1)/2} \log^{(v+1)/2}(n)}
  = \frac{n^{(1-k_1-k_2)/2}}{\log^{(v-1)/2}(n)}
  \rightarrow \infty,
\end{equation*}
which implies that $n^{-(k_1+k_2)} \log(n)$ dominates $n^{-1} \log^v(n)$ and $n^{-(k_1+k_2+1)/2} \log^{(v+1)/2}(n)$ asymptotically. Therefore, 
\begin{equation*}
    \frac{\max\limits_{1\le\ell\le L} \be[U_3^{\pell} \mid \cd^{\pmell}] \cdot \bone\{\cd\in\ca_n\}}{n^{-(k_1+k_2)} \log(n)} 
\le C
\end{equation*}
for some constant $C>0$ and all sufficiently large $n$. Then by using the similar arguments as in Lemma \ref{Lemma:Rate_U12}, we arrive at the desired result.
\end{Proof}

\subsection{Auxiliary Results for Proposition \ref{Prop:Double-robustness-boot}}

\begin{lemma}\label{Lemma:decomp-boot}
It holds that 
\begin{equation*}
    |\widehat{U}^b - U^{b\ast}| 
\le \max\limits_{1\le\ell\le L} R_1^{\pell} 
  + \max\limits_{1\le\ell\le L} R_2^{\pell} 
  + \max\limits_{1\le\ell\le L} R_3^{\pell}
  + \max\limits_{1\le\ell\le L} U_1^{\pell} \cdot |\bar\varepsilon| 
  + \max\limits_{1\le\ell\le L} U_2^{\pell} \cdot |\bar\varepsilon| 
  + \max\limits_{1\le\ell\le L} U_3^{\pell} \cdot |\bar\varepsilon|,
\end{equation*}
where $\bar\varepsilon = \frac{1}{n}\sum\limits_{i=1}^{n}\varepsilon_i$, $U_1^{\pell}, U_2^{\pell}, U_3^{\pell}$ are defined as in Lemma \ref{Lemma:decomp}, and 
\begin{eqnarray*}
    R_1^{\pell}
&=& \lrnorm{\frac{1}{n_0} \sum\limits_{i\in\ci_\ell} 
    \varepsilon_i \phi_{f,g,h}(X_{1i},Y_{1i}^{c\ast},\widehat{Y}_{1i},X_{2i},Y_{2i},Y_{2i}^{c\ast})}_{\Phi}, \\
    R_2^{\pell}
&=& \lrnorm{\frac{1}{n_0} \sum\limits_{i\in\ci_\ell} 
    \varepsilon_i \phi_{f,g,h}(X_{1i},Y_{1i},Y_{1i}^{c\ast},X_{2i},Y_{2i}^{c\ast},\widehat{Y}_{2i})}_{\Phi}, \\
    R_3^{\pell}
&=& \lrnorm{\frac{1}{n_0} \sum\limits_{i\in\ci_\ell} 
    \varepsilon_i \phi_{f,g,h}(X_{1i},Y_{1i}^{c\ast},\widehat{Y}_{1i},X_{2i},Y_{2i}^{c\ast},\widehat{Y}_{2i})}_{\Phi}.
\end{eqnarray*}
\end{lemma}
\begin{Proof}
By definition, we have that 
\BEqn
& & \widehat{U}^{b} \\
&=& \sup\limits_{\phi_{f,g,h}\in\Phi}
    \lrabs{
    \frac{1}{n} \sum\limits_{i=1}^{n} 
    \lrp{\phi_{f,g,h}(X_{1i},Y_{1i},\widehat{Y}_{1i},X_{2i},Y_{2i},\widehat{Y}_{2i})
         - \frac{1}{n} \sum\limits_{j=1}^{n} \phi_{f,g,h}(X_{1j},Y_{1j},\widehat{Y}_{1j},X_{2j},Y_{2j},\widehat{Y}_{2j})}
    \varepsilon_i } \\
&=& \sup\limits_{\phi_{f,g,h}\in\Phi}
    \lrabs{
    \frac{1}{n} \sum\limits_{i=1}^{n} 
    \phi_{f,g,h}(X_{1i},Y_{1i},\widehat{Y}_{1i},X_{2i},Y_{2i},\widehat{Y}_{2i}) \varepsilon_i
    - \bigp{ \frac{1}{n} \sum\limits_{j=1}^{n} \phi_{f,g,h}(X_{1j},Y_{1j},\widehat{Y}_{1j},X_{2j},Y_{2j},\widehat{Y}_{2j})} 
      \bigp{\frac{1}{n}\sum\limits_{i=1}^{n} \varepsilon_i }} \\
&=& \sup\limits_{\phi_{f,g,h}\in\Phi}
    \lrabs{
    \frac{1}{n} \sum\limits_{i=1}^{n} 
    \phi_{f,g,h}(X_{1i},Y_{1i},\widehat{Y}_{1i},X_{2i},Y_{2i},\widehat{Y}_{2i})
    \bigp{\varepsilon_i - \frac{1}{n}\sum\limits_{j=1}^{n} \varepsilon_j}} \\
&=& \sup\limits_{\phi_{f,g,h}\in\Phi}
    \lrabs{
    \frac{1}{n} \sum\limits_{i=1}^{n} 
    \phi_{f,g,h}(X_{1i},Y_{1i},\widehat{Y}_{1i},X_{2i},Y_{2i},\widehat{Y}_{2i}) 
    (\varepsilon_i - \bar\varepsilon)}.
\EEqn
Similarly, we have that 
\begin{equation*}
    U^{b\ast}
  = \sup\limits_{\phi_{f,g,h}\in\Phi}
    \lrabs{
    \frac{1}{n} \sum\limits_{i=1}^{n} 
    \phi_{f,g,h}(X_{1i},Y_{1i},Y_{1i}^{c\ast},X_{2i},Y_{2i},Y_{2i}^{c\ast}) 
    (\varepsilon_i - \bar\varepsilon)}.
\end{equation*}
It follows that
\BEqn
& & |\widehat{U}^b - U^{b\ast}| \\
&\le& \sup\limits_{\phi_{f,g,h}\in\Phi}
    \lrabs{
    \frac{1}{n} \sum\limits_{i=1}^{n} 
    \bigp{\phi_{f,g,h}(X_{1i},Y_{1i},\widehat{Y}_{1i},X_{2i},Y_{2i},\widehat{Y}_{2i}) 
          - \phi_{f,g,h}(X_{1i},Y_{1i},Y_{1i}^{c\ast},X_{2i},Y_{2i},Y_{2i}^{c\ast})}    
    \bigp{\varepsilon_i - \bar\varepsilon)}} \\
&\le& \sup\limits_{\phi_{f,g,h}\in\Phi}
    \lrabs{
    \frac{1}{n} \sum\limits_{i=1}^{n} 
    \bigp{\phi_{f,g,h}(X_{1i},Y_{1i},\widehat{Y}_{1i},X_{2i},Y_{2i},\widehat{Y}_{2i}) 
          - \phi_{f,g,h}(X_{1i},Y_{1i},Y_{1i}^{c\ast},X_{2i},Y_{2i},Y_{2i}^{c\ast})} 
    \varepsilon_i} \\
& & + \sup\limits_{\phi_{f,g,h}\in\Phi}
    \lrabs{
    \frac{1}{n} \sum\limits_{i=1}^{n} 
    \bigp{\phi_{f,g,h}(X_{1i},Y_{1i},\widehat{Y}_{1i},X_{2i},Y_{2i},\widehat{Y}_{2i}) 
          - \phi_{f,g,h}(X_{1i},Y_{1i},Y_{1i}^{c\ast},X_{2i},Y_{2i},Y_{2i}^{c\ast})}}
    \cdot |\bar\varepsilon|.
\EEqn
By applying the same three-term telescoping decomposition arguments in the proof of Lemma \ref{Lemma:decomp}, we have that
\BEqn
& & |\widehat{U}^b - U^{b\ast}| \\
&\le& \max\limits_{1\le\ell\le L}  
    \lrnorm{\frac{1}{n_0} \sum\limits_{i\in\ci_\ell} 
    \varepsilon_i \phi_{f,g,h}(X_{1i},Y_{1i}^{c\ast},\widehat{Y}_{1i},X_{2i},Y_{2i},Y_{2i}^{c\ast})}_{\Phi}
  + \max\limits_{1\le\ell\le L}  
    \lrnorm{\frac{1}{n_0} \sum\limits_{i\in\ci_\ell} 
    \varepsilon_i \phi_{f,g,h}(X_{1i},Y_{1i},Y_{1i}^{c\ast},X_{2i},Y_{2i}^{c\ast},\widehat{Y}_{2i})}_{\Phi} \\
& & + \max\limits_{1\le\ell\le L}  
    \lrnorm{\frac{1}{n_0} \sum\limits_{i\in\ci_\ell} 
    \varepsilon_i \phi_{f,g,h}(X_{1i},Y_{1i}^{c\ast},\widehat{Y}_{1i},X_{2i},Y_{2i}^{c\ast},\widehat{Y}_{2i})}_{\Phi} \\
& & + \max\limits_{1\le\ell\le L}  
    \lrnorm{\frac{1}{n_0} \sum\limits_{i\in\ci_\ell}
    \phi_{f,g,h}(X_{1i},Y_{1i}^{c\ast},\widehat{Y}_{1i},X_{2i},Y_{2i},Y_{2i}^{c\ast})}_{\Phi} 
    \cdot |\bar\varepsilon| \\
& & + \max\limits_{1\le\ell\le L}  
    \lrnorm{\frac{1}{n_0} \sum\limits_{i\in\ci_\ell} 
    \phi_{f,g,h}(X_{1i},Y_{1i},Y_{1i}^{c\ast},X_{2i},Y_{2i}^{c\ast},\widehat{Y}_{2i})}_{\Phi} 
    \cdot |\bar\varepsilon| \\
& & + \max\limits_{1\le\ell\le L}  
    \lrnorm{\frac{1}{n_0} \sum\limits_{i\in\ci_\ell} 
    \phi_{f,g,h}(X_{1i},Y_{1i}^{c\ast},\widehat{Y}_{1i},X_{2i},Y_{2i}^{c\ast},\widehat{Y}_{2i})}_{\Phi} 
    \cdot |\bar\varepsilon| \\
&=& \max\limits_{1\le\ell\le L} R_1^{\pell} 
  + \max\limits_{1\le\ell\le L} R_2^{\pell} 
  + \max\limits_{1\le\ell\le L} R_3^{\pell}
  + \max\limits_{1\le\ell\le L} U_1^{\pell} \cdot |\bar\varepsilon| 
  + \max\limits_{1\le\ell\le L} U_2^{\pell} \cdot |\bar\varepsilon| 
  + \max\limits_{1\le\ell\le L} U_3^{\pell} \cdot |\bar\varepsilon|,
\EEqn
which arrives at the desired result.
\end{Proof}

\begin{lemma}[Multiplier inequalities]\label{Lemma:Rad_bounds}
Let $\{X_i\}_{i=1}^{n}$ be a random sample from a common law $P$ and $\cf$ be a class of measurable functions satisfying that $\be^{\ast}[\|f(X_1)\|_{\cf}] < \infty$. Let $\{\xi_i\}_{i=1}^{n}$ be iid Rademacher variables independent of $\{X_i\}_{i=1}^{n}$. Then for any iid sample $\{\varepsilon_i\}_{i=1}^{n}$ of real, mean-zero random variables independent of $\{X_i\}_{i=1}^{n}$, and any $1\le k\le n$, it holds that 
\BEqn
& & \frac{1}{2}\|\varepsilon_1\|_1 
    \be^{\ast}\lrbk{\lrnorm{\frac{1}{\sqrt{n}}\sum\limits_{i=1}^{n}\xi_i f(X_i)}_{\cf}} \\
&\le& \be^{\ast}\lrbk{\lrnorm{\frac{1}{\sqrt{n}}\sum\limits_{i=1}^{n} \varepsilon_i f(X_i)}_{\cf}} \\
&\le& (2k-1) \be^{\ast}[\|f(X_1)\|_{\cf}] \blre{\max\limits_{1\le i\le n}\frac{|\varepsilon_i|}{\sqrt{n}}} 
+ 2\sqrt{2}\|\varepsilon_1\|_{2,1} \max\limits_{k\le t\le n} \be^{\ast} \lrbk{\lrnorm{\frac{1}{\sqrt{t}} \sum\limits_{i=k}^{t} \xi_i f(X_i)}_{\cf}}.
\EEqn
\end{lemma}
\begin{Proof}
See Lemma 2.9.1 of \cite{van1996weak} and Lemma 10.2 of \cite{kosorok2008introduction}. Here, $\|f(X_1)\|_{\cf} = \sup\limits_{f\in\cf} |f(X_1)|$ and $\|\varepsilon_1\|_{2,1} = \int_{0}^{\infty} \sqrt{\bp(|\varepsilon_1| \geq x)} dx$.
\end{Proof}

\begin{lemma}\label{Lemma:Gaussian_aux}
Let $\{\varepsilon_i\}_{i=1}^{n} \stsim{iid} \cn(0,1)$ be iid standard normal random variables, then it holds that $\|\varepsilon_1\|_{2,1} \le \sqrt{\pi} < \infty$ and $\blre{\max\limits_{1\le i\le n}|\varepsilon_i|} \le \sqrt{2\log(2n)}$.
\end{lemma}
\begin{Proof}
Note that for $\varepsilon_1\sim\cn(0,1)$, the density function is given by $\phi(s) = (2\pi)^{-1/2}\exp(-s^2/2)$ for any $s\in\br$, then it holds for any $t>0$ that
\BEqn
    \bp(\varepsilon_1 \geq t)
&=& \bp(\varepsilon_1 - t \geq 0)
 =  \int_{0}^{\infty} (2\pi)^{-1/2}\exp(-\frac{(s+t)^2}{2}) ds \\
&\le& \exp(-\frac{t^2}{2}) \int_{0}^{\infty} (2\pi)^{-1/2}\exp(-\frac{s^2}{2}) ds \\
&=& \frac{1}{2} \exp(-\frac{t^2}{2}),
\EEqn
which implies that $\bp(|\varepsilon_1|\geq t) \le \exp(-\frac{t^2}{2})$ by the symmetry.

It follows that 
\begin{equation*}
    \|\varepsilon_1\|_{2,1}
  = \int_{0}^{\infty} \sqrt{\blrp{|\varepsilon_1|\geq t}}dt 
  \le \int_{0}^{\infty} \exp(-\frac{t^2}{4}) dt
  = 2\int_{0}^{\infty} \exp(-s^2) ds
  = \int_{-\infty}^{\infty} \exp(-s^2) ds 
  = \sqrt{\pi}
  < \infty.
\end{equation*}
Also, it follows from the MGF of the standard normal random variable that $\be[\exp(t\varepsilon_1)]=\exp(t^2/2)$ for any $t>0$. Then for any $t>0$, it follows from Jensen's inequality that 
\BEqn
    \bbe{\max\limits_{1\le i\le n}\varepsilon_i} 
&=& \blre{\frac{1}{t} \log\exp\lrp{t \max\limits_{1\le i\le n}\varepsilon_i}}
\le \frac{1}{t} \log\blre{\exp\lrp{t\max\limits_{1\le i\le n}\varepsilon_i}} \\
&\le& \frac{1}{t} \log\blre{\sum\limits_{i=1}^{n} \exp\lrp{t\varepsilon_i}} 
 =  \frac{1}{t} \log\lrp{n\exp(t^2/2)} \\
&=& \frac{1}{t}\log(n) + \frac{t}{2}.
\EEqn
Note that this inequality holds for arbitrary $t>0$, by taking $t=\sqrt{2\log(n)}$, we have that $\bbe{\max\limits_{1\le i\le n}\varepsilon_i} \le \sqrt{2\log(n)}$. Since $|\varepsilon_i| = \max\{\varepsilon_i, -\varepsilon_i\}$, we have that
\begin{equation*}
    \max\limits_{1\le i\le n} |\varepsilon_i| 
  = \max\limits_{1\le i\le n} \{\varepsilon_i, -\varepsilon_i\}.
\end{equation*}
Since $-\varepsilon_i \sim \cn(0,1)$, applying the same log-sum-exp argument to the $2n$ variables $\{\varepsilon_1,\cdots,\varepsilon_n,-\varepsilon_1,\cdots,-\varepsilon_n\}$ yields $\bbe{\max\limits_{1\le i\le n}|\varepsilon_i|} \le \sqrt{2\log(2n)}$, which complete the proof.
\end{Proof}

\begin{lemma}\label{Lemma:Rate_R123}
Under Assumption \ref{Assumpt:kernel} and Assumption \ref{Assumpt:double_robustness}, it holds that
\BEqn
& & \max\limits_{1\le\ell\le L} R_1^{\pell}
  = O_p(n^{-(k_1+1)/2} \log^v(n)), \\
& & \max\limits_{1\le\ell\le L} R_2^{\pell}
  = O_p(n^{-(k_2+1)/2} \log^v(n)), \\
& & \max\limits_{1\le\ell\le L} R_3^{\pell}
  = O_p(n^{-(k_1+k_2+1)/2} \log^v(n)).
\EEqn
\end{lemma}
\begin{Proof}
Recall that for each $\ell=1,\cdots,L$,
\begin{equation*}
     R_1^{\pell}
 =  \lrnorm{\frac{1}{n_0} \sum\limits_{i\in\ci_\ell} 
    \varepsilon_i \phi_{f,g,h}(X_{1i},Y_{1i}^{c\ast},\widehat{Y}_{1i},X_{2i},Y_{2i},Y_{2i}^{c\ast})}_{\Phi}.
\end{equation*}
Conditioning on $\cd^{\pmell}$, with independent generator noises, the sample $\{(X_{1i},Y_{1i}^{c\ast},\widehat{Y}_{1i},X_{2i},Y_{2i},Y_{2i}^{c\ast})\}_{i\in\ci_\ell}$ is iid. Under Assumption \ref{Assumpt:kernel}\ref{Assumpt:kernel_bound}, it follows from similar arguments used for Lemma \ref{Lemma:Pf_mu12} that 
\BEqn
& & \blre{\|\phi_{f,g,h}(X_{1i}, Y_{1i}^{c\ast}, \widehat{Y}_{1i}, X_{2i}, Y_{2i}, Y_{2i}^{c\ast})\|_{\Phi} \bone\{i\in\ci_\ell\}
    \mid \cd^{\pmell}} \bone\{\cd\in\ca_n\} \\
&\le& 2u_{\ch,\ch'}^2 
\max\limits_{1\le\ell\le L}    
    \lrp{\bbe{ \sup\limits_{f\in\ch} \bigp{f(X_{1i},Y_{1i}^{c\ast}) - f(X_{1i},\widehat{Y}_{1i})}^2 \bone\{i\in\ci_\ell\}
    \mid \cd^{\pmell}}}^{1/2} 
    \bone\{\cd\in\ca_n\} \\
&\le& 2u_{\ch,\ch'}^2 M_0 n^{-k_1}\log^{1/2}(n).
\EEqn
For $1\le k\le n_0$, we use $\ci_\ell(k)$ to denote a random draw of $k$ indices from $\ci_\ell$. By Lemma \ref{Lemma:Rad_VC} and Lemma \ref{Lemma:Pf_sigma12}, we have that
\BEqn
& & \blre{\lrnorm{\sum\limits_{i\in\ci_\ell(k)}
    \xi_i \phi_{f,g,h}(X_{1i},Y_{1i}^{c\ast},\widehat{Y}_{1i},X_{2i},Y_{2i},Y_{2i}^{c\ast})}_{\Phi}
    \mid \cd^{\pmell}} 
    \bone\{\cd\in\ca_n\} \\
&\le& C \lrp{4u_{\ch,\ch'}^3 \log^v\lrp{\frac{12au_{\ch,\ch'}^2}{\sigma_1}} + \sqrt{k\sigma_1^2\log^v\lrp{\frac{12au_{\ch,\ch'}^2}{\sigma_1}}}} \\
&\le& C\lrp{\log^v(n) + k^{1/2} n^{-k_1} \log^{(v+1)/2}(n)},
\EEqn
where $\{\xi_i\}_{i=1}^{n_0}$ are iid Rademacher variables independent of $\{(X_{1i},Y_{1i}^{c\ast},\widehat{Y}_{1i},X_{2i},Y_{2i},Y_{2i}^{c\ast})\}_{i\in\ci_\ell}$, and $C$ is a positive constant that depends only on $p,q$ and may vary from line to line.

Recall that $n_0=n/L$. By Lemma \ref{Lemma:Rad_bounds} and Lemma \ref{Lemma:Gaussian_aux}, it holds for $1\le k\le n_0$ that
\BEqn
& & \blre{R_1^{\pell} \mid \cd^{\pmell}} \bone\{\cd\in\ca_n\} \\
&\le& (2k-1) \lrp{2u_{\ch,\ch'}^2 M_0 n^{-k_1}\log^{1/2}(n)} \lrp{n_0^{-1} \sqrt{2\log(2n_0)}} \\
& & + 2\sqrt{2\pi} \lrp{C n_0^{-1/2} \max\limits_{k\le t\le n_0} t^{-1/2} \lrp{\log^v(n) + (t-k+1)^{1/2} n^{-k_1} \log^{(v+1)/2}(n)}} \\
&\le& C \lrp{k n^{-(k_1+1)}\log(n) + k^{-1/2} n^{-1/2} \log^v(n) + n^{-(k_1+\frac{1}{2})}\log^{(v+1)/2}(n)},
\EEqn
where $C$ is a positive constant that depends only on $p,q$, and may vary from line to line again. By taking $k = \lrfl{n_0^{k_1}}$, we have that 
\begin{equation*}
    \blre{R_1^{\pell} \mid \cd^{\pmell}} \bone\{\cd\in\ca_n\} 
\le C\lrp{n^{-1} \log(n) + n^{-(k_1+1)/2} \log^v(n) + n^{-(k_1+\frac{1}{2})} \log^{(v+1)/2}(n)}.
\end{equation*}

Under Assumption \ref{Assumpt:double_robustness}, with $0<k_1,k_2<\frac{1}{2}$, we have that
\begin{equation*}
    \frac{n^{-(k_1+1)/2} \log^v(n)}{n^{-1} \log(n)}
  = n^{(1-k_1)/2}\log^{v-1}(n)
  \rightarrow \infty,
\end{equation*}
and
\begin{equation*}
    \frac{n^{-(k_1+1)/2} \log^v(n)}{n^{-(k_1+\frac{1}{2})} \log^{(v+1)/2}(n)}
  = n^{k_1/2}\log^{(v-1)/2}(n)
  \rightarrow \infty,
\end{equation*}
implying that $n^{-(k_1+1)/2} \log^v(n)$ dominates $n^{-1} \log(n)$ and $n^{-(k_1+\frac{1}{2})} \log^{(v+1)/2}(n)$ asymptotically. Then we repeat the same arguments in Lemma \ref{Lemma:Rate_U12}, and it follows from the tower property of conditional expectation and the fact that $\bp(\ca_n) \rightarrow 1$ that
\begin{equation*}
    \max\limits_{1\le\ell\le L} R_1^{\pell}
  = O_p(n^{-(k_1+1)/2} \log^v(n)).
\end{equation*}

Using similar arguments, we can show the results for $R_2^{\pell}, R_3^{\pell}$ and thus complete the proof.
\end{Proof}

\begin{lemma}\label{Lemma:Rate_R456}
Under Assumption~\ref{Assumpt:kernel}-\ref{Assumpt:X1X2} and Assumption~\ref{Assumpt:entropy}--\ref{Assumpt:double_robustness}, it holds that 
\BEqn
& & \max\limits_{1\le\ell\le L} U_1^{\pell} \cdot |\bar\varepsilon|
  = O_p\lrp{n^{-(k_1+\frac{1}{2})} \log^{1/2}(n)}, \\
& & \max\limits_{1\le\ell\le L} U_2^{\pell} \cdot |\bar\varepsilon|
  = O_p\lrp{n^{-(k_2+\frac{1}{2})} \log^{1/2}(n)}, \\
& & \max\limits_{1\le\ell\le L} U_3^{\pell} \cdot |\bar\varepsilon|
  = O_p\lrp{n^{-(k_1+k_2+\frac{1}{2})} \log(n)}.
\EEqn
\end{lemma}
\begin{Proof}
We have shown in Lemma \ref{Lemma:Expectation_U} that 
\begin{equation*}
    \max\limits_{1\le\ell\le L} \be[U_1^{\pell} \mid \cd^{\pmell}] \cdot \bone\{\cd\in\ca_n\}
\le C \lrp{4u_{\ch,\ch'}^3 n^{-1} \log^v\lrp{\frac{12au_{\ch,\ch'}^2}{\sigma_1}} + n^{-1/2}\sigma_1 \log^{v/2}\lrp{\frac{12au_{\ch,\ch'}^2}{\sigma_1}}} + \mu_1,
\end{equation*}
where $\mu_1 = 2u_{\ch,\ch'}^2 M_0 n^{-k_1} \log^{1/2}(n)$ and $\sigma_1 = 2u_{\ch,\ch'}^2 M_0 n^{-k_1} \log^{1/2}(n)$.

Since the random sample $\{\varepsilon_i\}_{i=1}^{n} \stsim{iid} \cn(0,1)$ is independent of $U_1^{\pell}$ and $\cd$, we have that $\bar\varepsilon = n^{-1} \sum_{i=1}^{n} \varepsilon_i \sim \cn(0,\frac{1}{n})$ and then $\be[|\bar\varepsilon|] = \sqrt{\frac{2}{n\pi}}$. It follows that
\BEqn
& & \max\limits_{1\le\ell\le L} \be[U_1^{\pell} \cdot |\bar\varepsilon| 
    \mid \cd^{\pmell}] \cdot \bone\{\cd\in\ca_n\} \\
&=& \max\limits_{1\le\ell\le L} \be[U_1^{\pell} \mid \cd^{\pmell}] 
    \cdot \be[|\bar\varepsilon|] 
    \cdot \bone\{\cd\in\ca_n\} \\
&=& \sqrt{\frac{2}{n\pi}} \max\limits_{1\le\ell\le L} \be[U_1^{\pell} \mid \cd^{\pmell}] \cdot \bone\{\cd\in\ca_n\}.
\EEqn
Then by using the similar arguments for Lemma \ref{Lemma:Rate_U12}, we obtain that 
\begin{equation*}
    \max\limits_{1\le\ell\le L} U_1^{\pell} \cdot |\bar\varepsilon|
  = O_p\lrp{n^{-(k_1+\frac{1}{2})} \log^{1/2}(n)}.
\end{equation*}
Similarly, we can verify the counterparts for $U_2^{\pell}, U_3^{\pell}$, for which we spare the details.
\end{Proof}

\section{Auxiliary Results for Bootstrap Validity}\label{Appendix:AuxBoot}

\begin{lemma}\label{Lemma:Donsker_Uast}
Under Assumption~\ref{Assumpt:kernel} and Assumption~\ref{Assumpt:entropy}, it holds that
\begin{equation*}
    \sqrt{n} \lrnorm{(\bp_n-P)(\phi_{f,g,h})}_{\Phi}
\stra{d} \lrnorm{\bg(\phi_{f,g,h})}_{\Phi},
\end{equation*}
where 
\BEqn
    \bp_n(\phi_{f,g,h}) 
&=& \frac{1}{n} \sum\limits_{i=1}^{n} \phi_{f,g,h}(X_{1i},Y_{1i},Y_{1i}^{c\ast},X_{2i},Y_{2i},Y_{2i}^{c\ast}), \\
    P(\phi_{f,g,h}) 
&=& \bbe{\phi_{f,g,h}(X_{1i},Y_{1i},Y_{1i}^{c\ast},X_{2i},Y_{2i},Y_{2i}^{c\ast})},
\EEqn
and $\bg$ is a tight Gaussian process with mean zero and the covariance function
\BEqn
& & \Cov\lrp{\bg(\phi_{f_1,g_1,h_1}), \bg(\phi_{f_2,g_2,h_2})} \\
&=& \blre{\phi_{f_1,g_1,h_1}(X_{1i},Y_{1i},Y_{1i}^{c\ast},X_{2i},Y_{2i},Y_{2i}^{c\ast})
          \phi_{f_2,g_2,h_2}(X_{1i},Y_{1i},Y_{1i}^{c\ast},X_{2i},Y_{2i},Y_{2i}^{c\ast})} \\
& & - \blre{\phi_{f_1,g_1,h_1}(X_{1i},Y_{1i},Y_{1i}^{c\ast},X_{2i},Y_{2i},Y_{2i}^{c\ast})}
      \blre{\phi_{f_2,g_2,h_2}(X_{1i},Y_{1i},Y_{1i}^{c\ast},X_{2i},Y_{2i},Y_{2i}^{c\ast})}
\EEqn
for any $\phi_{f_1,g_1,h_1}, \phi_{f_2,g_2,h_2} \in \Phi$.
\end{lemma}
\begin{Proof}
Under Assumption~\ref{Assumpt:kernel} and Assumption~\ref{Assumpt:entropy}, we have shown in Lemma \ref{Lemma:Phi_Donsker} that $\Phi$ is $P$-Donsker, i.e.
\begin{equation*}
    \bg_n = \sqrt{n} (\bp_n-P) \leadsto \bg \quad\mbox{in }\ell^{\infty}(\Phi),
\end{equation*}
where the random measure $\bg_n$ is given by
\BEqn
& & \bg_n(\phi_{f,g,h}) \\
&=& \sqrt{n} (\bp_n-P)(\phi_{f,g,h}) \\
&=& \frac{1}{\sqrt{n}} \sum\limits_{i=1}^{n} 
    \lrp{\phi_{f,g,h}(X_{1i},Y_{1i},Y_{1i}^{c\ast},X_{2i},Y_{2i},Y_{2i}^{c\ast})
        - \bbe{\phi_{f,g,h}(X_1,Y_1,Y_1^{c\ast},X_2,Y_2,Y_2^{c\ast})}},
\EEqn
and $\bg$ is a tight Gaussian process with zero mean and covariance function
\BEqn
& & \Cov\lrp{\bg(\phi_{f_1,g_1,h_1}), \bg(\phi_{f_2,g_2,h_2})} \\
&=& \Cov\lrp{\phi_{f_1,g_1,h_1}(X_{1i},Y_{1i},Y_{1i}^{c\ast},X_{2i},Y_{2i},Y_{2i}^{c\ast}),
             \phi_{f_2,g_2,h_2}(X_{1i},Y_{1i},Y_{1i}^{c\ast},X_{2i},Y_{2i},Y_{2i}^{c\ast})} \\
&=& \blre{\phi_{f_1,g_1,h_1}(X_{1i},Y_{1i},Y_{1i}^{c\ast},X_{2i},Y_{2i},Y_{2i}^{c\ast})
          \phi_{f_2,g_2,h_2}(X_{1i},Y_{1i},Y_{1i}^{c\ast},X_{2i},Y_{2i},Y_{2i}^{c\ast})} \\
& & - \blre{\phi_{f_1,g_1,h_1}(X_{1i},Y_{1i},Y_{1i}^{c\ast},X_{2i},Y_{2i},Y_{2i}^{c\ast})}
      \blre{\phi_{f_2,g_2,h_2}(X_{1i},Y_{1i},Y_{1i}^{c\ast},X_{2i},Y_{2i},Y_{2i}^{c\ast})}.
\EEqn
Note that the supremum map $\|\cdot\|_{\Phi}$ is uniformly continuous in $\ell^{\infty}(\Phi)$, i.e. for any $x,y\in\ell^{\infty}(\Phi)$, we have that
\begin{equation*}
    \lrabs{\sup\limits_{\phi_{f,g,h}\in\Phi} |x(\phi_{f,g,h})|
  - \sup\limits_{\phi_{f,g,h}\in\Phi} |y(\phi_{f,g,h})|}
\le \sup\limits_{\phi_{f,g,h}\in\Phi} |x(\phi_{f,g,h}) - y(\phi_{f,g,h})|
  = \|x-y\|_{\Phi},
\end{equation*}
By the continuous mapping theorem, we have that
\begin{equation*}
    \sqrt{n} \lrnorm{(\bp_n-P)(\phi_{f,g,h})}_{\Phi}
  = \|\bg_n(\phi_{f,g,h})\|_{\Phi} 
\stra{d} \|\bg(\phi_{f,g,h})\|_{\Phi},
\end{equation*}
which arrives at the desired results.
\end{Proof}

\begin{lemma}\label{Lemma:Donsker_Ubast}
Under Assumption~\ref{Assumpt:kernel} and Assumption~\ref{Assumpt:entropy}, it holds that
\begin{equation*}
    \sqrt{n} U^{b\ast} \leadsto_{\varepsilon}^{p} \|\bg(\phi_{f,g,h})\|_{\Phi},
\end{equation*}
where $\bg$ is the tight Gaussian process given in Lemma \ref{Lemma:Donsker_Uast}.
\end{lemma}
\begin{Proof}
Under Assumption~\ref{Assumpt:kernel} and Assumption~\ref{Assumpt:entropy}, we have shown in Lemma~\ref{Lemma:Phi_Donsker} that $\Phi$ is $P$-Donsker, where $P$ is the law of $(X_1,Y_1,Y_1^{c\ast},X_2,Y_2,Y_2^{c\ast})$. Then it follows from Lemma \ref{Lemma:Donsker_boot} that $\bg_n'' \leadsto_{\varepsilon}^{p} \bg$, where the empirical process $\bg_n''$ is given by
\begin{equation*}
    \bg_n''(\phi_{f,g,h})
 =  \frac{1}{\sqrt{n}} \sum\limits_{i=1}^{n} 
    \phi_{f,g,h}(X_{1i},Y_{1i},Y_{1i}^{c\ast},X_{2i},Y_{2i},Y_{2i}^{c\ast}) \lrp{\varepsilon_i-\bar\varepsilon}.
\end{equation*}
By using the continuous mapping theorem, we obtain that
\begin{equation*}
    \|\bg_n''(\phi_{f,g,h})\|_{\Phi} \leadsto_{\varepsilon}^{p} \|\bg(\phi_{f,g,h})\|_{\Phi}.
\end{equation*}
By noting that 
\begin{equation*}
    \|\bg_n''(\phi_{f,g,h})\|_{\Phi}
  = \sup\limits_{\phi_{f,g,h}\in\Phi} \lrabs{\bg_n''(\phi_{f,g,h})}
  = \sqrt{n} U^{b\ast},
\end{equation*}
we arrive at the desired result.
\end{Proof}

To facilitate the subsequent analysis, we introduce some useful results below. 
\begin{defi}\label{Def:Pre-Gaussian}
A function class $\cf$ is called $P$-pre-Gaussian if there exists a tight centered Gaussian process $G_P$ in $\ell^{\infty}(\cf)$ with mean zero and covariance function 
\begin{equation*}
    \be[G_P(f) G_P(g)]
  = P(fg) - P(f) P(g)
  = \be[f(X_1)g(X_1)] - \be[f(X_1)]~\be[g(X_1)], \qquad \forall~f,g\in\cf.
\end{equation*}
\end{defi}

\begin{lemma}\label{Lemma:S1_aux}
Let $\bg=\{\bg(\phi_{f,g,h}):\phi_{f,g,h}\in\Phi\}$ denote the tight Gaussian process defined in Lemma \ref{Lemma:Donsker_Uast}. Under Assumption~\ref{Assumpt:kernel} and Assumption~\ref{Assumpt:entropy}, it holds that 
\begin{equation*}
    \be[\lrnorm{\bg(\phi_{f,g,h})}_{\Phi} ] < \infty.
\end{equation*}
\end{lemma}
\begin{Proof}
For any $\phi_{f_1,g_1,h_1}, \phi_{f_2,g_2,h_2}\in\Phi$, it follows from the definition of Gaussian process that $\bg(\phi_{f_1,g_1,h_1}) - \bg(\phi_{f_2,g_2,h_2})$ is a Gaussian random variable with mean zero and the covariance 
\BEqn
& & \Var(\bg(\phi_{f_1,g_1,h_1}) - \bg(\phi_{f_2,g_2,h_2})) \\
&=& \Var(\bg(\phi_{f_1,g_1,h_1})) 
  - 2\Cov(\bg(\phi_{f_1,g_1,h_1}), \bg(\phi_{f_2,g_2,h_2})) 
  + \Var(\bg(\phi_{f_2,g_2,h_2})) \\
&=& \Var(\phi_{f_1,g_1,h_1})
  - 2\Cov(\phi_{f_1,g_1,h_1}, \phi_{f_2,g_2,h_2})
  + \Var(\phi_{f_2,g_2,h_2}) \\
&=& \Var(\phi_{f_1,g_1,h_1} - \phi_{f_2,g_2,h_2}) \\
&\le& \blre{(\phi_{f_1,g_1,h_1} - \phi_{f_2,g_2,h_2})^2} \\
&=& \|\phi_{f_1,g_1,h_1} - \phi_{f_2,g_2,h_2}\|_{L_2(P)}^2.
\EEqn
It follows from the Gaussian tail inequality that
\BEqn
    \blrp{|\bg(\phi_{f_1,g_1,h_1}) - \bg(\phi_{f_2,g_2,h_2})| \geq u} 
&\le& 2\exp\lrp{-\frac{u^2}{2 \Var(\bg(\phi_{f_1,g_1,h_1}) - \bg(\phi_{f_2,g_2,h_2}))}} \\
&\le& 2\exp\lrp{-\frac{u^2}{2\|\phi_{f_1,g_1,h_1} - \phi_{f_2,g_2,h_2}\|_{L_2(P)}^2}}.
\EEqn
This implies that $\{\bg(\phi_{f,g,h}):\phi_{f,g,h}\in\Phi\}$ is sub-Gaussian with respect to the metric space $(\Phi, \|\cdot\|_{L_2(P)})$. By Lemma \ref{Lemma:Dudley}, for any fixed $\phi_{f_0,g_0,h_0}\in\Phi$, we have that
\begin{equation*}
    \blre{\sup\limits_{\phi_{f,g,h}\in\Phi} |\bg(\phi_{f,g,h}) - \bg(\phi_{f_0,g_0,h_0})|}
\le C\int_{0}^{\infty} \sqrt{\log N(\Phi, \|\cdot\|_{L_2(P)}, \varepsilon)} d\varepsilon
< \infty,
\end{equation*}
where the last inequality follows from the uniform entropy integral established in Lemma \ref{Lemma:Unif_Entropy}.

Also note that $\bg(\phi_{f_0,g_0,h_0})$ is a normal random variable, thus $|\bg(\phi_{f_0,g_0,h_0})|$ is a half-normal random variable with a finite expectation. Therefore, we can conclude that
\begin{equation*}
    \be[\lrnorm{\bg(\phi_{f,g,h})}_{\Phi} ] 
\le \blre{\sup\limits_{\phi_{f,g,h}\in\Phi} |\bg(\phi_{f,g,h}) - \bg(\phi_{f_0,g_0,h_0})|} + \blre{|\bg(\phi_{f_0,g_0,h_0})|}
< \infty.
\end{equation*}
\end{Proof}

\begin{lemma}\label{Lemma:ContinuousCDF}
Under Assumption~\ref{Assumpt:kernel} and Assumption~\ref{Assumpt:entropy},\ref{Assumpt:continuity}, it holds that $\|\bg\|_{\Phi}$ has a continuous distribution on $\br$.
\end{lemma}
\begin{Proof}
Under Assumption~\ref{Assumpt:kernel} and Assumption~\ref{Assumpt:entropy}, it follows from Lemma \ref{Lemma:Donsker_Uast} that $\mathbb G$ is a tight centered Gaussian process in $\ell^{\infty}(\Phi)$. In Lemma \ref{Lemma:Unif_Entropy}, we have shown that the metric space $(\Phi,\|\cdot\|_{L_2(P)})$ has a finite entropy integral, which implies that $(\Phi,\|\cdot\|_{L_2(P)})$ is totally bounded and hence separable. In addition, Lemma \ref{Lemma:Unif_Entropy} and Lemma \ref{Lemma:DudleyContinuity} jointly imply that $\bg$ admits a version with almost surely uniformly continuous sample paths on $(\Phi, \|\cdot\|_{L_2(P)})$.

By separability, there exists a countable dense subset $\Phi_0 = \{\phi_{f_i,g_i,h_i}\in\Phi\}_{i\geq1}$, such that for any $\phi_{f_0,g_0,h_0} \in \Phi$, there exists a subset $\{\phi_{f_{i_m},g_{i_m},h_{i_m}} \in \Phi_0\}_{m\geq1}$, such that
\begin{equation*}
    \|\phi_{f_{i_m},g_{i_m},h_{i_m}} - \phi_{f_0,g_0,h_0}\|_{L_2(P)} \rightarrow 0.
\end{equation*}
Then it follows from the uniform continuity that almost surely,
\begin{equation*}
    \bg(\phi_{f_{i_m},g_{i_m},h_{i_m}}) \rightarrow \bg(\phi_{f_0,g_0,h_0}),
\end{equation*}
and
\begin{equation*}
    \lrabs{\bg(\phi_{f_{i_m},g_{i_m},h_{i_m}})} \rightarrow \lrabs{\bg(\phi_{f_0,g_0,h_0})}.
\end{equation*}
By the arbitrariness of $\phi_{f_0,g_0,h_0}\in\Phi$, we obtain that
\begin{equation*}
    \|\bg\|_{\Phi} \le \|\bg\|_{\Phi_0}.
\end{equation*}
Note that $\Phi_0 \subseteq \Phi$, it is trivial that $\|\bg\|_{\Phi_0} \le \|\bg\|_{\Phi}$, which further implies that $\|\bg\|_{\Phi} = \|\bg\|_{\Phi_0}$.

For each $m\geq1$, let $\Phi_m = \{\phi_{f_i,g_i,h_i}\in\Phi\}_{1\le i\le m}$, then $\Phi_m$ is a finite subset of $\Phi_0$ and $\|\bg\|_{\Phi_m} \uparrow \|\bg\|_{\Phi_0} = \|\bg\|_{\Phi}$. Consequently, for every $\eta > 0$ and $\delta > 0$, there exists $m\geq1$, such that
\begin{equation*}
    \blrp{\lrabs{ \|\bg\|_{\Phi} - \|\bg\|_{\Phi_m}} > \delta} < \eta.
\end{equation*}
On the other hand, for each finite set $\Phi_m$, $\|\bg\|_{\Phi_m} = \max_{1\le i\le m} |\bg(\phi_{f_i,g_i,h_i})|$. For any $x>0$,
\begin{equation*}
    \{\|\bg\|_{\Phi_m} = x\}
\subseteq \bigcup\limits_{i=1}^{m} 
    \lrp{\{\bg(\phi_{f_i,g_i,h_i}) = x\} 
         \cup
         \{\bg(\phi_{f_i,g_i,h_i}) = -x\}}.
\end{equation*}
For each $1\le i\le m$, $\bg(\phi_{f_i,g_i,h_i})$ is a centered Gaussian random variable. Since a centered degenerate Gaussian can only be degenerate at zero, we have
\begin{equation*}
    \bp(\bg(\phi_{f_i,g_i,h_i}) = x) 
  = \bp(\bg(\phi_{f_i,g_i,h_i}) = -x)
  = 0,
  \qquad \forall x>0.
\end{equation*}
Therefore, $\bp(\|\bg\|_{\Phi_m} = x) = 0$ for all $x>0$, that is, $\|\bg\|_{\Phi_m}$ has no atoms on $(0,\infty)$.

Next, we investigate the atoms of $\|\bg\|_{\Phi}$. For any fixed $x > 0$ and any $\delta > 0$,
\begin{equation*}
    \blrp{\|\bg\|_{\Phi} = x}
\le \blrp{\lrabs{ \|\bg\|_{\Phi} - \|\bg\|_{\Phi_m}} > \delta} 
  + \blrp{\lrabs{ \|\bg\|_{\Phi_m} - x} \le \delta}.
\end{equation*}
For arbitrary $\eta > 0$, taking $m$ sufficiently large yields
\begin{equation*}
    \blrp{\lrabs{ \|\bg\|_{\Phi} - \|\bg\|_{\Phi_m}} > \delta} < \eta.
\end{equation*}
For such $m$, since $\|\bg\|_{\Phi_m}$ has no atom at $x > 0$, we have that
\begin{equation*}
    \blrp{\lrabs{\|\bg\|_{\Phi_m} - x} \le \delta} \rightarrow 0
    \qquad \mbox{ as }\delta\downarrow0.
\end{equation*}
By the arbitrariness of $\eta, \delta > 0$, we conclude that $\blrp{\|\bg\|_{\Phi} = x} = 0$ for any $x > 0$.

It remains to verify that zero is not an atom of $\|\bg\|_{\Phi}$. Under Assumption \ref{Assumpt:continuity}, there exists $\phi_{f_0,g_0,h_0} \in \Phi$ such that
\begin{equation*}
    \Var(\phi_{f_0,g_0,h_0}(X_1,Y_1,Y_1^{c\ast},X_2,Y_2,Y_2^{c\ast})) 
 =: \underline\sigma^2 > 0.
\end{equation*}
By Lemma \ref{Lemma:Donsker_Uast}, $\bg(\phi_{f_0,g_0,h_0}) \sim \cn(0, \underline\sigma^2)$, which is non-degenerate. It follows that
\begin{equation*}
    \blrp{\|\bg\|_{\Phi} = 0}
\le \blrp{\lrabs{\bg(\phi_{f_0,g_0,h_0})} = 0}
  = 0,
\end{equation*}
which implies that the distribution of $\|\bg\|_{\Phi}$ has no atoms on $\br$, and hence its cumulative distribution function is continuous.
\end{Proof}

\begin{lemma}\label{Lemma:S1}
Under Assumption~\ref{Assumpt:kernel} and Assumption~\ref{Assumpt:entropy},\ref{Assumpt:continuity}, it holds that
\begin{equation*}
    S_1
 := \sup\limits_{x\in\br} 
    \lrabs{  \blrp{\sqrt{n}\lrnorm{(\bp_n-P)(\phi_{f,g,h})}_{\Phi} \le x} 
           - \blrp{\lrnorm{\bg(\phi_{f,g,h})}_{\Phi} \le x}} 
  = o(1),
\end{equation*}
where $\bg$ is the tight Gaussian process defined in Lemma \ref{Lemma:Donsker_Uast}.
\end{lemma}
\begin{Proof}
Under Assumption \ref{Assumpt:kernel} and Assumption~\ref{Assumpt:entropy}, we have shown in Lemma \ref{Lemma:Donsker_Uast} that
\begin{equation*}
    \sqrt{n}\lrnorm{(\bp_n-P)(\phi_{f,g,h})}_{\Phi} \stra{d} \lrnorm{\bg(\phi_{f,g,h})}_{\Phi},
\end{equation*}
which implies that for any $x\in\br$, it holds under the null that
\begin{equation*}
    \lrabs{
    \blrp{\sqrt{n}\lrnorm{(\bp_n-P)(\phi_{f,g,h})}_{\Phi} \le x}
  - \blrp{\lrnorm{\bg(\phi_{f,g,h})}_{\Phi} \le x}} 
  = o(1).
\end{equation*}
Additionally, with Assumption~\ref{Assumpt:continuity}, we have shown in Lemma \ref{Lemma:ContinuousCDF} that $\lrnorm{\bg(\phi_{f,g,h})}_{\Phi}$ has a continuous distribution function over $\br$, then it follows from P\'{o}lya's uniform convergence theorem that the result above is uniform in $x$, i.e. $S_1 = o(1)$.
\end{Proof}

\begin{lemma}\label{Lemma:S2}
Under Assumption \ref{Assumpt:kernel}--\ref{Assumpt:X1X2} and Assumption \ref{Assumpt:entropy}--\ref{Assumpt:continuity}, it holds under $H_0$ that
\begin{equation*}
    S_2
 := \sup\limits_{x\in\br} 
    \lrabs{  \blrp{\sqrt{n}\widehat{U} \le x \mid H_0} 
           - \blrp{\sqrt{n} U^{\ast} \le x \mid H_0}} = o(1).
\end{equation*}
\end{lemma}
\begin{Proof}
Note that for any $x\in\br$, we have that
\begin{eqnarray}
& & \blrp{\sqrt{n}\widehat{U} \le x \mid H_0} \nonumber \\
&=& \blrp{\sqrt{n}\widehat{U} \le x, \sqrt{n}|\widehat{U}-U^{\ast}|\geq\varepsilon\log^{-1/2}(n) \mid H_0} \nonumber\\
& & +  \blrp{\sqrt{n}\widehat{U} \le x, \sqrt{n}|\widehat{U}-U^{\ast}|<\varepsilon\log^{-1/2}(n) \mid H_0} \nonumber \\
&\le& \blrp{\sqrt{n}|\widehat{U}-U^{\ast}|\geq\varepsilon\log^{-1/2}(n) \mid H_0} 
 +  \blrp{\sqrt{n}U^{\ast}\le x+\varepsilon\log^{-1/2}(n) \mid H_0}. \label{Equ:S2_aux_1}
\end{eqnarray}
Similarly, we can also show that
\begin{equation*}
    \blrp{\sqrt{n}U^{\ast} \le x \mid H_0}
\le \blrp{\sqrt{n}|\widehat{U}-U^{\ast}|\geq\varepsilon\log^{-1/2}(n) \mid H_0} 
 +  \blrp{\sqrt{n}\widehat{U}\le x+\varepsilon\log^{-1/2}(n) \mid H_0}.
\end{equation*}
Since $x\in\br$ is arbitrary, we can replace $x$ with $x-\varepsilon\log^{-1/2}(n)$ and obtain that 
\begin{eqnarray}
& & \blrp{\sqrt{n}U^{\ast} \le x-\varepsilon\log^{-1/2}(n) \mid H_0} \nonumber\\
&\le& \blrp{\sqrt{n}|\widehat{U}-U^{\ast}|\geq\varepsilon\log^{-1/2}(n) \mid H_0} 
 +  \blrp{\sqrt{n}\widehat{U}\le x \mid H_0}, \label{Equ:S2_aux_2}
\end{eqnarray}
Equation (\ref{Equ:S2_aux_1}) and Equation (\ref{Equ:S2_aux_2}) jointly imply that
\BEqn
& & \lrabs{\blrp{\sqrt{n}\widehat{U}\le x \mid H_0} - \blrp{\sqrt{n}U^{\ast} \le x \mid H_0}} \\
&\le& \blrp{\sqrt{n}|\widehat{U}-U^{\ast}|\geq\varepsilon\log^{-1/2}(n) \mid H_0} \\
& & + \max\lrcp{
    \bbp{x<\sqrt{n}U^{\ast}\le x+\varepsilon\log^{-1/2}(n) \mid H_0},
    \bbp{x-\varepsilon\log^{-1/2}(n)<\sqrt{n}U^{\ast} \le x \mid H_0}} \\
&\le& \blrp{\sqrt{n}|\widehat{U}-U^{\ast}|\geq\varepsilon\log^{-1/2}(n) \mid H_0} 
    + \blrp{|\sqrt{n}U^{\ast} - x| \le \varepsilon\log^{-1/2}(n) \mid H_0}.
\EEqn
Note that this inequality holds for any $x\in\br$, then we further have that
\BEqn
& & \sup\limits_{x\in\br}
    \lrabs{ \blrp{\sqrt{n}\widehat{U} \le x \mid H_0} - \blrp{\sqrt{n} U^{\ast} \le x \mid H_0}} \\
&\le& \blrp{\sqrt{n}|\widehat{U}-U^{\ast}|\geq\varepsilon\log^{-1/2}(n) \mid H_0} 
    + \sup\limits_{x\in\br}\blrp{|\sqrt{n}U^{\ast} - x| \le \varepsilon\log^{-1/2}(n) \mid H_0} \\
&\le& \blrp{\sqrt{n}|\widehat{U}-U^{\ast}|\geq\varepsilon\log^{-1/2}(n) \mid H_0} \\    
& & + \sup\limits_{x\in\br}\lrabs{\blrp{|\sqrt{n}U^{\ast} - x| \le \varepsilon\log^{-1/2}(n) \mid H_0} - \blrp{\lrabs{\lrnorm{\bg(\phi_{f,g,h})}_{\Phi} - x} \le \varepsilon\log^{-1/2}(n)}} \\
& & + \sup\limits_{x\in\br} \blrp{\lrabs{\lrnorm{\bg(\phi_{f,g,h})}_{\Phi} - x} \le \varepsilon\log^{-1/2}(n)}.
\EEqn
It suffices to analyze each individual term.

We first consider $\blrp{\sqrt{n}|\widehat{U}-U^{\ast}|\geq\varepsilon\log^{-1/2}(n) \mid H_0}$. Under the stated assumptions, we have shown in Theorem \ref{Thm:Double-robustness} that under $H_0$,
\begin{equation*}
    \sqrt{n}|\widehat{U} - U^{\ast}| = O_p(n^{-(k_1+k_2-\frac{1}{2})}\log(n)).
\end{equation*}
If $k_1+k_2>\frac{1}{2}$, it holds that $n^{(k_1+k_2-\frac{1}{2})}$ dominates $\log^{3/2}(n)$ asymptotically and we further have that $\sqrt{n}|\widehat{U} - U^{\ast}| = o_p(\log^{-1/2}(n))$, i.e.
\begin{equation*}
    \blrp{\sqrt{n}|\widehat{U}-U^{\ast}| \geq \varepsilon\log^{-1/2}(n) \mid H_0} = o(1), \quad\forall\varepsilon>0.
\end{equation*}
For the second term, note that under Assumption \ref{Assumpt:kernel}--\ref{Assumpt:X1X2}, Theorem \ref{Thm:Validity} justifies that $\|P(\phi_{f,g,h})\|_{\Phi} = 0$ under the null, which implies that $U^{\ast} = \|\bp_n(\phi_{f,g,h})\|_{\Phi} = \|(\bp_n-P)(\phi_{f,g,h})\|_{\Phi}$ under the null. Then Lemma \ref{Lemma:S1} yields that the second term also has an order of $o(1)$, that is,
\begin{equation*}
    \sup\limits_{x\in\br}
    \lrabs{  \blrp{|\sqrt{n}U^{\ast} - x| 
                    \le \varepsilon\log^{-1/2}(n) \mid H_0}
           - \blrp{\lrabs{\lrnorm{\bg(\phi_{f,g,h})}_{\Phi} - x} 
                    \le \varepsilon\log^{-1/2}(n)}} 
  \le 2S_1 = o(1).
\end{equation*}
For the third term, it follows from Lemma \ref{Lemma:ContinuousCDF} that $\|\bg\|_{\Phi}$ has a continuous distribution function, which is naturally uniformly continuous on $\br$. Therefore
\BEqn
& & \sup\limits_{x\in\br}
    \blrp{\lrabs{\lrnorm{\bg(\phi_{f,g,h})}_{\Phi} - x} \le \varepsilon\log^{-1/2}(n) } \\
&=& \sup\limits_{x\in\br}
    \lrcp{  \blrp{\lrnorm{\bg(\phi_{f,g,h})}_{\Phi} \le x + \varepsilon\log^{-1/2}(n) }
          - \blrp{\lrnorm{\bg(\phi_{f,g,h})}_{\Phi} \le x - \varepsilon\log^{-1/2}(n) } } \\
&=& o(1).
\EEqn
Putting the three terms together, we conclude that $S_2 = o(1)$.
\end{Proof}

\begin{lemma}\label{Lemma:S3}
Under Assumption~\ref{Assumpt:kernel} and Assumption~\ref{Assumpt:entropy},\ref{Assumpt:continuity}, it holds that
\begin{equation*}
    S_3
 := \sup\limits_{x\in\br} 
    \lrabs{  \blrp{\sqrt{n}U^{b\ast} \le x \mid \cd, \cz^{c \ast}} 
           - \blrp{\lrnorm{\bg(\phi_{f,g,h})}_{\Phi} \le x}} = o_p(1),
\end{equation*}
where $\bg$ is the tight Gaussian process defined in Lemma \ref{Lemma:Donsker_Uast}.    
\end{lemma}
\begin{Proof}
Under Assumption \ref{Assumpt:kernel} and Assumption~\ref{Assumpt:entropy}, we have shown in Lemma \ref{Lemma:Donsker_Ubast} that $\sqrt{n}U^{b\ast} \leadsto_{\varepsilon}^{p} \lrnorm{\bg(\phi_{f,g,h})}_{\Phi}$, where $\bg$ denotes the tight Gaussian process given in Lemma \ref{Lemma:Donsker_Uast}. This implies that
\begin{equation*}
    \lrabs{\blrp{\sqrt{n}U^{b\ast} \le x \mid \cd, \cz^{c \ast}} 
    - \blrp{\lrnorm{\bg(\phi_{f,g,h})}_{\Phi} \le x}}
  = o_p(1), 
  \quad \forall~x\in\br.
\end{equation*}
Recall that we have shown in Lemma \ref{Lemma:ContinuousCDF} that $\lrnorm{\bg(\phi_{f,g,h})}_{\Phi}$ has a continuous distribution function over $\br$ if Assumption~\ref{Assumpt:continuity} is additionally satisfied, then by the conditional version of P\'{o}lya's theorem, the aforementioned result is uniform in $x\in\br$, i.e.
\begin{equation*}
    S_3
  = \sup\limits_{x\in\br} 
    \lrabs{ \blrp{\sqrt{n}U^{b\ast} \le x \mid \cd, \cz^{c \ast}} 
    - \blrp{\lrnorm{\bg(\phi_{f,g,h})}_{\Phi} \le x}} 
  = o_p(1),
\end{equation*}
which completes the proof.
\end{Proof}

\begin{lemma}\label{Lemma:S4}
Under Assumption~\ref{Assumpt:kernel} and Assumption \ref{Assumpt:entropy}--\ref{Assumpt:continuity}, it holds that
\begin{equation*}
    S_4
 := \sup\limits_{x\in\br} 
    \lrabs{  \blrp{\sqrt{n} U^{b\ast} \le x \mid \cd, \cz^{c \ast}} 
           - \blrp{\sqrt{n} \widehat{U}^{b} \le x \mid \cd, \cz}}
  = o_p(1).
\end{equation*} 
\end{lemma}
\begin{Proof}
Since $U^{b\ast}$ is measurable with respect to $(\cd,\cz^{c\ast},\varepsilon)$, and the multipliers $\varepsilon$ are generated independently of $(\cd,\cz,\cz^{c\ast})$, the conditional law of $U^{b\ast}$ given $(\cd,\cz^{c\ast})$ remains the same as that given $(\cd,\cz,\cz^{c\ast})$, that is,
\begin{equation*}
    \blrp{\sqrt{n} U^{b\ast} \le x \mid \cd, \cz^{c\ast}} 
  = \blrp{\sqrt{n} U^{b\ast} \le x \mid \cd, \cz, \cz^{c\ast}}.
\end{equation*}
Analogously, we have $\blrp{\sqrt{n} \widehat{U}^{b} \le x \mid \cd, \cz} = \blrp{\sqrt{n} \widehat{U}^{b} \le x \mid \cd, \cz, \cz^{c\ast}}$. Then it follows from simple calculations that
\BEqn
& & S_4 \\
&=& \sup\limits_{x\in\br} 
    \lrabs{ \blrp{\sqrt{n} U^{b\ast} \le x \mid \cd, \cz^{c \ast}} 
    - \blrp{\sqrt{n} \widehat{U}^{b} \le x \mid \cd, \cz}} \\
&=& \sup\limits_{x\in\br} 
    \lrabs{ \blrp{\sqrt{n} U^{b\ast} \le x \mid \cd, \cz, \cz^{c \ast}} 
    - \blrp{\sqrt{n} \widehat{U}^{b} \le x \mid \cd, \cz, \cz^{c \ast}}} \\
&\le& \blrp{\sqrt{n} |\widehat{U}^{b} - U^{b\ast}| \geq \varepsilon\log^{-1/2}(n) \mid \cd, \cz, \cz^{c \ast}} \\
& & + \sup\limits_{x\in\br}\lrabs{\blrp{|\sqrt{n} U^{b\ast} - x| \le \varepsilon \log^{-1/2}(n) \mid \cd, \cz, \cz^{c \ast}} - \blrp{\lrabs{\|\bg(\phi_{f,g,h})\|_{\Phi} - x} \le \varepsilon \log^{-1/2}(n)}} \\
& & + \sup\limits_{x\in\br} \blrp{\lrabs{\|\bg(\phi_{f,g,h})\|_{\Phi} - x} \le \varepsilon \log^{-1/2}(n)} \\
&=& \blrp{\sqrt{n} |\widehat{U}^{b} - U^{b\ast}| \geq \varepsilon \log^{-1/2}(n) \mid \cd, \cz, \cz^{c \ast}} + o_p(1),
\EEqn
where the first step leverages the fact that $U^{b\ast} \indept \cz \mid \cd, \cz^{c \ast}$ and $\widehat{U}^{b} \indept \cz^{c \ast} \mid \cd, \cz$, and the last step follows from Lemma \ref{Lemma:ContinuousCDF} and Lemma \ref{Lemma:S3}.

Under the stated assumptions, it follows from Proposition \ref{Prop:Double-robustness-boot} that 
\begin{equation*}
    \sqrt{n} \lrabs{\widehat{U}^b - U^{b\ast}} 
  = O_p(n^{-\min\{k_1,k_2\}/2} \log^{v}(n))
  = o_p(\log^{-1/2}(n)).
\end{equation*}
Then for any $\varepsilon > 0$, it follows from the law of total probability that
\begin{equation*}
    \blre{\blrp{\sqrt{n} |\widehat{U}^{b} - U^{b\ast}| \geq \varepsilon \log^{-1/2}(n) \mid \cd, \cz, \cz^{c \ast}}}
  = \blrp{\sqrt{n} |\widehat{U}^{b} - U^{b\ast}| \geq \varepsilon \log^{-1/2}(n)}
  = o(1).
\end{equation*}
By the Markov inequality, we obtain that
\begin{equation*}
    \blrp{\sqrt{n} |\widehat{U}^{b} - U^{b\ast}| \geq \varepsilon \log^{-1/2}(n) \mid \cd, \cz, \cz^{c \ast}} 
  = o_p(1),
\end{equation*}
which completes the proof.
\end{Proof}

\bibliographystyle{agsm}
\bibliography{ref}

\end{document}